\documentclass[twocolumn]{aastex62}

\usepackage{ulem}
\usepackage{natbib}
\usepackage{amsmath}
\usepackage{color}
\newcommand{\fdir}{./}
\newcommand{\teff}{\log(T_{\rm eff}/\mbox{K})}

\newcommand{\mzams}{M_{\rm ZAMS}}
\newcommand{\amin}{a_{\min}}
\newcommand{\qmin}{q_{\min}}
\newcommand{\mi}{m_{\rm 1,i}}
\newcommand{\qi}{q_{\rm i}}
\newcommand{\mbi}{m_{\rm bin,i}}
\newcommand{\ai}{a_{\rm i}}
\newcommand{\ei}{e_{\rm i}}
\newcommand{\rpi}{r_{\rm p,i}}
\newcommand{\mche}{M_{\rm c,He}}
\newcommand{\mbp}{m_{\rm b,p}}
\newcommand{\mbs}{m_{\rm b,s}}
\newcommand{\vecz}{\overrightarrow{0}}
\newcommand{\vecl}{\overrightarrow{L}}
\newcommand{\xbp}{\overrightarrow{\chi}_{\rm p}}
\newcommand{\xbs}{\overrightarrow{\chi}_{\rm s}}
\newcommand{\xeff}{\chi_{\rm eff}}
\newcommand{\msun}{M_\odot}
\newcommand{\lsun}{L_\odot}
\newcommand{\rsun}{R_\odot}
\newcommand{\zsun}{Z_\odot}
\newcommand{\td}{t_{\rm d}}
\newcommand{\tdi}{t_{\rm d,i}}
\newcommand{\tdf}{t_{\rm d,f}}

\begin{document}

\title{Merger rate density of Population III binary black holes below,
  above, and in the pair-instability mass gap}

\correspondingauthor{Ataru Tanikawa}
\email{tanikawa@ea.c.u-tokyo.ac.jp}

\author{Ataru Tanikawa}
\affiliation{Department of Earth Science and Astronomy, College of
  Arts and Sciences, The University of Tokyo, 3-8-1 Komaba, Meguro-ku,
  Tokyo 153-8902, Japan}

\author{Hajime Susa}
\affiliation{Department of Physics, Konan University, Kobe, Japan}

\author{Takashi Yoshida}
\affiliation{Department of Astronomy, Graduate School of Science, The
  University of Tokyo, Bunkyo-ku, Tokyo, Japan}

\author{Alessandro A. Trani}
\affiliation{Department of Earth Science and Astronomy, College of
  Arts and Sciences, The University of Tokyo, 3-8-1 Komaba, Meguro-ku,
  Tokyo 153-8902, Japan}

\author{Tomoya Kinugawa}
\affiliation{Institute for Cosmic Ray Research, The University of
  Tokyo, Kashiwa, Chiba}

\begin{abstract}

We present the merger rate density of Population (Pop.) III binary
black holes (BHs) by means of a widely-used binary population
synthesis code {\tt BSE} with extensions to very massive and extreme
metal-poor stars. We consider not only low-mass BHs (lBHs:
$5-50\msun$) but also high-mass BHs (hBHs: $130-200\msun$), where lBHs
and hBHs are below and above the pair-instability mass gap ($50-130
\msun$), respectively. Pop.~III BH-BHs can be categorized into three
subpopulations: BH-BHs without hBHs (hBH0s: $m_{\rm tot} \lesssim
100\msun$), with one hBH (hBH1s: $m_{\rm tot} \sim 130-260\msun$), and
with two hBHs (hBH2s: $m_{\rm tot} \sim 270-400\msun$), where $m_{\rm
  tot}$ is the total mass of a BH-BH. Their merger rate densities at
the current universe are $\sim 0.1$~yr$^{-1}$~Gpc$^{-3}$ for hBH0s,
and $\sim 0.01$~yr$^{-1}$~Gpc$^{-3}$ for the sum of hBH1s and hBH2s,
provided that the mass density of Pop.~III stars is $\sim 10^{13}
\msun$~Gpc$^{-3}$. These rates are modestly insensitive to initial
conditions and single star models. The hBH1 and hBH2 mergers can
dominate BH-BHs with hBHs discovered in near future. They have low
effective spins $\lesssim 0.2$ in the current universe. The number
ratio of the hBH2s to the hBH1s is high, $\gtrsim 0.1$. We also find
BHs in the mass gap (up to $\sim 85 \msun$) merge. These merger rates
can be reduced to nearly zero if Pop.~III binaries are always wide
($\gtrsim 100 \rsun$), and if Pop.~III stars always enter into
chemically homogeneous evolution. The presence of close Pop.~III
binaries ($\sim 10 \rsun$) are crucial for avoiding the worst
scenario.

\end{abstract}


\section{Introduction}
\label{sec:Introduction}

In 2015, gravitational wave (GW) observatories, LIGO, have first
detected GWs, and the GWs have come from a merger of two black holes
(BHs) \citep{2016PhRvL.116f1102A}. Since then, GW observatories, LIGO
and Virgo, have continuously discovered a large number of BH-BH
mergers, and have published $\sim 50$ BH-BH mergers with detail
information, such as their masses, distances, spin magnitudes, and
spin-orbit misalignments as of December 2020
\citep{2019PhRvX...9c1040A,2020PhRvD.102d3015A,2020ApJ...896L..44A,2020PhRvL.125j1102A,2020arXiv201014527A,2020arXiv201014533T}.
Moreover, research groups other than the LIGO-Virgo collaboration have
found other BH-BHs
\citep{2019arXiv191009528Z,2020PhRvD.101h3030V}. Many of them contain
BHs with $\sim 30 \msun$, significantly more massive than BHs in X-ray
binaries discovered previously, $\lesssim 15 \msun$
\citep[e.g.][]{2017hsn..book.1499C}. Thus, the origin(s) of merging
BH-BHs have been controversial.

Many formation scenarios of merging BH-BHs have been currently
suggested. Population (Pop.) I/II binaries evolve to merging BH-BHs
through common envelope evolution
\citep{1998ApJ...506..780B,2002ApJ...572..407B,2014ApJ...789..120B,2016Natur.534..512B,2020A&A...636A.104B,2012ApJ...759...52D,2013ApJ...779...72D,2014A&A...564A.134M,2015MNRAS.451.4086S,2019MNRAS.485..889S,2016MNRAS.462.3302E,2017PASA...34...58E,2019MNRAS.482..870E,2017MNRAS.472.2422M,2018MNRAS.479.4391M,2019MNRAS.487....2M,2017NatCo...814906S,2018MNRAS.480.2011G,2018MNRAS.481.1908K}. Close
binaries can also form merging BH-BHs through chemically homogeneous
evolution (CHE)
\citep{2009A&A...497..243D,2016MNRAS.458.2634M,2016A&A...588A..50M,2020MNRAS.499.5941D,2020arXiv201000002R,2020MNRAS.498.4287H}. Dynamical
interactions in globular clusters (GCs) raise merging BH-BHs
\citep{2000ApJ...528L..17P,2010MNRAS.407.1946D,2013MNRAS.435.1358T,2016PhRvD..93h4029R,2014MNRAS.440.2714B,2016ApJ...824L...8R,2018PhRvL.120o1101R,2018PhRvD..98l3005R,2017PASJ...69...94F,2017MNRAS.464L..36A,2017MNRAS.469.4665P,2017ApJ...840L..14S,2018PhRvD..97j3014S,2018ApJ...855..124S,2018MNRAS.480.5645H}. In
open clusters, merging BH-BHs are formed by combination of common
envelope evolution and dynamical interactions
\citep{2014MNRAS.441.3703Z,2016MNRAS.459.3432M,2017MNRAS.467..524B,2018MNRAS.473..909B,2018MNRAS.481.5123B,2019MNRAS.486.3942K,2020MNRAS.495.4268K,2019MNRAS.487.2947D,2020MNRAS.497.1043D,2020MNRAS.498..495D,2019MNRAS.483.1233R}. Dynamical
interactions can form merging BH-BHs also in galactic centers
\citep{2009MNRAS.395.2127O,2012ApJ...757...27A,2016ApJ...828...77V,2016ApJ...831..187A,2017ApJ...835..165B,2017ApJ...846..146P,2017MNRAS.464..946S,2018ApJ...856..140H,2018ApJ...865....2H,2018ApJ...866...66M,2018MNRAS.474.5672L,2019ApJ...876..122Y,2019ApJ...881...20R,2019ApJ...885..135T,2020ApJ...898...25T,2020ApJ...899...26T,2020MNRAS.498.4088M,2020ApJ...891...47A,2020arXiv200715022M}. Merging
BH-BHs can emerge through secular evolution in stable triple systems
(TSs), and quadruple systems (QSs)
\citep{2014ApJ...781...45A,2017ApJ...836...39S,2017ApJ...841...77A,2018ApJ...863....7R,2019MNRAS.483.4060L,2019MNRAS.486.4781F,2020ApJ...898...99H,2020ApJ...895L..15F}. Although
merging BH-BHs originate from astrophysical objects in the above
scenarios, even primordial BHs, made from dark matter, can generate
merging BH-BHs \citep{2016PhRvL.117f1101S}.

Pop.~III binary stars, which are formed from primordial gas at the
high-redshift universe, can also raise merging BH-BHs. The typical BH
mass is $\sim 30 \msun$, larger than Pop.~I/II BH-BHs
\citep{2014MNRAS.442.2963K}. This is not because Pop.~III stars have a
top-heavy initial mass function (IMF), but because Pop.~III binaries
tend to experience stable mass transfer owing to their radiative
envelopes \citep{2016MNRAS.456.1093K,2017MNRAS.468.5020I}. This
argument is also robust against uncertainties of binary evolutions,
such as efficiencies of mass transfer, mass accretion, common
envelope, and tidal interaction \citep{2020MNRAS.498.3946K}. Since the
typical BH mass is consistent with observed BH-BH masses, Pop.~III
binaries can be one of their promising origins.

It is uncertain whether Pop.~III BH-BHs are a dominant origin of
observed BH-BHs
\citep{2016MNRAS.460L..74H,2017MNRAS.471.4702B,2020MNRAS.498.3946K}. Moreover,
observed BH-BHs may be mainly formed through dynamical interactions
\citep{2020arXiv201109570C}. However, even if Pop.~III BH-BHs are not
dominant, it should be worth while to identify Pop.~III BH-BHs from
observed BH-BHs. Pop.~III stars are key ingredients of the cosmic
dawn, the reionization of the universe, the beginning of stellar
nucleosynthesis, and so on. Despite of their importance, Pop.~III
stars are not directly observed \citep{2013MNRAS.429.3658R}. Since
Pop.~III stars are typically massive stars, $10-1000 \msun$
\citep{1998ApJ...508..141O,2002Sci...295...93A,2004ARA&A..42...79B,2008Sci...321..669Y,2011Sci...334.1250H,2011MNRAS.413..543S,2012MNRAS.422..290S,2013RPPh...76k2901B,2013ApJ...773..185S,2014ApJ...792...32S,2015MNRAS.448..568H},
they are short-lived, and located at the high-redshift universe. They
can be observed only by ultimately large telescopes
\citep[e.g.][]{2020ApJ...904..145S}. Although some of Pop.~III stars
can be formed as low-mass and long-lived stars
\citep{2001ApJ...548...19N,2008ApJ...677..813M,2011ApJ...727..110C,2011Sci...331.1040C,2011ApJ...737...75G,2012MNRAS.424..399G,2013MNRAS.435.3283M,2014ApJ...792...32S,2016MNRAS.463.2781C},
they have not yet been observed
\citep{2015ARA&A..53..631F,2018MNRAS.473.5308M,2019MNRAS.487..486M}. Although
extreme metal-poor (EMP) stars might be Pop.~III stars metal-polluted
by accreting interstellar medium \citep{2015ApJ...808L..47K}, many
studies have suggested EMP stars are not polluted Pop.~III stars,
since such accretion can be blocked by stellar winds of low-mass
Pop.~III stars
\citep{2017ApJ...844..137T,2018PASJ...70...34S}. Interstellar
asteroids cannot pollute Pop.~III stars up to EMP stars
\citep{2018PASJ...70...80T,2019MNRAS.486.5917K}.

In this paper, we predict properties of Pop.~III BH-BHs to identify
them from BH-BHs observed by current and future GW
observatories. Although the current GW observatories can observe
BH-BHs within redshift ($z$) of $\lesssim 1$, future ground-based GW
observatories, such as Einstein telescope
\citep{2010CQGra..27s4002P,2020JCAP...03..050M} and Cosmic explorer
\citep{2019BAAS...51g..35R}, will detect BH-BHs up to $z \sim 10$. A
future space-borne GW observatory LISA \citep{2017arXiv170200786A} is
expected to catch Pop.~III BH-BHs
\citep[e.g.][]{2009ApJ...698L.129S}. Another space-borne GW
observatory DECIGO \citep{2006CQGra..23S.125K} can follow the cosmic
evolution of BH-BHs, and distinguish Pop.~III BH-BHs
\citep{2016PTEP.2016i3E01N}.

So far, \cite{2014MNRAS.442.2963K} have shown that Pop.~III BH-BHs
have typically $\sim 30 \msun$, and \cite{2020MNRAS.498.3946K} have
shown its robustness against binary star evolution models. In this
paper, we focus on the dependence of Pop.~III BH-BHs on initial
conditions and single star evolution models. Previous studies have
assumed that Pop.~III binaries can have initial separations comparable
to their stellar radii at their zero-age main-sequence (ZAMS)
times. As well as this case, we also consider Pop.~III binaries with
the minimum separation of $\sim 100 \rsun$. Pop.~III binaries may not
be initially compact, since Pop.~III stars can expand to $\gtrsim 100
\rsun$ at their protostellar phases
\citep{1986ApJ...302..590S,2001ApJ...561L..55O,2003ApJ...589..677O}. Although
previous studies have not taken into account stellar winds, we account
for stellar winds, which are excited by stellar rotations
\citep{2012A&A...542A.113Y}. Moreover, we set the maximum ZAMS mass to
$300 \msun$, while previous studies do to $150 \msun$. This maximum
mass is reasonable according to recent numerical simulations
\citep{2014ApJ...792...32S,2014ApJ...781...60H,2015MNRAS.448..568H}.
Since such massive stars can overcome pair instability supernovae
(PISNe)
\citep{1967PhRvL..18..379B,1968Ap&SS...2...96F,1984ApJ...280..825B,1986A&A...167..274E,2001ApJ...550..372F,2002ApJ...567..532H,2002ApJ...565..385U},
we can get BHs with $\gtrsim 100 \msun$. Eventually, we show
properties of Pop.~III BH-BHs with several $10$ and $100\msun$, and
discuss how to identify them from other BH-BHs by current and future
GW observatories.

\cite{2004ApJ...608L..45B} have investigated the formation of BH-BHs
from Pop.~III stars with $100-500 \msun$. Due to the IMF, they have
not investigated a BH-BH consisting of a BH overcoming PISN, and a BH
not overcoming PISN. Moreover, studies of Pop.~III stars have rapidly
developed in this decade \citep[see][for
  review]{2018PhR...780....1D}. We can reflect these achievements.
\cite{2020MNRAS.495.2475L} and \cite{2020MNRAS.tmp.3458L} have studied
Pop.~III BH-BHs formed by dynamical interactions. On the other hand,
we focus on Pop.~III BH-BHs formed through binary evolution. These
studies should be complementary.

The structure of this paper is as follows. In
section~\ref{sec:Method}, we describe our method. In
section~\ref{sec:Results}, we show the calculation results. In
section~\ref{sec:Identification}, we consider how to identify Pop.~III
BH-BHs from other BH-BHs observed by current and future GW
observatories. In section~\ref{sec:Discussion}, we discuss about
effects we do not examine in section~\ref{sec:Results}. In
section~\ref{sec:Summary}, we summarize this paper.

\section{Method}
\label{sec:Method}

We perform calculations of binary population synthesis by means of the
{\tt BSE} code \citep{2000MNRAS.315..543H,2002MNRAS.329..897H} with
extensions to very massive and EMP stars including Pop.~III stars. In
sections~\ref{sec:SingleStarModel} and \ref{sec:BinaryStarModel}, we
present single and binary star models in our {\tt BSE} code,
respectively. In section~\ref{sec:InitialConditions}, we show our
initial conditions of binaries. In section~\ref{sec:ParameterSets}, we
describe parameter sets of our calculation runs. In
section~\ref{sec:PopIIIFormationModel}, we give our Pop.~III formation
model to derive the merger rate of Pop.~III BH-BHs.

\subsection{Single star model}
\label{sec:SingleStarModel}

\begin{figure}[ht!]
  \plotone{\fdir/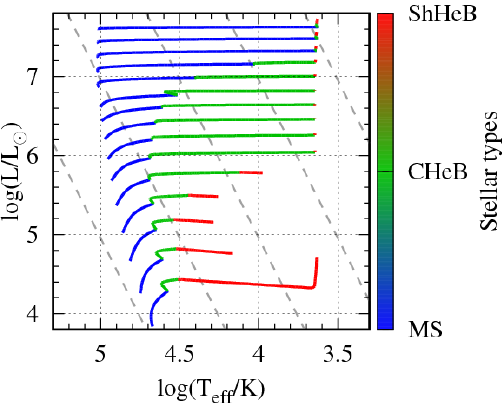}
  \caption{Hertzsprung-Russell diagram of $10^{-8}\zsun$ stars with
    $M=10-1280\msun$ at intervals of $2^{1/2}$ from bottom to
    top. Stellar winds are not taken into account. Gray dashed lines
    indicate stellar radii of $1, 10, 10^2, 10^3$, and $10^4 \rsun$
    from left to right. Colors indicate stellar phases: MS (blue),
    CHeB (green), and ShHeB (red).}
  \label{fig:hrd}
\end{figure}

We use fitting formulae for stars with stellar metallicity $Z=10^{-8}
\zsun$ incorporated into {\tt BSE} \citep{2020MNRAS.495.4170T}. The
fitting formulae are based on simulation results of $Z=10^{-8} \zsun$
stars from $8 \msun$ to $1280 \msun$ by means of the {\tt HOSHI} code
\citep{2016MNRAS.456.1320T,2018ApJ...857..111T,Takahashi19,Yoshida19}. They
consist of luminosity ($L$), radius ($R$), helium (He) core mass
($M_{\rm c,He}$), and carbon-oxygen (CO) core mass ($M_{\rm c,CO}$) as
functions of time and stellar mass ($M$). They follow stellar
evolutions from the ZAMS time to the carbon ignition time during which
stars experience main-sequence (MS), core He-burning (CHeB), and shell
He-burning (ShHeB) phases. If stars lose their hydrogen (H) envelopes
through stellar wind mass loss and binary interactions, we adopt
fitting formulae of naked He (nHe) stars derived in
\cite{2000MNRAS.315..543H} for their luminosities, and set
$R=0.2239(\mche/\msun)^{0.62} \rsun$ for their radii
\citep{1997MNRAS.291..732T}.  Figure~\ref{fig:hrd} shows
Hertzsprung-Russell diagram of $10^{-8}\zsun$ stars with $M=10-1280
\msun$ at intervals of $2^{1/2}$, where we do not account for stellar
winds. Stars with $10 \lesssim M/\msun \lesssim 50$ end their lives
with blue-supergiant (BSG) stars with temperature of $\gtrsim
10^{3.65}$~K. Stars with $50 \lesssim M/\msun \lesssim 640$ become
red-supergiant (RSG) stars when they are in their CHeB/ShHeB (or
post-MS) phases. Stars with $M/\msun \gtrsim 640$ are still in their
MS phases, when they become RSG stars.

\begin{figure}[ht!]
  \plotone{\fdir/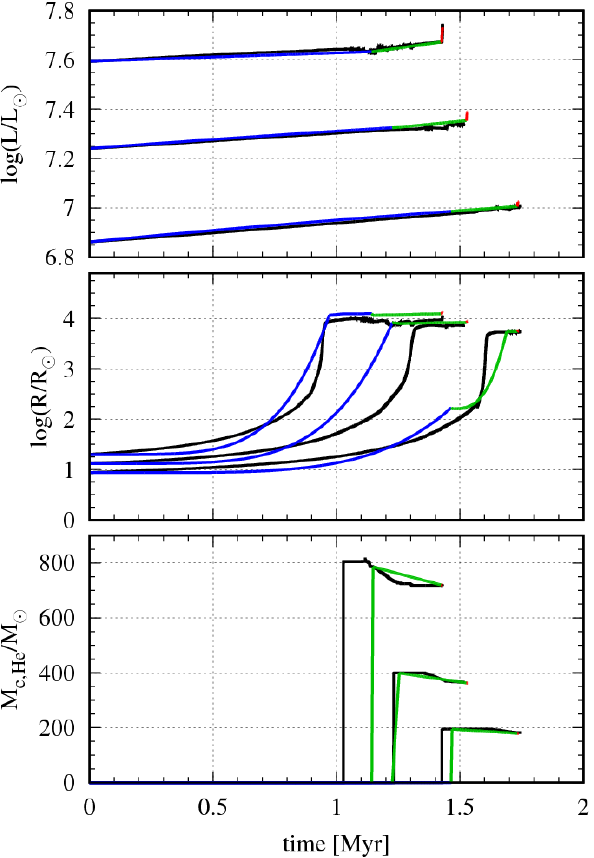}
  \caption{Time evolution of luminosities, radii, and He core masses
    for stars with $M=320, 640$, and $1280\msun$. Black curves
    indicate data of detail simulations, and colored curves indicate
    our fitting formulae. The color codes are the same as in
    Figure~\ref{fig:hrd}. More massive stars have larger values for
    all the quantities.}
  \label{fig:comparison}
\end{figure}

\cite{2020MNRAS.495.4170T} have devised the fitting formulae of stars
with $M \le 160 \msun$, and have compared the fitting formulae with
detail simulation results obtained by the {\tt HOSHI} code. Here, we
make similar comparisons for $M/\msun$ = 320, 640, and 1280
in Figure~\ref{fig:comparison}. The luminosities and He core masses
are in good agreement with each other. The He core masses of the
fitting formulae grow later than those of the simulation results. This
is because we assume the He core masses are zero when stars are in
their MS phases. The radii of the fitting formulae deviate from those
of the simulation results in the middle of the evolution. However,
these deviations should have small effects on binary evolutions, since
the minimum and maximum radii for each MS, CHeB and ShHeB phase are
consistent with each other. In particular, it is important that the
maximum radii of stars (i.e. the maximum radii of ShHeB phases) are
consistent between the fitting formulae and simulation results. The
maximum radii decide whether binary stars interact with each other, or
not.

We take into account stellar wind mass loss in a post-processing
way. The wind model is based on \cite{2010ApJ...714.1217B} with
modification of luminous-blue-variable (LBV) wind, and enhancement
through stellar rotations. For no rotating stars, the stellar wind
mass loss is expressed as
\begin{align}
  &\dot{M}_{\rm wind} \nonumber \\
  &= \left\{
  \begin{array}{ll}
    \max(\dot{M}_{\rm NJ},\dot{M}_{\rm OB}) & (\mbox{MS}) \\
    \max(\dot{M}_{\rm NJ},\dot{M}_{\rm OB},\dot{M}_{\rm
      R},\dot{M}_{\rm WR},\dot{M}_{\rm LBV}) & (\mbox{CHeB}) \\
    \max(\dot{M}_{\rm NJ},\dot{M}_{\rm OB},\dot{M}_{\rm
      R},\dot{M}_{\rm WR},\dot{M}_{\rm LBV},\dot{M}_{\rm VW}) &
    (\mbox{ShHeB}) \\
    \max(\dot{M}_{\rm R},\dot{M}_{\rm WR}) & (\mbox{nHe}) \\
  \end{array}
  \right.,
\end{align}
where $\dot{M}_{\rm NJ},\dot{M}_{\rm OB},\dot{M}_{\rm R},\dot{M}_{\rm
  WR},\dot{M}_{\rm LBV}$, and $\dot{M}_{\rm VW}$ are mass loss of
luminous stars \citep{Nieuwenhuijzen90,1989A&A...219..205K}, hot
massive H-rich stars \citep{Vink01}, giant-branch stars
\citep{1978A&A....70..227K,1983ARA&A..21..271I}, Wolf-Rayet stars
\citep{1998A&A...335.1003H,2005A&A...442..587V}, LBV stars
\citep{1994PASP..106.1025H}, and asymptotic giant branch (AGB) stars
\citep{1993ApJ...413..641V}. The expressions of $\dot{M}_{\rm
  NJ},\dot{M}_{\rm OB},\dot{M}_{\rm R},\dot{M}_{\rm WR}$, and
$\dot{M}_{\rm VW}$ are the same as Eqs.~(56), (57), (58), (59), and
(61) in \cite{2020MNRAS.495.4170T}. We modify the expression of
$\dot{M}_{\rm LBV}$ as
\begin{align}
  &\frac{\dot{M}_{\rm LBV}}{\msun~\mbox{yr}^{-1}} \nonumber \\
  &= \left\{
  \begin{array}{ll}
    1.5 \times 10^{-4} (Z/\zsun)^{0.86} & (L > 6
    \times 10^5\lsun, \mbox{~and~} x_{\rm LBV} > 1) \\
    0 & (\mbox{otherwise}),
  \end{array}
  \right.,
\end{align}
where $x_{\rm LBV} = 10^{-5}(R/\rsun)(L/\lsun)^{0.5}$. The expression
is the same as \cite{2010ApJ...714.1217B} except for the metallicity
dependence, and the metallicity dependence is the same as {\tt MOBSE2}
\citep{2018MNRAS.474.2959G}.

The stellar wind can be enhanced by stellar rotation. We include the
enhancement as follows:
\begin{align}
  &\dot{M}_{\rm wind,rot} = \min \left[ \dot{M}_{\rm wind} \left( 1 -
    v_{\rm rot}/v_{\rm crit} \right)^{-0.43}, 0.1M/t_{\rm KH} \right] \\
  &v_{\rm crit} = \sqrt{GM(1-L/L_{\rm Edd})/R}
\end{align}
\citep{2010ApJ...725..940Y,2012A&A...542A.113Y,2018ApJ...857..111T},
where G, $v_{\rm rot}$, $t_{\rm KH}$, and $L_{\rm Edd}$ are the
gravitational constant, the stellar surface rotation velocity,
Kelvin-Helmholtz timescale, and Eddington luminosity, respectively. We
adopt $L_{\rm Edd} = 3.7 \times 10^4 (M/\msun) \lsun$ for stars in MS,
CHeB, and ShHeB phases.  We put an upper limit on $L$ for nHe stars,
such that $L \le 0.99L_{\rm Edd}$, where we adopt $L_{\rm Edd} = 6.3
\times 10^4 (M/\msun) \lsun$ for nHe stars. Note that $L_{\rm Edd}$ of
nHe stars is larger than those of MS, CHeB, and ShHeB stars due to
smaller opacity of electron scattering.

We put such an upper limit on luminosities of nHe stars for the
following reason. We infer luminosity of nHe stars from our fitting
formulae of ShHeB stars. Since a ShHeB star with $M=320 \msun$ has a
He core with $180 \msun$, and its luminosity is at most $10^7 \lsun$
(see Figure~\ref{fig:comparison}), a nHe star with $180 \msun$ should
have luminosity of $10^7 \lsun$. Thus, this nHe star should not exceed
the Eddington luminosity of nHe stars. When we apply this inference to
stars with various $M$, we find that nHe stars with $\lesssim 300
\msun$ do not exceed the Eddington luminosity according to our fitting
formulae. We have confirmed this inference by performing simulations
of non-rotating nHe stars with $50$, $100$, $200$, and $320
\msun$ and a $200 \msun$ star with the initial rotating
  velocity of $0.1 v_k$, where $v_k = \sqrt{GM/R}$ is the Kepler
  velocity, with the {\tt HOSHI} code. These stars do not exceed
their Eddington limits, and lose little mass through rotationally
enhanced stellar winds.  On the other hand, the fitting formulae of
nHe stars in \cite{2000MNRAS.315..543H} are made from nHe star models
with $M=0.32-10 \msun$, and nHe stars with $\gtrsim 100 \msun$ exceed
their Eddington luminosity.  If we adopt this luminosity, we
overestimate the rotationally enhanced mass loss. Hence, we constrain
luminosity of nHe stars, such that $L \le 0.99L_{\rm Edd}$.

Stars are assumed to experience supernovae or direct collapse
immediately after the carbon ignition, and leave stellar remnants like
NSs and BHs, or no remnants. We adopt the rapid model in
\cite{2012ApJ...749...91F} modified by pulsational pair instability
(PPI)
\citep{2002ApJ...567..532H,2007Natur.450..390W,2016MNRAS.457..351Y,2017ApJ...836..244W,2019ApJ...887...72L}
and pair instability supernovae (PISNe)
\citep{1967PhRvL..18..379B,1968Ap&SS...2...96F,1984ApJ...280..825B,1986A&A...167..274E,2001ApJ...550..372F,2002ApJ...567..532H,2002ApJ...565..385U}. Then,
stellar remnant masses can be written as
\begin{align}
  &M_{\rm rem} \nonumber \\
  &= \left\{
  \begin{array}{ll}
    M_{\rm rapid}    & (M_{\rm c,He} \le M_{\rm c,He,PPI}) \\
    M_{\rm c,He,PPI} & (M_{\rm c,He,PPI} < M_{\rm c,He} \le M_{\rm c,He,PISN}) \\
    0                & (M_{\rm c,He,PISN} < M_{\rm c,He} \le M_{\rm c,He,DC}) \\
    M_{\rm rapid}    & (M_{\rm c,He} > M_{\rm c,He,DC}) \\
  \end{array}
  \right., \label{eq:PairInstability}
\end{align}
and
\begin{align}
  &\frac{M_{\rm rapid}}{\msun} \nonumber \\
  &= \left\{
  \begin{array}{ll}
    1.2 & (M_{\rm c,CO}/\msun \le 2.5) \\
    0.286M_{\rm c,CO} + 0.486 & (2.5 < M_{\rm c,CO}/\msun \le 6) \\
    M_{\rm t} & (6 < M_{\rm c,CO}/\msun \le 7) \\
    M_{\rm t} - M_{\rm ej} & (7 < M_{\rm c,CO}/\msun \le 11) \\
    M_{\rm t} & (M_{\rm c,CO}/\msun > 11)
  \end{array}
  \right. \label{eq:RapidModel} \\
  &\frac{M_{\rm ej}}{\msun} = [0.25(M_{\rm t}-1) - 1.275](11-M_{\rm
    c,CO}). \label{eq:RapidEjection}
\end{align}
Here, Eq.~(\ref{eq:PairInstability}) indicates PPI/PISN
corrections. The PPI/PISN corrections consist of four regimes:
core-collapse supernovae (or direct collapse), PPI, PISNe, and direct
collapse in the ascending order of He core mass. We adopt the
boundaries of these regimes, such that the minimum He core mass of PPI
($M_{\rm c,He,PPI}=45\msun$), the minimum He core mass of PISN
($M_{\rm c,He,PISN}=65\msun$), and the minimum He core mass of direct
collapse ($M_{\rm c,He,DC}=135\msun$) in the same way as
\cite{2016A&A...594A..97B}. We assume stars undergoing PPI leave BHs
with masses of $M_{\rm c,He,PPI}$. Eqs.~(\ref{eq:RapidModel}) and
(\ref{eq:RapidEjection}) show the rapid model, where $M_{\rm t}$ is
the total stellar mass. We regard stellar remnants as NSs and BHs if
their masses are below and above $3\msun$, respectively. The lower
mass limit of NSs is defined to be $1.2\msun$. If a stellar remnant
has its mass of $< 1.2 \msun$, it is a white dwarf, however such a
stellar remnant does not appear in this paper.

We account for a natal kick when massive stars leave NSs and BHs. The
natal kicks are randomly oriented, and have Maxwellian velocity
distribution with $\sigma_{\rm k}$. We assume that $\sigma_{\rm k}$
does not depend on the remnant mass, contrary to the often used
prescription which attributes to higher mass remnant a lower kick
velocity. In this paper, NSs and BHs acquire the same kick as
motivated in section~\ref{sec:ParameterSets}.

We model BH spins as follows. BH progenitors get and lose their spin
angular momenta through single and binary evolutions, such as stellar
winds, wind accretion, tidal interactions, Roche lobe overflow, and so
on. Then, the spin vectors are always parallel to the binary orbit
vectors. BH progenitors keep their spin angular momenta when they
become BHs, except that they experience PPI. If BH progenitors
experience PPI, we force the spins of their remnants to be zero, since
these progenitors lose large amounts of masses. Frequently, the BH
progenitors have spin angular momenta larger than those of extreme
Kerr BHs. In this case, we force the spin parameters of their remnants
to be unity, and assume that the BH masses are equal to their
progenitor masses. Although BHs can get spin angular momenta through
mass transfer, we do not account for this process. In summary, we can
express a normalized BH spin vector as follows:
\begin{align}
  \displaystyle \overrightarrow{\chi} = \left\{
  \begin{array}{ll}
    \vecz & \mbox{if PPI} \\
    \min \left( |\overrightarrow{S}|/(GM^2/c), 1 \right) \vecl/|\vecl|
    & \mbox{otherwise}
  \end{array}
  \right.,
\end{align}
where $c$ is the speed of light, $\overrightarrow{S}$ is the spin
angular momentum of a BH progenitor just before its collapse, and
$\vecl$ is the binary orbit vector. Natal kicks can tilt
$\overrightarrow{\chi}$ from $\vecl$ as described in
section~\ref{sec:BinaryStarModel}.

\subsection{Binary star model}
\label{sec:BinaryStarModel}

Since we use the {\tt BSE} code, our binary star model is basically
similar to that of \cite{2002MNRAS.329..897H} with some
modifications. In this section, we briefly describe our binary star
model.

Before describing our binary star model, we define names of binary
members. We call the heavier and lighter stars at their ZAMS times
``1st-evolving star'' and ``2nd-evolving star'', respectively, since
the former evolves earlier than the latter. We attach subscripts of
$1$ and $2$ with quantities indicating 1st- and 2nd-evolving stars,
respectively. For example, masses of 1st- and 2nd-evolving stars are
$m_1$ and $m_2$. We define the 1st and 2nd BHs as BHs which the 1st-
and 2nd-evolving stars leave, respectively. The 1st BH is not always
heavier than the 2nd BH. We call the heavier and lighter BHs ``primary
BH'' and ``secondary BH'', respectively. We attach subscripts of ``p''
and ``s'' with quantities indicating primary and secondary BHs. Thus,
masses of primary and secondary BHs are $\mbp$ and $\mbs$,
respectively.

We take into account wind accretion through which a star gets mass
outflowing from its companion through stellar winds. In the original
{\tt BSE} code, the accretion rate is estimated as a Bondi-Hoyle
mechanism \citep{1944MNRAS.104..273B} with an upper limit of a
fraction of the mass loss rate of the stellar winds. We find that the
accretion rate sometimes exceeds the Eddington limit due to the
presence of massive stellar winds and massive BHs. Thus, we put an
upper limit on the wind accretion rate, and set the limit to the
Eddington limit expressed by \cite{1967Natur.215..464C}. Through
stellar winds and wind accretion, stars lose and gain their spin
angular momenta, and binaries change their semi-major axes and
eccentricities. Our treatment for these processes is the same as that
of the original {\tt BSE} code.

Our prescriptions for tidal evolution are the same as the original
{\tt BSE} code. We adopt the equilibrium tide with convective damping
for stars with convective envelopes, and the dynamical tide with
radiative damping for stars with radiative envelopes. However, the
difference between the original {\tt BSE} code and our binary star
model is which post-MS stars (here, CHeB and ShHeB stars) have which
types of envelopes. In the original {\tt BSE} code, CHeB and ShHeB
stars have radiative and convective envelopes, respectively, where
ShHeB stars are called AGB stars in the original {\tt BSE} code. In
our binary model, stars with $\teff \ge 3.65$ and $<3.65$ have
radiative and convective envelopes, respectively. We need the above
treatment, since CHeB stars can have convective envelopes, and ShHeB
stars can have radiative envelopes in our single star model.

We also treat mass transfer in the same way as the original {\tt BSE}
code. If mass transfer is stable, a binary experiences stable mass
transfer (or stable Roche-lobe overflow), and otherwise common
envelope evolution. Its stability strongly depends on whether a donor
star has radiative or convective envelopes. If a donor star has a
radiative (convective) envelope, mass transfer tends to be stable
(unstable). We determine which post-MS stars have which types of
envelopes in the same way as the case of tidal evolution described
above. The common envelope evolution is modeled as the $\alpha$
formalism \citep[e.g.][]{1984ApJ...277..355W}. Thus, we have to decide
the common envelope efficiency $\alpha_{\rm CE}$, and the structural
binding energy parameter of a star $\lambda_{\rm CE}$. We set
$\alpha_{\rm CE}=1$ and $\lambda_{\rm CE}=1$, if unspecified.

We consider orbital decay through GW radiation in the same way as the
original {\tt BSE} code. On the other hand, we switch off magnetic
braking. It is difficult to expect the B-field configuration of
  Pop~III binaries. In the state-of-art Pop~III formation simulation
  \citep{2020MNRAS.497..336S}, the dominant component of the magnetic
  field is tangled.  In such a case, the effect of magnetic breaking
  is weak, so we omit the process.

BH spin vectors can be misaligned to binary orbit vectors due to the
natal kicks. We record angles between BH spin and binary orbit
vectors, where the binary orbit vectors are the ones when BH-BHs are
formed. The angle between the 2nd BH spin and binary orbit vectors
($\theta_2$) can be decided only by the kick vector and binary
parameters at the birth time. On the other hand, to determine the
angle between the 1st BH spin and binary orbit vectors ($\theta_1$),
we need to know differences between binary phases at the birth times
of the 1st and 2nd BHs. Although we can know them, we choose the
differences by Monte Carlo technique. Since binary periods are
generally much smaller than time intervals between the birth times, it
is good approximation to choose the differences of phases randomly.
Then, the angles $\theta_1$ and $\theta_2$ can be given by
\begin{align}
  \cos \theta_1 &=  \overrightarrow{\chi_1} \cdot
  \vecl/|\vecl| \\
  \cos \theta_2 &= \overrightarrow{\chi_2} \cdot
  \vecl/|\vecl|,
\end{align}
where the ``$\cdot$'' operator means the inner product,
$\overrightarrow{\chi_1}$ and $\overrightarrow{\chi_2}$ are normalized
BH spin vectors of the 1st and 2nd BHs, respectively, and $\vecl$ is
the binary orbit vector of the final BH-BHs. Note that $\vecl$ can be
obtained from the natal kick velocity of the 2nd BH and the binary
parameter at its birth time as described above. We suppose the
coordinate in which $z$-axis is parallel to the binary orbit vector
just before the 2nd BH is born. Then, $\overrightarrow{\chi_1}$ and
$\overrightarrow{\chi_2}$ can be written as
\begin{align}
  \overrightarrow{\chi_1} &= (\sin \theta_1' \cos \phi'_1 , \sin
  \theta_1' \sin \phi'_2, \cos \theta_1') \\
  \overrightarrow{\chi_2} &= (0, 0, 1),
\end{align}
where $\theta'_1$ is the angle between the 1st BH spin vector and
binary orbit vector just before the 2nd BH is born, and $\phi'_1$ is
randomly chosen between $0$ and $2\pi$. The spin vector of the 2nd BH
should be parallel to the binary orbit vector just before its birth,
since we do not account for spin-orbit misalignment mechanism other
than natal kicks. Note that $\theta'_1$ can be determined by the natal
kick vector of the 1st BH and binary parameters at the birth time of
the 1st BH.

\subsection{Initial conditions}
\label{sec:InitialConditions}

To generate a bunch of binaries, we set distribution of 1st-evolving
star's masses ($\mi$), mass ratios of 2nd-evolving stars to
1st-evolving stars ($\qi=m_{\rm 2,i}/\mi$), semi-major axes ($\ai$),
and eccentricities ($\ei$) at the initial time as follows. The
distribution of $\mi$ is given by
\begin{align}
  f(\mi) \propto \mi^{-1} \; (10 \le \mi/\msun \le 300).
\end{align}
This is based on logarithmically flat mass distribution of Pop.~III
stars in the range from about $10 \msun$ to several $100 \msun$
\citep{2014ApJ...792...32S,2014ApJ...781...60H,2015MNRAS.448..568H}.
Note that although we set the maximum mass to $300 \msun$, the maximum
mass can be smaller \citep[e.g.][]{2020ApJ...897...58T}.

We set the distribution of $\qi$, such that
\begin{align}
  f(\qi) &\propto {\rm const} \; (q'_{\min} \le \qi \le 1) \\
  q'_{\min} &= \max(\qmin, 10\msun/\mi).
\end{align}
The second equation means the minimum 2nd-evolving star's mass is $10
\msun$, the same as the minimum mass of 1st-evolving stars. We
determine $\qmin$ in section~\ref{sec:ParameterSets}.

The average mass of binaries at the initial time, $\langle \mbi
\rangle$, can be calculated as
\begin{align}
  \frac{\langle \mbi \rangle}{\msun} &= \int_{10}^{300} d\mi f(\mi)
  \int_{\qmin}^{1} d\qi f(\qi) \mi (1+\qi) \nonumber \\
&\sim 43 (3 + \qmin). \label{eq:AverageBinaryMass}
\end{align}

The semi-major axis distribution can be written as
\begin{align}
  f(\ai) &\propto \ai^{-1} \; (a'_{\min} \le \ai/\rsun \le 2000) \\
  a'_{\min} &= \max \left[ \amin,
    \frac{0.6\qi^{2/3}+\log(1+\qi^{1/3})}{0.49\qi^{2/3}} R_{\rm
      1,i}/\rsun \right],
\end{align}
where $R_{\rm 1,i}$ is the initial radius of the 1st-evolving star.
The maximum value 2000$\rsun$ can be obtained from the simulation
results of Figure~5 in \cite{TanakaPTEP20} in preparation. The maximal
final semi-major axis of the binaries at $\gtrsim 2\times 10^4$ yr is
$\sim 10$ AU, after which the radiative feedback quenches further gas
accretion. This is a spatially coarse calculation, thus this could be
used as the upper bound of the separation of Pop~III binaries. Owing
to the second equation, binaries can avoid Roche-lobe overflow from
the beginning time. We choose $\amin$ in
section~\ref{sec:ParameterSets}.

The eccentricity distribution is the thermal distribution with
modification as
\begin{align}
  &f(\ei) \propto \ei \; (0 \le \ei \le e_{\max}) \\
  &e_{\max} = 1-a'_{\min}/\ai
\end{align}
The modification makes pericenter distance larger than $a'_{\min}$ at
the initial time. Note that $e_{\max}=1$ if the eccentricity
distribution is completely thermal.

Each star has nearly no rotation at the initial time.

\subsection{Parameter sets}
\label{sec:ParameterSets}

We prepare a bunch of parameter sets summarized in
Table~\ref{tab:ParameterSets}. We mainly focus on $16$ parameter sets
(hereafter, the main $16$ models) in which we consider non- and
$Z=10^{-3} \zsun$ stellar winds, and $\sigma_{\rm k}=0,
265$~km~s$^{-1}$, $\qmin=0.0, 0.9$, and $\amin=10, 200\rsun$. We
suppose that these stellar winds are weak and strong extremes. The
natal kick velocities of $\sigma_{\rm k}=0$ and $265$~km~s$^{-1}$ are
also weak and strong extremes. We adopt $\sigma_{\rm
  k}=265$~km~s$^{-1}$ obtained in \cite{2005MNRAS.360..974H}.  We set
$\qmin=0.0$ and $0.9$. We can get variety of mass combinations for
$\qmin=0.0$, while we restrict the mass combinations for $\qmin=0.9$
due to the fact that the mass ratios of binaries tend to be close to
unity \citep{2019ApJ...877...99S}. We use $\amin=200\rsun$, since
Pop.~III stars can expand to $\gtrsim 100 \rsun$ at their protostellar
phases
\citep{1986ApJ...302..590S,2001ApJ...561L..55O,2003ApJ...589..677O},
and binaries with $a \lesssim 100 \rsun$ can merge before they enter
into their MS phases. Nevertheless, we also choose $\amin = 10\rsun$.
This is because they may avoid their mergers for unknown reasons.  In
fact, the evolution of the protostellar radius in the proto-binary
system is not known, although only a single star evolution has been
investigated so far\, assuming spherically symmetric mass accretion
\citep{1986ApJ...302..590S,2001ApJ...561L..55O,2003ApJ...589..677O}.
It is also worth noting that such close binaries with $a \lesssim 100
\rsun$ may be formed through dynamical processes after they enter into
their MS phases.  The model names are named after w/ and w/o wind, w/
and w/o kick, and the values of $\qmin$ and $\amin$. The ``opti'' and
``pess'' stand for ``optimistic'' and ``pessimistic'',
respectively. The ``opti'' and ``pess'' models are thought to form
large and small numbers of merging BH-BHs, respectively.

Note that the natal kick velocities are smaller for larger remnant
masses in the prescription that is usually adopted in the
literature. However, we do not adopt here this prescription. In the
commonly used prescription, kick velocities are set to $\sigma_{\rm k}
(M_{\rm rem}/M_{\rm rem,0})^{-1}$, where $\sigma_{\rm k} =
265$~km~s$^{-1}$ and $M_{\rm rem,0} = 3.0 \msun$
\citep[e.g.][]{2016MNRAS.457.1015W}. BHs with $\sim 10 \msun$ receive
kick velocities of $\sim 80$~km~s$^{-1}$. On the other hand, BH-BHs
with two $\sim 10 \msun$ BHs have orbital velocities of $\gtrsim
400$~km~s$^{-1}$ if they merge within the Hubble time. Thus, such BH
natal kicks have small effects on merging BH-BHs within the Hubble
time. The effects becomes smaller for more massive BHs. Even if we
adopt the usual prescription, we will get similar results to those of
models without BH natal kicks.  Moreover, the usual prescription may
be hard to explain several X-ray binaries with BHs
\citep{2005ApJ...625..324W,2012ApJ...747..111W,2015MNRAS.453.3341R,2017MNRAS.467..298R,2016MNRAS.456..578M,2019MNRAS.485.2642G,2020MNRAS.496L..22G}. The
spin-orbit misalignment of GW~190412 needs natal kicks of several
$100$~km~s$^{-1}$, if GW~190412 is formed through binary evolution
\citep{2020ApJ...901L..39O,2020ApJ...897L...7S}. There are several
theoretical studies for natal kicks with high velocities
\citep{2019arXiv190404835B,2010MNRAS.401.1644B}.

\begin{deluxetable}{l|rrrr}
  \tablecaption{Summary of parameter sets. \label{tab:ParameterSets}}
  \tablehead{Model & Wind & $\sigma_{\rm k}$ & $\qmin$ & $\amin$}
\startdata
optiq0.0a1e1 & w/o              & $0$   & $0.0$ & $10$  \\
optiq0.0a2e2 & w/o              & $0$   & $0.0$ & $200$ \\
optiq0.9a1e1 & w/o              & $0$   & $0.9$ & $10$  \\
optiq0.9a2e2 & w/o              & $0$   & $0.9$ & $200$ \\
kickq0.0a1e1 & w/o              & $265$ & $0.0$ & $10$  \\
kickq0.0a2e2 & w/o              & $265$ & $0.0$ & $200$ \\
kickq0.9a1e1 & w/o              & $265$ & $0.9$ & $10$  \\
kickq0.9a2e2 & w/o              & $265$ & $0.9$ & $200$ \\
windq0.0a1e1 & $10^{-3} \zsun$ & $0$   & $0.0$ & $10$  \\
windq0.0a2e2 & $10^{-3} \zsun$ & $0$   & $0.0$ & $200$ \\
windq0.9a1e1 & $10^{-3} \zsun$ & $0$   & $0.9$ & $10$  \\
windq0.9a2e2 & $10^{-3} \zsun$ & $0$   & $0.9$ & $200$ \\
pessq0.0a1e1 & $10^{-3} \zsun$ & $265$ & $0.0$ & $10$  \\
pessq0.0a2e2 & $10^{-3} \zsun$ & $265$ & $0.0$ & $200$ \\
pessq0.9a1e1 & $10^{-3} \zsun$ & $265$ & $0.9$ & $10$  \\
pessq0.9a2e2 & $10^{-3} \zsun$ & $265$ & $0.9$ & $200$ \\
\hline
half-kick    & w/o             & $130$ & $0.0$ & $10$  \\
weak-wind    & $10^{-5} \zsun$ &   $0$ & $0.0$ & $10$  \\
no-rot-wind  & $10^{-3} \zsun$ &   $0$ & $0.0$ & $10$  \\
\hline
K14          & w/o              & $0$   & $0.0$ & $0$
\enddata
\tablecomments{The units of $\sigma_{\rm k}$ and $\amin$ are
  km~s$^{-1}$ and $\rsun$, respectively. In the wind column,
  metallicity adopted for our wind model is described. In the
  ``no-rot-wind'' model, we do not account for rotationally enhanced
  stellar winds, $\dot{M}_{\rm wind,rot}$. In the ``K14'' model, we
  set the $\mi$ distribution as $f(\mi) \propto {\rm const} \; (10 \le
  \mi/\msun \le 100)$, and do not take into account PPI effects.}
\end{deluxetable}

Additionally, we make $4$ parameter sets. For the ``half-kick'' model,
we investigate the dependence of BH-BH populations on natal kick
velocities. We examine effects of metallicity and rotational
enhancement on stellar winds, investigating the ``weak-wind'' and
``no-rot-wind' models, respectively. In order to compare our results
with results in \cite{2020MNRAS.498.3946K}, we prepare the K14 model,
which has initial condition and single/binary star models similar to
the K14 model in \cite{2020MNRAS.498.3946K}.

For each parameter set, we follow time evolutions of $10^6$ binaries
until $100$~Gyr. The initial conditions of $10^6$ binaries are the
same if $\qmin$ and $\amin$ are identical. For each parameter set, we
spend $1-2$ hours calculating on $1$ CPU core of Intel Core i9-7980XE
CPU with turbo boost enabled.

\subsection{Pop.~III formation model}
\label{sec:PopIIIFormationModel}

In this paper, we adopt Pop.~III formation model as follows. All
Pop.~III stars are formed in minihalos at the same time. Only one
Pop.~III binary is born in one minihalo, $\eta_{\rm bin}=1$. The
number density of the minihalos is $n_{\rm DM} =
10^{11}$~Gpc$^{-3}$. Then, the number density of Pop.~III binaries is
$10^{11}$~Gpc$^{-3}$. Using Eq.~(\ref{eq:AverageBinaryMass}), we can
obtain the mass density of Pop.~III binaries as $1.3 \times
10^{13}$~$\msun$~Gpc$^{-3}$ for $\qmin=0.0$ and $1.7 \times
10^{13}$~$\msun$~Gpc$^{-3}$ for $\qmin=0.9$. Our mass density is a
fraction of the Pop.~III star density of $\sim
10^{14}$~$\msun$~Gpc$^{-3}$ (see Figure~2 of
\cite{2016MNRAS.462.3591M}), and $\sim 3 \times
10^{13}$~$\msun$~Gpc$^{-3}$ (see Figure~4 of
\cite{2020MNRAS.492.4386S}). On the other hand, our mass density is
smaller than in \cite{2020MNRAS.498.3946K} by $2$ orders of
magnitude. They have estimated Pop.~III mass density from results of
\cite{2011A&A...533A..32D} modified by the argument of
\cite{2016MNRAS.461.2722I}. We can interpret that our formation rate
is the lower limit.

According to \cite{2016MNRAS.462.3591M}, Pop.~III stars stop forming
at $z \sim 5$, or at the lookback time of $\sim 12.5$~Gyr. Thus,
Pop.~III BH-BHs which merge in the local universe should have the
delay time of $\sim 12.5$-$13.8$~Gyr. Note that the upper bound should
be less than $13.8$~Gyr, since Pop.~III stars are not formed at the
beginning of the universe. However, we assume that Pop.~III BH-BHs
with the delay time of $\sim 10$-$15$~Gyr merge in the local
universe. This assumption enables us to obtain many Pop.~III BH-BHs
from our binary population synthesis calculations, and investigate
their mass and spin distributions in the local universe.

\section{Results}
\label{sec:Results}

We describe the results of binary population synthesis
calculations. In section~\ref{sec:SingleStarEvolution}, we show the
relation between ZAMS and remnant masses through single star
evolution. It is instructive to investigate the dependence of the
relation on stellar winds. In section~\ref{sec:ComparisonWithK14}, we
compare our results with those of \cite{2020MNRAS.498.3946K},
examining the K14 model. In section~\ref{sec:MergingBH-BHs}, we
present populations of merging Pop.~III BH-BHs.

\subsection{Single star evolution}
\label{sec:SingleStarEvolution}

\begin{figure}[ht!]
  \plotone{\fdir/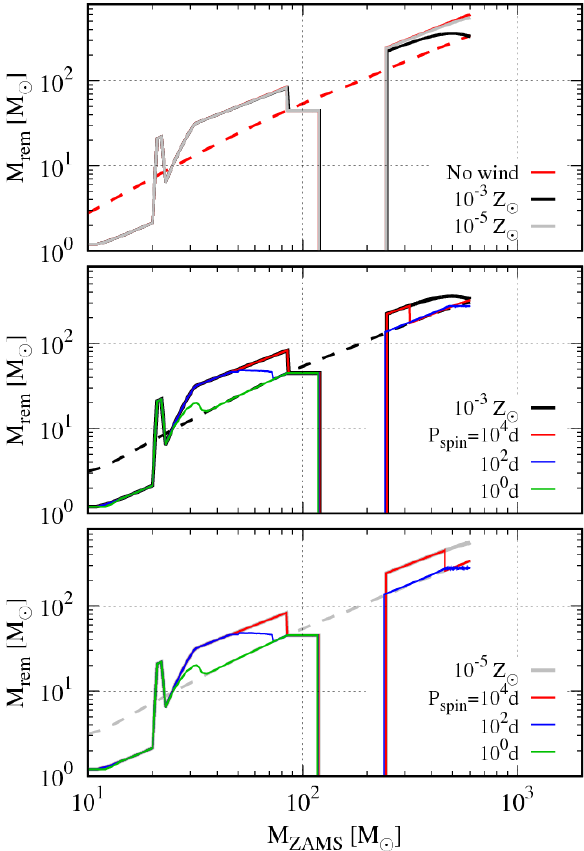}
  \caption{Relation between ZAMS and remnant masses. In the top panel,
    stars receive no wind (red), $10^{-3} \zsun$ winds (black), and
    $10^{-5} \zsun$ winds (gray). The stars receive no wind, or winds
    without $\dot{M}_{\rm wind,rot}$. In the middle and bottom panels,
    stars receive winds with $Z=10^{-3}$ and $10^{-5} \zsun$,
    respectively.  The black and gray curves in the middle and bottom
    panels are identical with those in the top panel, respectively. In
    the middle and bottom panels, stars receive winds with
    $\dot{M}_{\rm wind,rot}$ except for stars indicated by the black
    and gray curves. They are forced to keep stellar rotations with
    periods of $10^4$ (red), $10^2$ (blue), and $1$ days (green). As
    guides, He core masses are indicated by dashed curves in stars
    receiving no wind (top), $10^{-3} \zsun$ winds without
    $\dot{M}_{\rm wind,rot}$ (middle), and $10^{-5} \zsun$ winds
    without $\dot{M}_{\rm wind,rot}$ (bottom).}
  \label{fig:remnantMass}
\end{figure}

In Figure~\ref{fig:remnantMass}, we show the relation between ZAMS and
remnant masses. The maximum ZAMS mass is $600 \msun$, which is the
possible maximum mass of binaries in our calculations. Moreover, a
star with this maximum ZAMS mass creates a He core with $\sim 300
\msun$, and can form a nHe star with $\sim 300 \msun$. As described in
section~\ref{sec:SingleStarModel}, we confirm that our stellar wind
model is valid for nHe stars with $\lesssim 300 \msun$. We can see the
metallicity dependence without $\dot{M}_{\rm wind,rot}$ in the top
panel, and the dependence on stellar rotation speeds for $Z=10^{-3}$
and $10^{-5} \zsun$ in the middle and bottom panels,
respectively. We forcibly fix stellar rotation periods for stars
  in the middle and bottom panels. There are two reasons for this
  treatment. First, we simply attempt to investigate effects of
  stellar rotations on stellar wind mass loss. Second, stars could
  keep their rotation periods when they are synchronized with binary
  motions.

We first focus on the top panel. For any winds without $\dot{M}_{\rm
  wind,rot}$, stars with $\mzams \lesssim 20 \msun$ and $\gtrsim 20
\msun$ leave NSs and BHs, respectively. We can see that BH masses once
decrease at $\mzams \sim 25 \msun$, which is a feature of the rapid
model we adopt (see Eq.~(\ref{eq:RapidModel}) in detail). Stars with
$\mzams \sim 90-120 \msun$ have $45 \msun$ BHs due to the PPI effect,
stars with $\mzams \sim 120-250 \msun$ leave no remnant due to the
PISN effect, and stars with $\mzams \gtrsim 250 \msun$ undergo direct
collapse to BHs. PPI/PISN effects make ``the pair-instability (PI)
mass gap'' from $45 \msun$ to $135 \msun$. Nevertheless, some of stars
can have BHs in the PI mass gap. Their maximum mass is $\sim 85
\msun$. They can avoid PPI/PISN effects due to small He core masses,
and can grow to $>45 \msun$ owing to massive H envelopes. For $\mzams
\lesssim 250 \msun$, the relations has weak dependence on
metallicity. For $\mzams \gtrsim 250 \msun$, stars receiving $10^{-3}
\zsun$ winds leave BHs with distinctly smaller masses than the other
cases. This is due to luminous-star winds ($\dot{M}_{\rm NJ}$).

In the middle panel, we can find that the boundary between NSs and BHs
as well as the PPI/PISN effects are similar to the case of winds
without $\dot{M}_{\rm wind,rot}$. For $\mzams \lesssim 120 \msun$, BH
masses become smaller with rotation periods smaller. Stars with
rotation periods of $10^4$ days keep their H envelopes to the ends of
their lives. On the other hand, stars with rotation periods of $1$ and
$100$ days become nHe stars when their ZAMS masses are $\mzams \gtrsim
35 \msun$ and $\gtrsim 70 \msun$, respectively. For $\mzams \gtrsim
250 \msun$, stars with smaller periods also leave BHs with smaller
masses. For periods of $1$ day, stars undergo PISNe, and leave no
remnants, since they lose large amounts of masses in their MS phases,
and create their He core with $65-135 \msun$ entering into their CHeB
phases. For periods of $10^2$ days, they become nHe stars. For periods
of $10^4$ days, BH masses are suddenly decreased at $\mzams \sim 300
\msun$. Stars with $\mzams \lesssim 300 \msun$ keep their H envelopes,
while those with $\mzams \gtrsim 300 \msun$ become nHe stars. This is
because luminosities of the latter stars exceed their Eddington limit
in the middle of their evolution. Thus, they lose large amounts of
their masses, and become nHe stars.

In many cases, we can see that stars lose large amounts of masses in
MS, CHeB, and ShHeB phases, and stop losing masses in nHe phases. This
is because the Eddington luminosity is increased from the former
phases to the latter phase. Then, rotationally enhanced mass loss
becomes inactive when stars reach to nHe stars.

We find that the relations are weakly dependent on metallicity when
$\dot{M}_{\rm wind,rot}$ is taken into account (see the middle and
bottom panels). Nevertheless, for rotation periods of $10^4$ days, BH
masses are suddenly decreased at different $\mzams$: $\mzams \sim 300
\msun$ for $Z=10^{-3} \zsun$, and $\mzams \sim 450 \msun$ for
$Z=10^{-5} \zsun$. Stars with $Z=10^{-5} \zsun$ are harder to lose
mass than those with $Z=10^{-3} \zsun$, since they have larger masses,
i.e. larger Eddington limit, owing to slightly weaker stellar winds.

\subsection{Comparison with K14}
\label{sec:ComparisonWithK14}

\begin{figure}[ht!]
  \plotone{\fdir/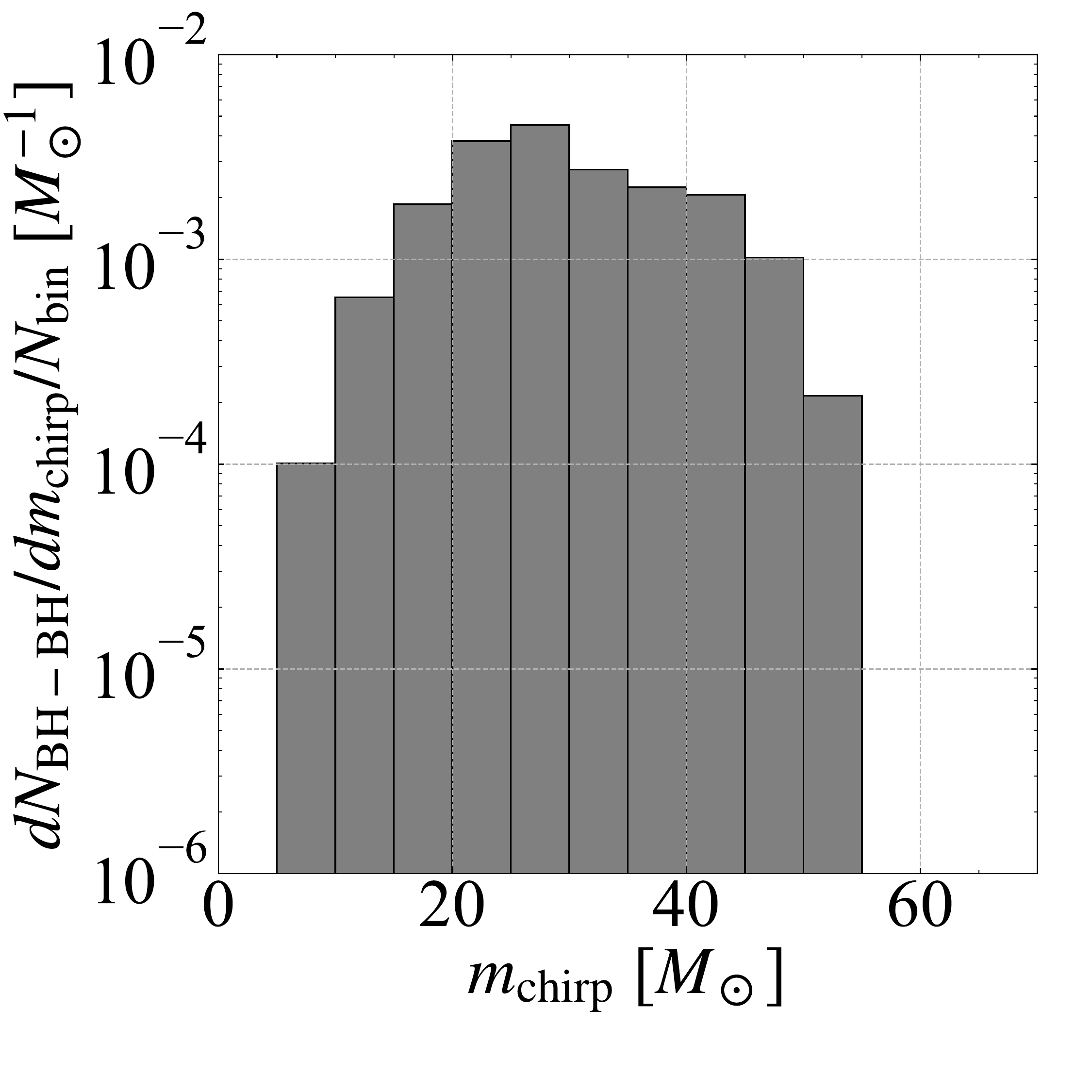}
  \caption{Chirp mass distribution of merging BH-BHs within $15$~Gyr
    in our K14 model. The number of merging BH-BHs ($N_{\rm BH-BH}$)
    is normalized by the total number of binaries, $N_{\rm
      bin}=10^6$.}
  \label{fig:chirpDistK14}
\end{figure}

We describe calculation results of our K14 model to compare them with
the K14 model in \cite{2020MNRAS.498.3946K} (hereafter, the original
K14 model). Figure~\ref{fig:chirpDistK14} shows the chirp mass
distribution of merging BH-BHs within $15$~Gyr in our K14 model.
  Hereafter, we use $15$~Gyr as the Hubble time, since it's just a
  round number. The chirp mass of BH-BHs can be expressed as
\begin{align}
  m_{\rm chirp} = \frac{(\mbp m_{\rm b,s})^{3/5}}{(\mbp+m_{\rm
      b,s})^{1/5}},
\end{align}
where $\mbp$ and $m_{\rm b,s}$ are the primary and secondary BH masses
of BH-BHs. We find that it is quite consistent with the results of the
original K14 model (see the purple curve in their
Figure~3). Both the K14 models have sharp peaks at $m_{\rm
  chirp} \sim 30 \msun$. Our K14 model has sharp drop at $m_{\rm
  chirp} \sim 60 \msun$, while the original K14 model do so at $m_{\rm
  chirp} \sim 80 \msun$. This comes from different treatments of H
envelopes of post-MS stars whose H envelopes are almost stripped. If
the treatments are the same, the sharp drop is at $m_{\rm chirp} \sim
60 \msun$ even in the original K14 model (see the light-blue curve of
Figure~1 in \cite{2016MNRAS.456.1093K}). We emphasize these agreements
are very surprising. This is because they and we use fitting formulae
of Pop.~III stars based on different stellar evolution models
\citep[][respectively]{2001A&A...371..152M,2020MNRAS.495.4170T}. These
results reinforce the argument of \cite{2014MNRAS.442.2963K} that
typical Pop.~III BH-BH mergers have $\sim 30 \msun$ BHs.

\begin{figure}[ht!]
  \plotone{\fdir/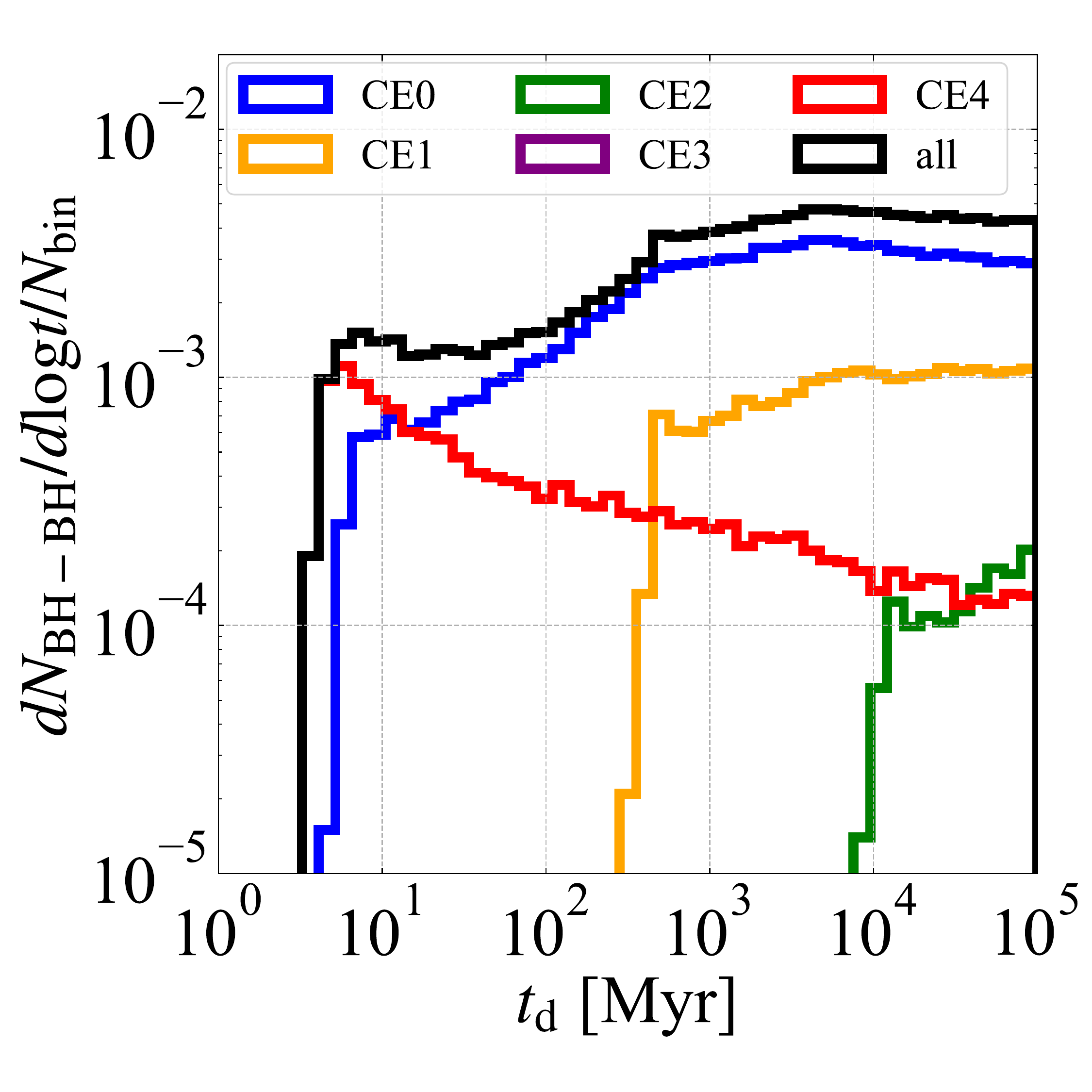}
  \caption{Delay time distribution of BH-BHs in our K14 model. The
    normalization is the same as in Figure~\ref{fig:chirpDistK14}. The
    black curve indicates the distribution of all the BH-BHs. All
    colored curves indicates BH-BHs formed through channels related to
    common envelope evolutions defined in the main text. There is no
    BH-BH formed through the CE3 channel.}
  \label{fig:allBhLogDelayTimeK14}
\end{figure}

We can see the delay time ($\td$) distribution of BH-BHs in our K14
model in Figure~\ref{fig:allBhLogDelayTimeK14}, where $\td$ is delay
time, or defined as time interval from the ZAMS time to the BH-BH
merger time. The black curve indicates the distribution of all the
BH-BHs, while the other curves indicates the distribution of BH-BHs
formed through channels of CE0, CE1, CE2, CE3, and CE4. In the CE0
channel, binary systems experience no common envelope evolution. In
the CE1 and CE2 channels, they undergo common envelope evolution,
ejecting H envelopes of the 1st- and 2nd-evolving stars,
respectively. In the CE3 channel, they have both the CE1 and CE2
channels. In the CE4 channel, they go through common envelope
evolution, ejecting H envelopes of both the 1st- and 2nd-evolving
stars at the same time.

The delay time distribution of all the BH-BHs is similar to the
original K14 model (see the purple curve in their Figure~4). Note that
their values in the vertical axis are $6$ digits larger than ours,
since they do not normalize the number of BH-BHs by the total number
of binaries. The number of mergers are sharply increased at several
Myr, and are gradually decreased until several $10$ Myr. After that,
the number turns to rise, and reaches to a few $10^{-3}$ at $10$~Gyr.

\begin{figure}[ht!]
  \plotone{\fdir/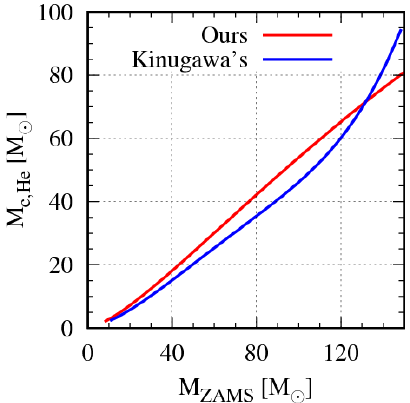}
  \caption{Relation between ZAMS and He core masses at the carbon
    ignition time in fitting formulae of \cite{2020MNRAS.498.3946K}
    (blue) and \cite{2020MNRAS.495.4170T} (red).}
  \label{fig:heCoreMass}
\end{figure}

We recognize common and different features, comparing the delay time
distribution of BH-BHs formed through each channel (see their
Figure~5(e)). The common feature is that the dominant channels are CE4
and CE0 at earlier and later times than a few $10$~Myr,
respectively. Another common feature is the distribution of BH-BHs
formed through the CE1 channel. The BH-BHs start formed from $\sim
10^2$~Myr, and keep the number $\sim 10^{-3}$ after that.

The numbers of BH-BHs formed through the CE2 and CE3 channels in our
K14 model are much smaller than in the original K14 model. This is
because our fitting formulae of single star models have larger He core
masses than theirs (see Figure~\ref{fig:heCoreMass}). As the ratios of
He core masses to the stellar masses become larger, common envelope
evolution is harder to happen. Although He core masses in our fitting
formulae are smaller than Kinugawa's ones for $\mzams \gtrsim 130
\msun$, stars with $\mzams \gtrsim 130 \msun$ have small chance to
undergo the CE2 and CE3 channels in both the K14 models.

\begin{figure}[ht!]
  \plotone{\fdir/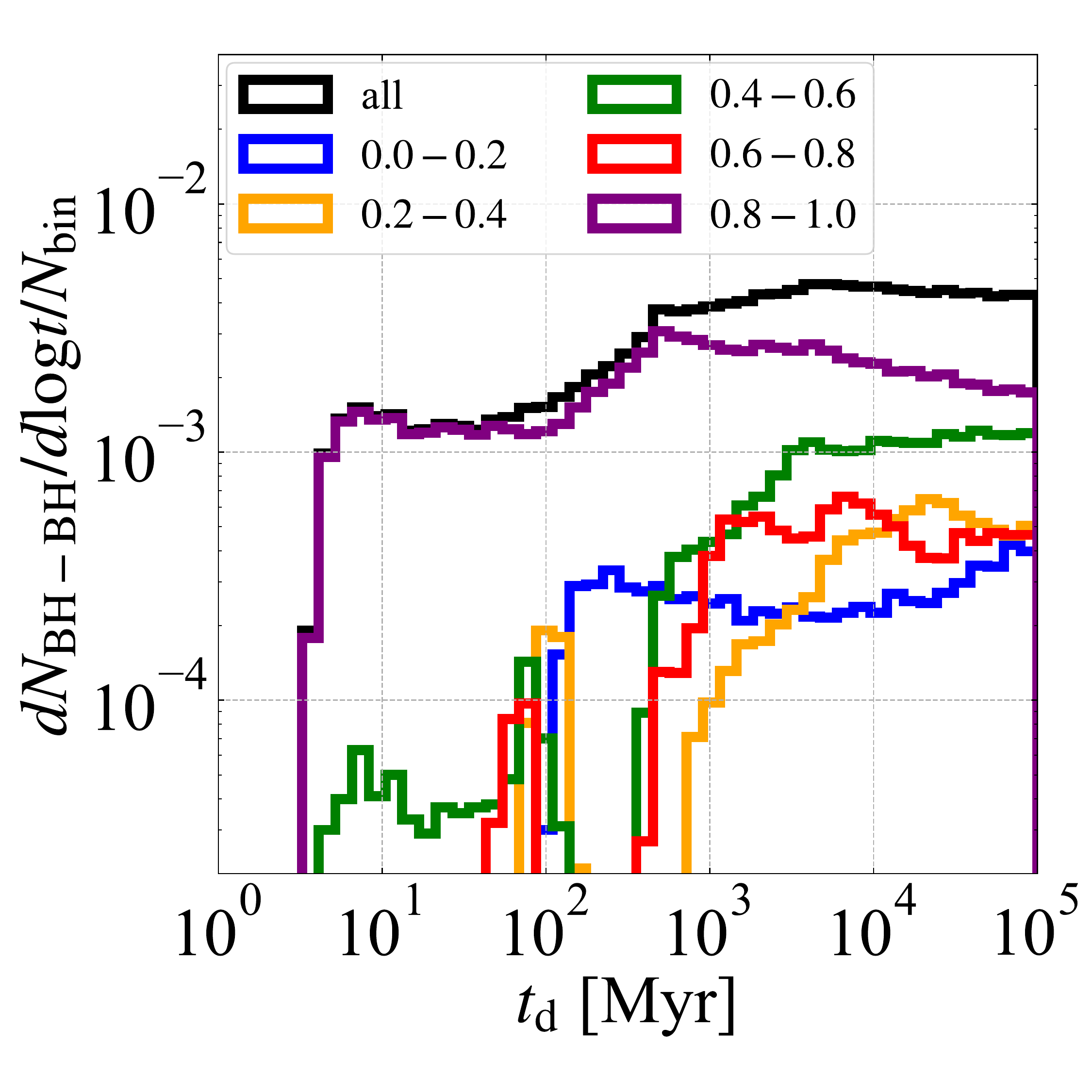}
  \caption{Delay time distribution of BH-BHs in our K14 model. The
    normalization is the same as in
    Figure~\ref{fig:chirpDistK14}. The black curve indicates
    the distribution of all the BH-BHs. Other curves indicate the
    distribution of BH-BHs with effective spins ($\chi_{\rm eff}$)
    described in the legend. The definition of $\chi_{\rm eff}$ is in
    the main text.}
  \label{fig:xiLogDelayTimeK14}
\end{figure}

Figure~\ref{fig:xiLogDelayTimeK14} shows the delay time distribution
of BH-BHs with effective spins, $\chi_{\rm eff}$. The definition of
$\chi_{\rm eff}$ is given by
\begin{align}
  \chi_{\rm eff} = \left( \frac{\mbp \xbp+ \mbs \xbs}{\mbp + \mbs}
  \right) \cdot \frac{\vecl}{|\vecl|},
\end{align}
where $\xbp$ and $\xbs$ are normalized spin vectors of the primary and
secondary BHs. We are going to making comparison between $\chi_{\rm
  eff}$ evolution in both the K14 models. However, we should note that
we cannot strictly compare Figure~\ref{fig:xiLogDelayTimeK14} with
$\chi_{\rm eff}$ evolution shown in the original K14 model (see their
Figure~7(e)). We assume that all the stars are instantaneously born at
$t=0$, while they adopt a Pop.~III star formation history in which
Pop.~III stars have a width of their formation times. In addition, the
horizontal axes of our and their figures are time and redshift,
respectively.

There are several common features. We find BH-BHs with $0.8<\chi_{\rm
  eff}<1.0$ are always dominant in both the results. When we compare
the numbers at $10$~Gyr in our K14 model with those at $z=0$ in the
original K14 model, BH-BHs with $\chi_{\rm eff}<0.8$ have significant
contributions to the numbers in both the K14 models. However, the
numbers are quite different at an early time. In our K14 model, only
BH-BHs with $0.8 < \chi_{\rm eff} < 1.0$ merge at time $<10^2$~Myr. On
the other hand, BH-BHs with $0.4 < \chi_{\rm eff} < 0.6$ and $0.8 <
\chi_{\rm eff} < 1.0$ have comparable contributions in the original
K14 model. These differences can be explained by the difference
between spin models. In our model, BHs keep their spin angular momenta
even when they lose their masses at their collapse. On the other hand,
in the original K14 model, BHs lose their spin angular
momenta. Nevertheless, spin distributions in both the K14 models are
consistent in the current universe.

\subsection{Merging BH-BHs}
\label{sec:MergingBH-BHs}

\begin{figure}[ht!]
  \plotone{\fdir/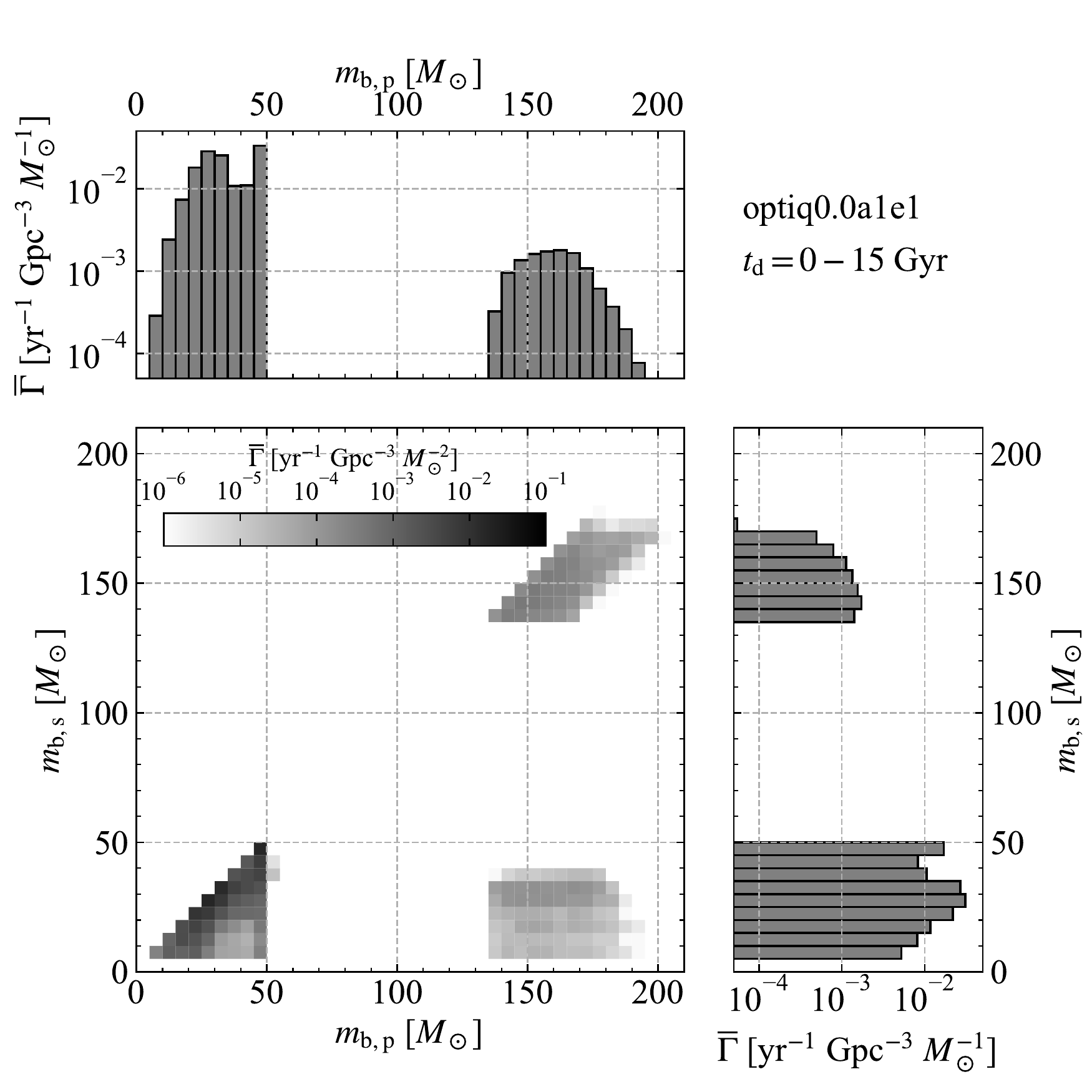}
  \caption{Average merger rate density of BH-BHs for $\td =
    0-15$~Gyr.}
  \label{fig:sampleMassDist}
\end{figure}

Figure~\ref{fig:sampleMassDist} shows the average merger rate density
of BH-BHs for $\td = 0-15$~Gyr in the optiq0.0a1e1 model. We define
the merger rate density as
\begin{align}
  \Gamma = \frac{d(N_{\rm BH-BH}/N_{\rm bin})}{d\td} \left(
  \frac{\eta_{\rm bin}}{1} \right) \left( \frac{n_{\rm
      DM}}{10^{11}{\rm Gpc}^{-3}} \right) \; [{\rm yr}^{-1}~{\rm
      Gpc}^{-3}]. \label{eq:MergerRateDensity}
\end{align}
As seen in the above formula, the average merger rate density is
defined in the local comoving volume at the source frame time. Using
Eq.~(\ref{eq:MergerRateDensity}), we can express the average merger
rate density as
\begin{align}
  \overline{\Gamma} = \frac{1}{\tdf-\tdi} \int_{\tdi}^{\tdf} \Gamma
  d\td, \label{eq:AverageMergerRateDensity}
\end{align}
where $\tdi$ and $\tdi$ are the beginning and ending times for the
averaging.

We can clearly see there are three subpopulations of BH-BHs: those
with two low-mass BHs (lBHs), those with one lBH and one high-mass BH
(hBH), and those with two hBHs, where we call BHs with $\lesssim 50
\msun$ ``lBHs'', and BHs with $\gtrsim 130 \msun$ ``hBHs''. There is a
large mass gap between lBHs and hBHs due to PPI/PISN effects, the PI
mass gap, as seen also in Figure~\ref{fig:remnantMass}. The mass gap
ranges from $\sim 50 \msun$ to $\sim 130 \msun$. Hereafter, we name
BH-BHs with two lBHs, those with one lBH and one hBH, and those with
two hBHs ``hBH0'', ``hBH1'', and ``hBH2'', respectively, after the
numbers of hBHs. The mass ranges of lBHs and hBHs are quite different;
the maximum mass of lBHs is $\sim 50 \msun$, and the minimum mass of
hBHs is $\sim 135 \msun$. Stars with $\mche < 45 \msun$ and H
envelopes can leave $> 45 \msun$ BHs avoiding PPI, and leave BHs with
$\sim 50 \msun$ at most. Thus, we can identify hBH0, hBH1, and hBH2 as
distinct subpopulations.

\begin{figure*}[ht!]
  \plotone{\fdir/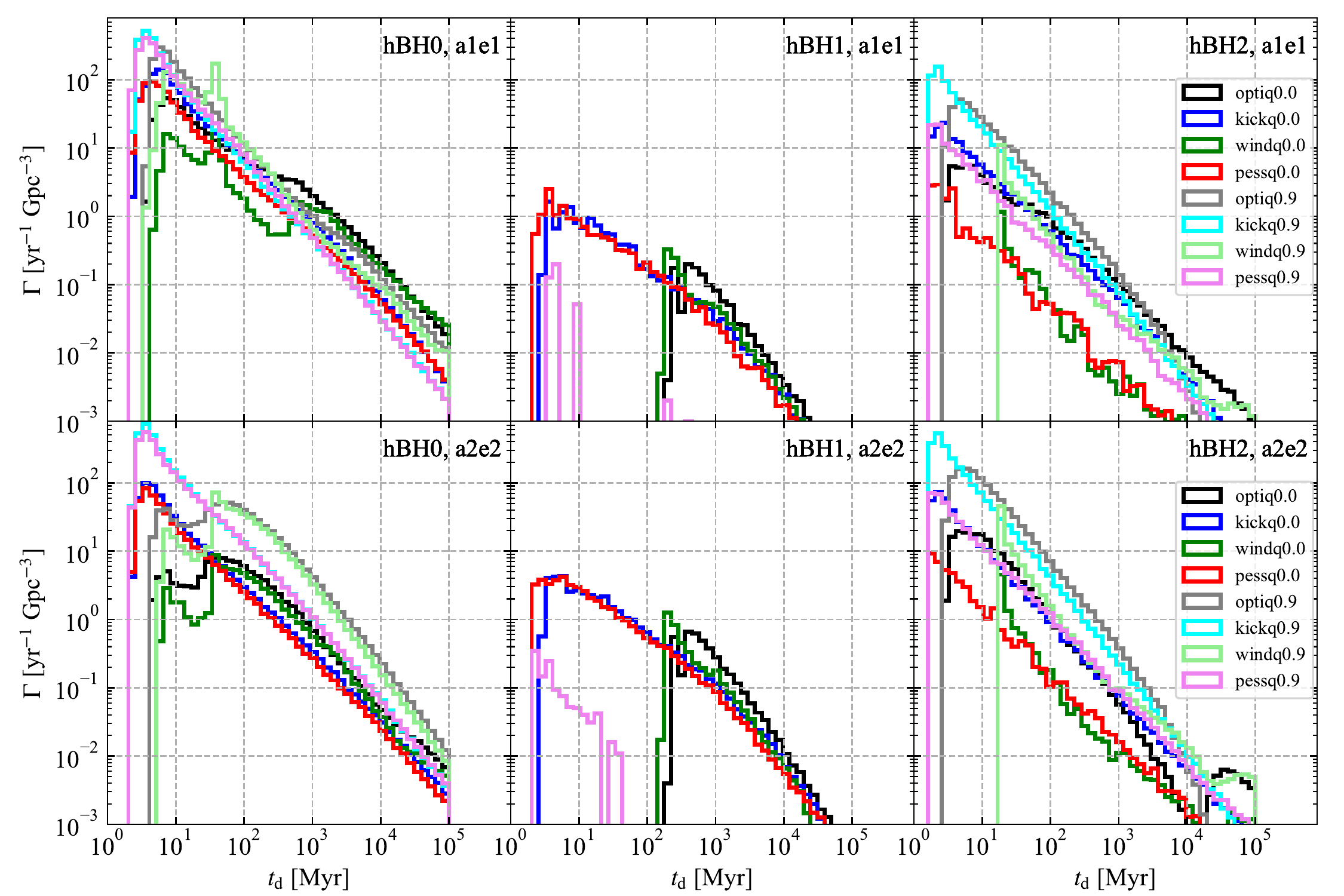}
  \caption{Merger rate densities of hBH0, hBH1, and hBH2 as a function
    of the delay time in the main 16 models. The left, middle, and
    right panels indicate the distributions of hBH0s, hBH1s, and
    hBH2s, respectively. The top and bottom panels indicate models
    with $\amin=10$ and $200\rsun$, respectively. Colors of the curves
    show single star models and $\qmin$.}
  \label{fig:delayTimeDist}
\end{figure*}

In Figure~\ref{fig:delayTimeDist}, we show the merger rate density of
hBH0, hBH1, and hBH2 as a function of $\td$ in the main 16 models.
The merger rate densities are not necessarily proportional to
$\td^{-1}$. Some of merger rate densities suddenly increase, since the
dominant formation channels of BH-BHs are switched. For example, for
the hBH2s of the optiq0.0a2e2 model, the dominant formation channels
are switched from CE4 to CE0 at $\td \sim 10$~Gyr. For $\td \lesssim
0.1$~Gyr, some of merger rate densities decrease more slowly than
$\td^{-1}$.  This is because a part of BH-BH progenitors with $\td
\gtrsim 0.1$~Gyr shrink their orbits due to mechanisms other than
common envelope evolutions. One mechanism is BH natal kicks. We can
see that the BH natal kicks yield hBH1s with $\td \lesssim 0.1$~Gyr,
comparing merger rate densities of hBH1s in the optiq0.0a1e1 and
kickq0.0a1e1 models. Another mechanism is stable mass transfer in
which the mass ratio of a donor star to an accretor star is
sufficiently larger than unity. This can sometimes strip H envelopes
of post-MS stars. This can be seen in hBH0s of the optiq0.0a1e1 model.

In the present day ($\sim 10$~Gyr), the merger rates of hBH0s are
$\sim 0.1$~yr$^{-1}$~Gpc$^{-3}$, independently of single star models,
$\qmin$, and $\amin$. This is much smaller than the merger rate
density inferred by LIGO/Virgo observations, $\sim
10-100$~yr$^{-1}$~Gpc$^{-3}$ \citep{2019PhRvX...9c1040A}. Although
this is smaller than estimated by \cite{2020MNRAS.498.3946K} by $2$
orders of magnitude, this is consistent with their results, since the
total Pop.~III mass in our model is smaller than theirs by $2$ orders
of magnitude (see section~\ref{sec:PopIIIFormationModel}).  If our
formation model is close to the actual formation rate, we may not
expect that Pop.~III hBH0 are a dominant origin of GW sources
discovered currently. However, it should be worth examining properties
of Pop.~III hBH0s for the following reason. It is important to
identify Pop.~III hBH0s with respect to the star formation history in
the universe, and moreover future GW observatories, such as Einstein
telescope \citep{2010CQGra..27s4002P,2020JCAP...03..050M} and Cosmic
explorer \citep{2019BAAS...51g..35R}, may detect a large number of
Pop.~III hBH0s.

We also inspect hBH1s and hBH2s for the following
reason. \cite{2020arXiv200602211M} have shown that hBH1s and hBH2s can
be detected by the current GW observatories if they merge within
detectable distances, although they have not yet been
detected. Moreover, space-borne GW observatories, such as LISA and
DECIGO
\citep[][respectively]{2017arXiv170200786A,2006CQGra..23S.125K}, may
be helpful for searching merging hBHs. Especially, LISA will detect
hBHs months to years before hBH mergers in the local universe
\citep{2016PhRvL.116w1102S}.

\begin{figure}[ht!]
  \plotone{\fdir/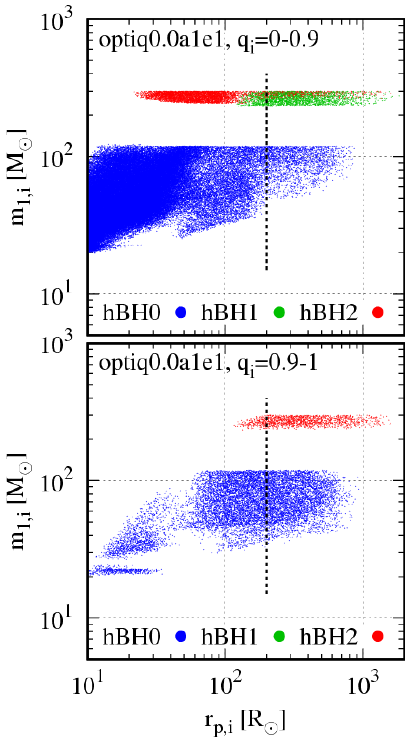}
  \caption{Initial pericenter distances ($\rpi$) and 1st-evolving
    star's masses ($\mi$) of BH-BHs with $\td = 0-15$~Gyr in the
    optiq0.0a1e1 model. In the top and bottom panels, the initial mass
    ratios ($\qi$) are $0-0.9$ and $0.9-1$, respectively. The vertical
    dashed lines indicate $\rpi = 200 \rsun$. The color codes show
    hBH0, hBH1, and hBH2.}
  \label{fig:icMtMergBh}
\end{figure}

Before we examine hBH0s, hBH1s, and hBH2s separately, we investigate
their initial conditions and formation channels. By this, we make
clear the reason why the merger rates of hBH0s and the sum of hBH1s
and hBH2s do not depend on initial conditions and single star
models. First, we mainly focus on the dependence on initial
conditions. Second, we mention the dependence on single star models.

Figure~\ref{fig:icMtMergBh} shows initial conditions of Pop.~III
binaries which form BH-BHs merging within $15$~Gyr for the
optiq0.0a1e1 model. We separate the initial conditions into two panels
by whether $\qi<0.9$ or $\qi \ge 0.9$., and mark an initial pericenter
distance ($\rpi$) of $200 \rsun$ by vertical dashed lines, so as to
imagine results of models with $\qmin=0.9$ and $\amin=200\rsun$.  We can
see a gap between $\mi \sim 120 - 250 \msun$ in which BH-BHs are not
formed. This is due to PISN effects. All hBH0s are under this gap, and
all hBH1s and hBH2s are above this gap. In other words, binaries with
1st-evolving stars with $\mi \lesssim 120 \msun$ can form only hBH0s,
and binaries with 1st-evolving stars with $\mi \gtrsim 250 \msun$ can
form only hBH1s and hBH2s. In both the panels, there is no BH-BH on
the top-left and right-bottom corners. The reason for their absence on
the top-left corner is that binaries merge before they form BH-BHs,
and the reason for their absence on the bottom-right corner is that
BH-BHs have too wide separation to merge within $15$~Gyr.

Even binaries with $\rpi \sim 10^3 \rsun$ can form hBH0s, hBH1s, and
hBH2s. This is because Pop.~III stars with $M \gtrsim 50 \msun$ expand
to $\gtrsim 10^3 \rsun$ (see Figure~\ref{fig:hrd}), and interact with
their companions. Thus, the merger rates of hBH0s, hBH1s, and hBH2s in
models with $\amin=200\rsun$ are comparable to those in models with
$\amin=10\rsun$ as seen in Figure~\ref{fig:delayTimeDist}.

There is no hBH1 in the bottom panel. Binaries with high $\qi$ can not
form a hBH1, since their members have similar masses. This is
consistent with the merger rate density as a function of $\td$ in the
middle panels of Figure~\ref{fig:delayTimeDist}. The merger rate of
hBH1s is nearly zero in models with $\qmin=0.9$. Nevertheless, the
sums of the merger rates of hBH1s and hBH2s are independent of
$\qmin$, since the increase of the hBH2 merger rate compensates for
the decrease of the hBH1 merger rate in the case of $\qmin=0.9$.

\begin{figure}[ht!]
  \plotone{\fdir/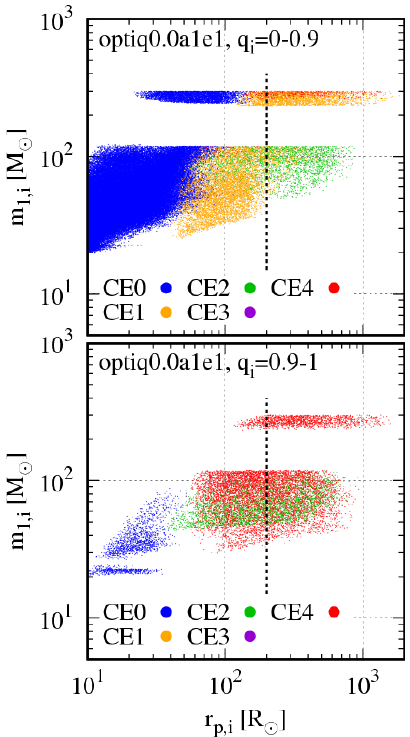}
  \caption{The same as in Figure~\ref{fig:icMtMergBh}, except for
    color codes that indicate CE channels forming BH-BHs.}
  \label{fig:icCeMergBh}
\end{figure}

We have to keep in mind that, although the merger rates of hBH0s and
the sum of hBH1s and hBH2s are insensitive to $\amin$ and $\qmin$,
their formation channels can depend on $\amin$ and $\qmin$.
Figure~\ref{fig:icCeMergBh} is the same plot as
Figure~\ref{fig:icMtMergBh}, except for color codes that indicate CE
channels forming BH-BHs. We can see that hBH0s are formed from
binaries with $\rpi \lesssim 50 \rsun$ and $\gtrsim 50 \rsun$ through
stable mass transfer (the CE0 channel), and common envelope evolution
(the CE1, CE2, and CE4 channels), respectively. The boundary of $\rpi$
increases with $\mi$, since BSG stars can have larger radii with
$\mi$. The reason for the absence of the CE3 channel is described in
section~\ref{sec:ComparisonWithK14}. Thus, only the CE1, CE2, and CE4
channels work for forming hBH0s in the cases of $\amin=200\rsun$.

\begin{figure}[ht!]
  \plotone{\fdir/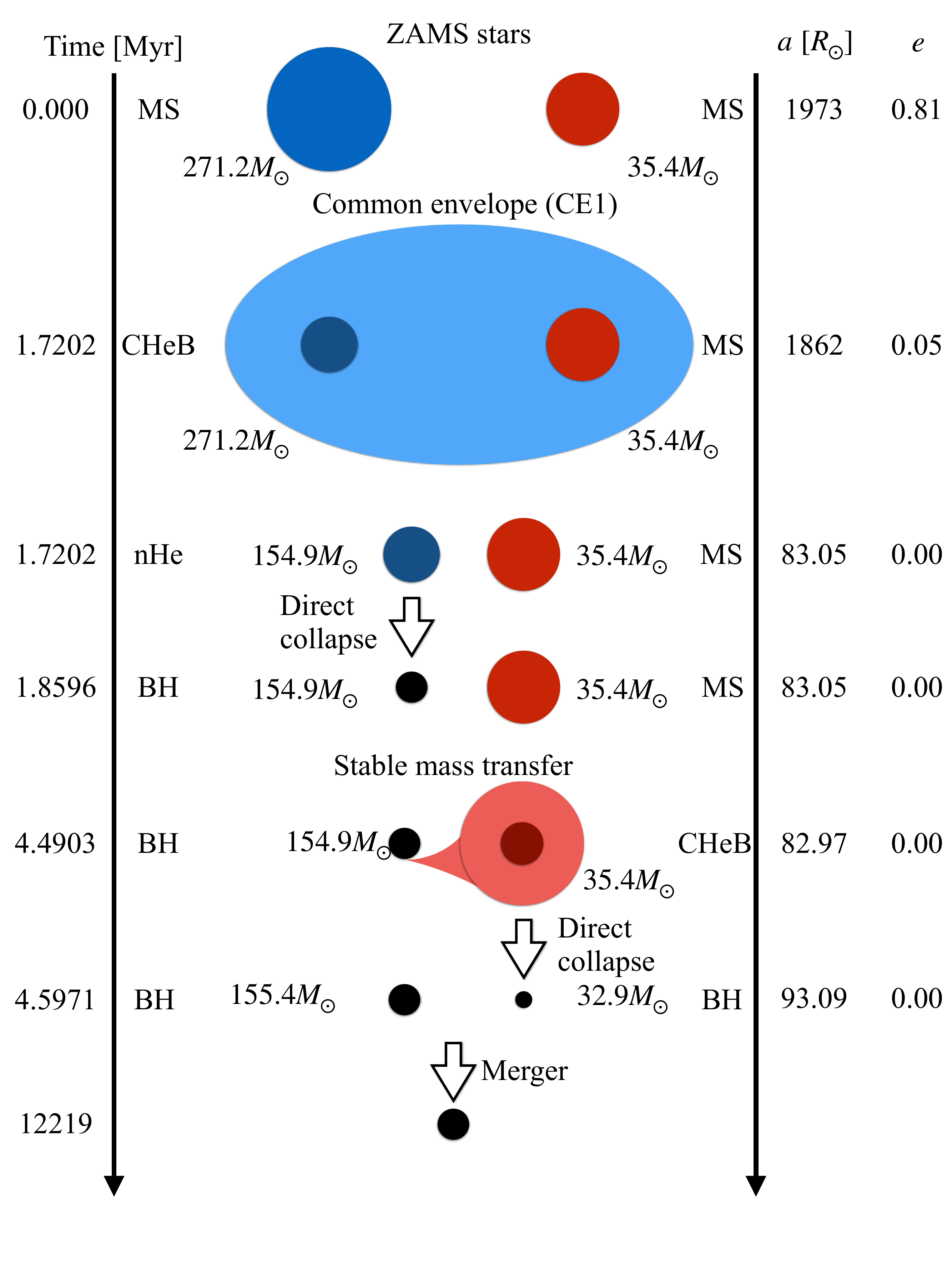}
  \caption{Example binary evolution leading to one hBH1 through the
    CE1 channel.}
  \label{fig:hbh1channel}
\end{figure}

All the hBH1s are formed from binaries with $\rpi \gtrsim 10^2 \rsun$
through the CE1 channel. We draw an example binary evolution leading
to one hBH1 in Figure~\ref{fig:hbh1channel}. To form hBH1s, binaries
should have a small $\qi$. Due to the small $\qi$, a binary
experiences merger or common envelope evolution when the 1st-evolving
star fills its Roche lobe. If the 1st-evolving star is in its MS and
CHeB/ShHeB phases, the binary experiences merger and common envelope
evolution, respectively. The common envelope evolution reduces the
binary separation down to several $10\rsun$. Then, the 2nd-evolving
star expands to $\sim 10\rsun$, fills its Roche lobe, and drives
stable mass transfer, since it is a BSG star, and the mass ratio of it
to the 1st BH is small. Eventually, binaries forming hBH1s experience
common envelope evolution driven by the 1st-evolving star, i.e. the
CE1 channel.

\begin{figure}[ht!]
  \plotone{\fdir/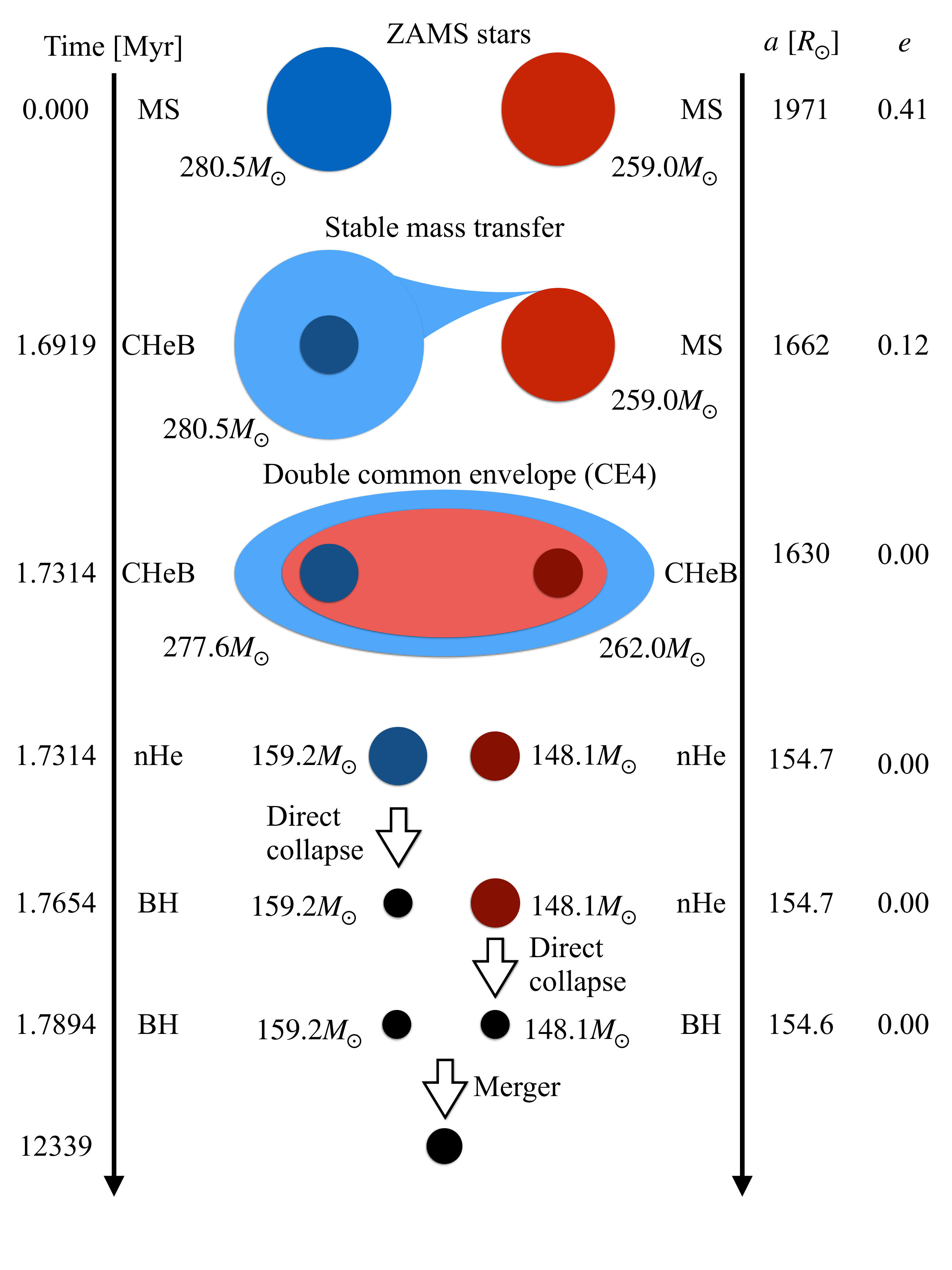}
  \caption{Example binary evolution leading to one hBH2 through the
    CE4 channel.}
  \label{fig:hbh2channel}
\end{figure}

A part of hBH2s are formed from binaries with $\rpi \lesssim 10^2
\rsun$ through the CE0 channel, and the rest of them from binaries
with $\rpi \gtrsim 10^2 \rsun$ through the CE4 channel. For binaries
with $\rpi \lesssim 10^2\rsun$, both the 1st- and 2nd-evolving stars
fill their Roche lobes when they are BSG stars. Then, they drive
stable mass transfer, i.e. the CE0 channel. For binaries with $\rpi
\gtrsim 10^2 \rsun$, they enter into their CHeB/ShHeB phases almost at
the same time due to their similar masses, and fill their Roche lobes.
Then, the binaries experience the CE4 channel. We draw an example
binary evolution leading to one hBH2 through the CE4 channel in
Figure~\ref{fig:hbh2channel}, which can occur regardless of $\qmin$
and $\qmin$ as seen in Figure~\ref{fig:icMtMergBh}.

The merger rates of hBH0s, hBH1s, and hBH2s are insensitive to single
star models for the following reasons. Natal kicks have small effects
on binary parameters of BH-BHs. This is because the natal kick
velocities are comparable to or smaller than internal velocities of
BH-BHs with $\td < 15$~Gyr. Stellar winds do not prevent stars from
filling their Roche lobe and interacting with their companions. Since
they are driven by stellar rotations, they are active only when stars
nearly fill their Roche lobe. They can strip H envelopes from stars,
and decrease stellar masses (see
Figure~\ref{fig:remnantMass}). Nevertheless, stars also lose their H
envelopes through common envelope evolution (see
Figure~\ref{fig:icMtMergBh}) even if stellar winds are switched
off. Thus, stellar winds do not have dominant roles in stripping H
envelopes from stars.

\begin{figure*}[ht!]
  \begin{center}
  \includegraphics[width=0.65\textwidth]{\fdir/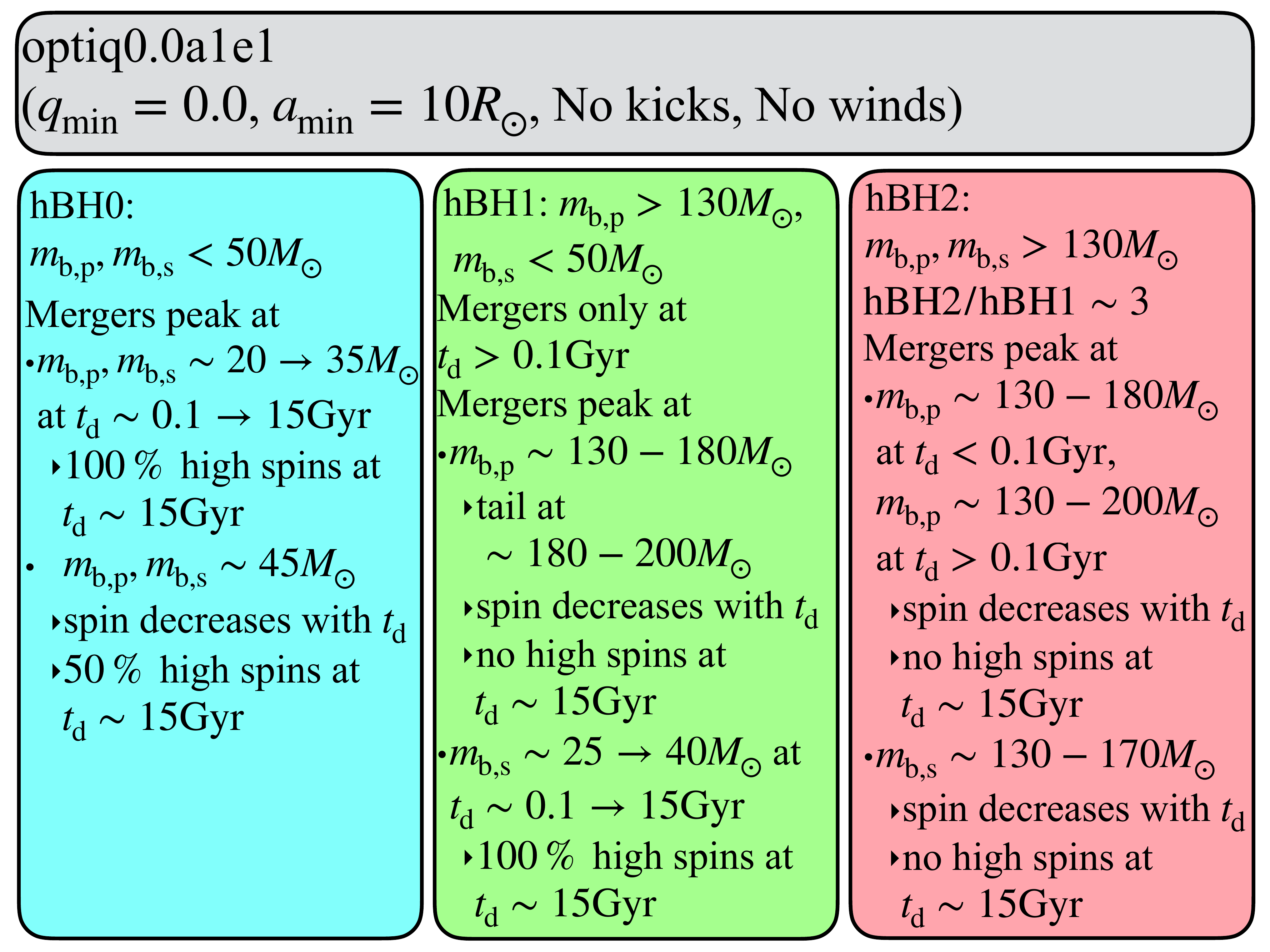}
  \end{center}
  \vspace{-4mm}
  \includegraphics[width=0.5\textwidth]{\fdir/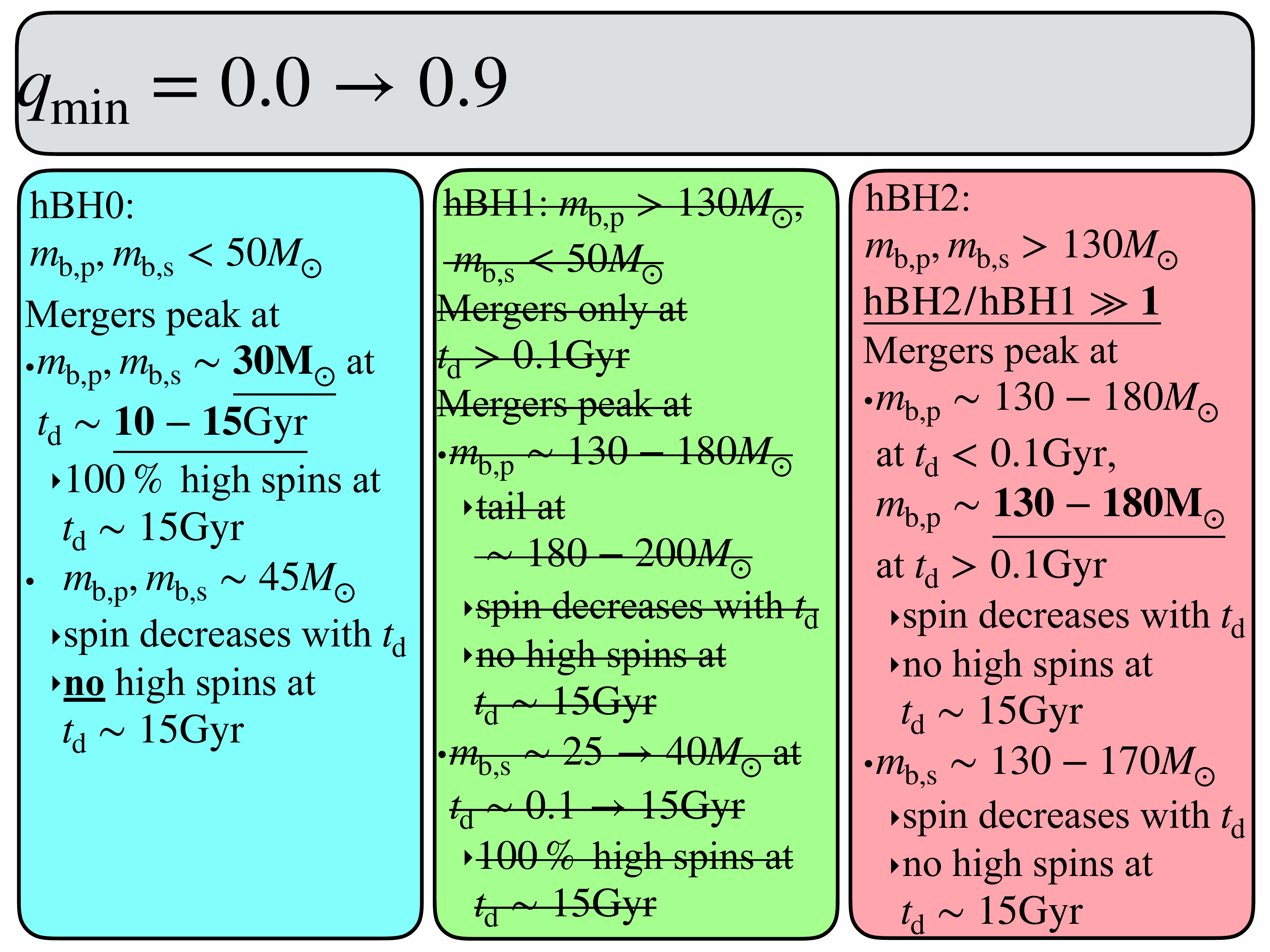}
  \includegraphics[width=0.5\textwidth]{\fdir/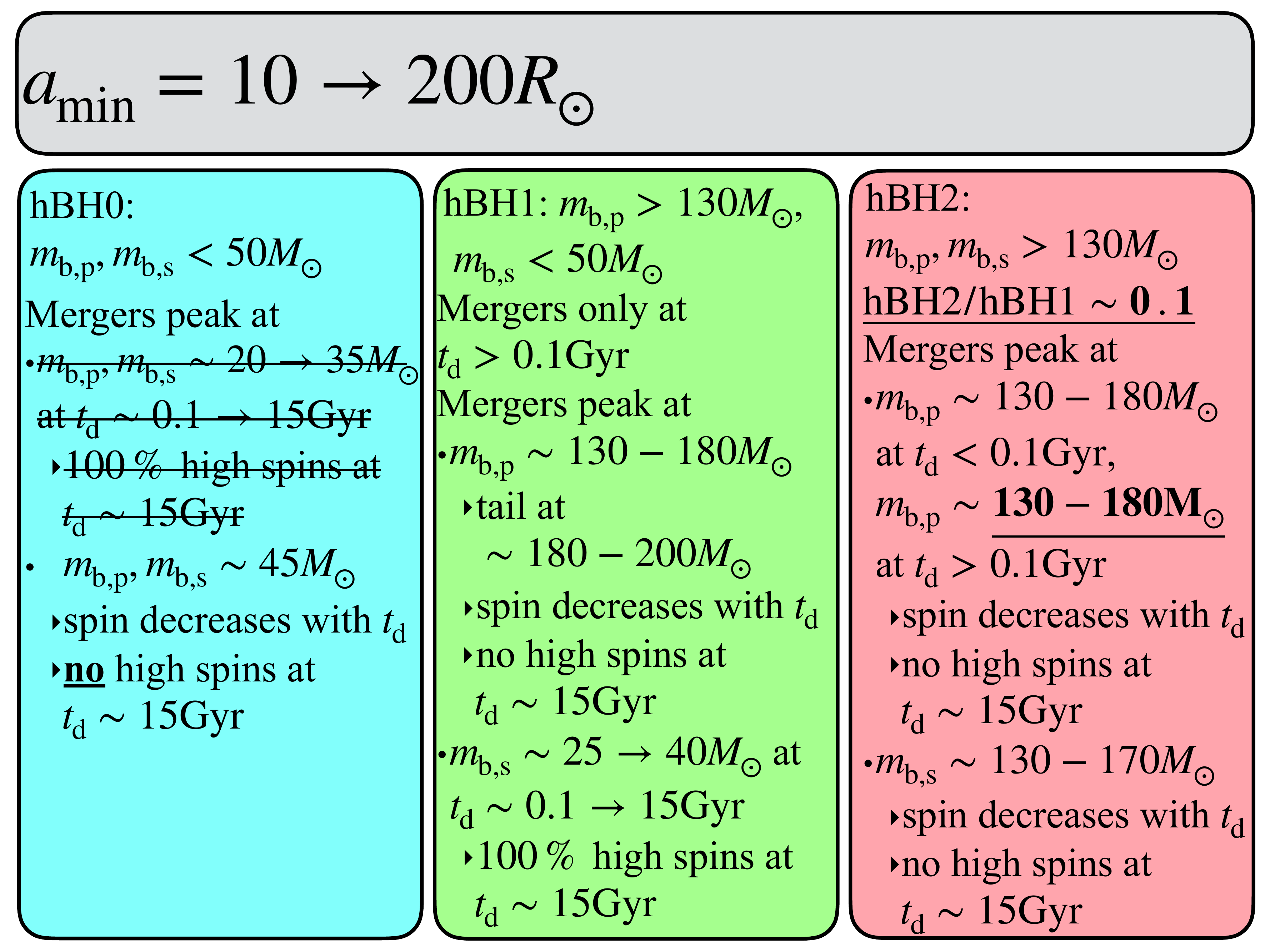}
  \includegraphics[width=0.5\textwidth]{\fdir/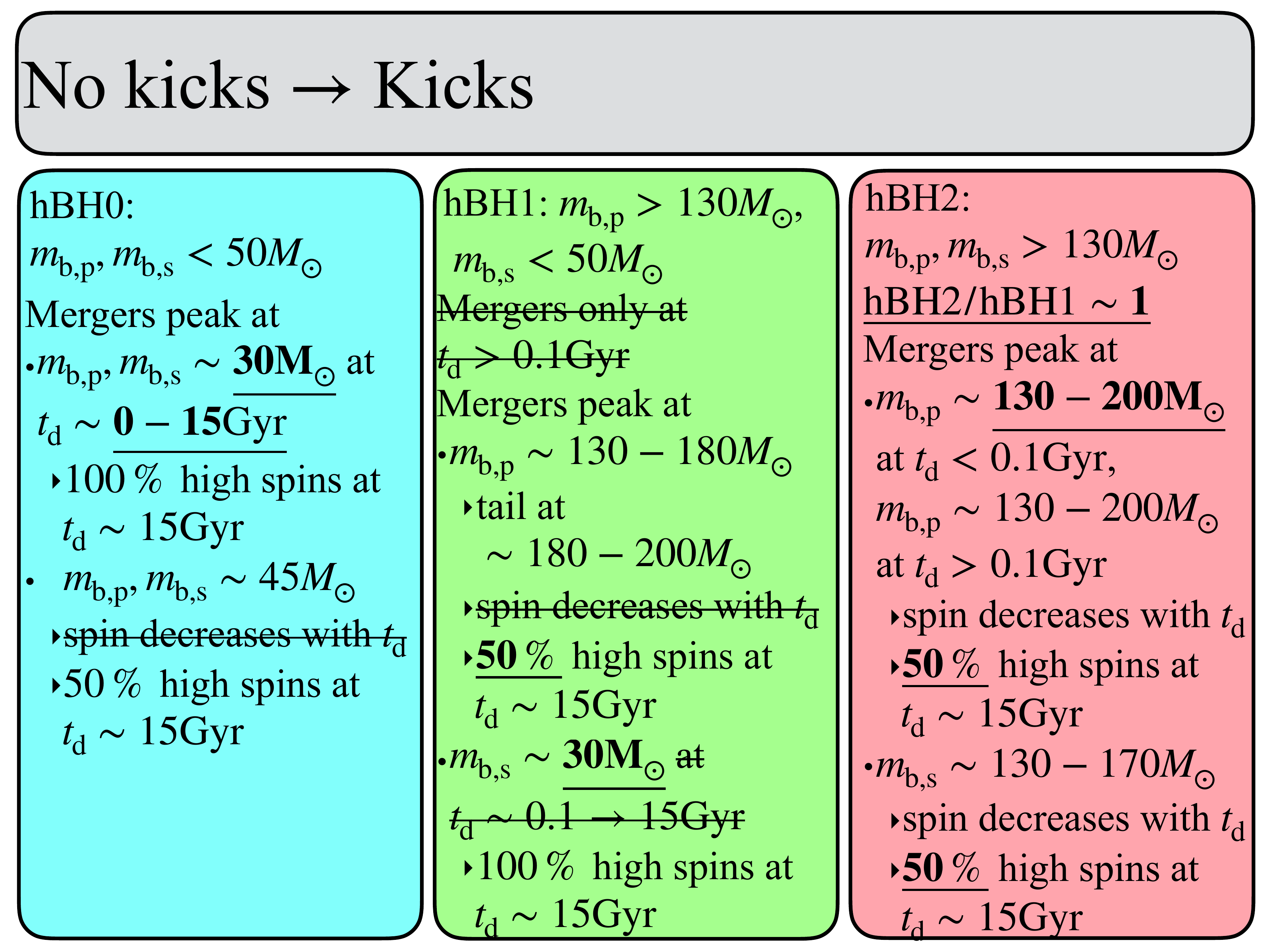}
  \includegraphics[width=0.5\textwidth]{\fdir/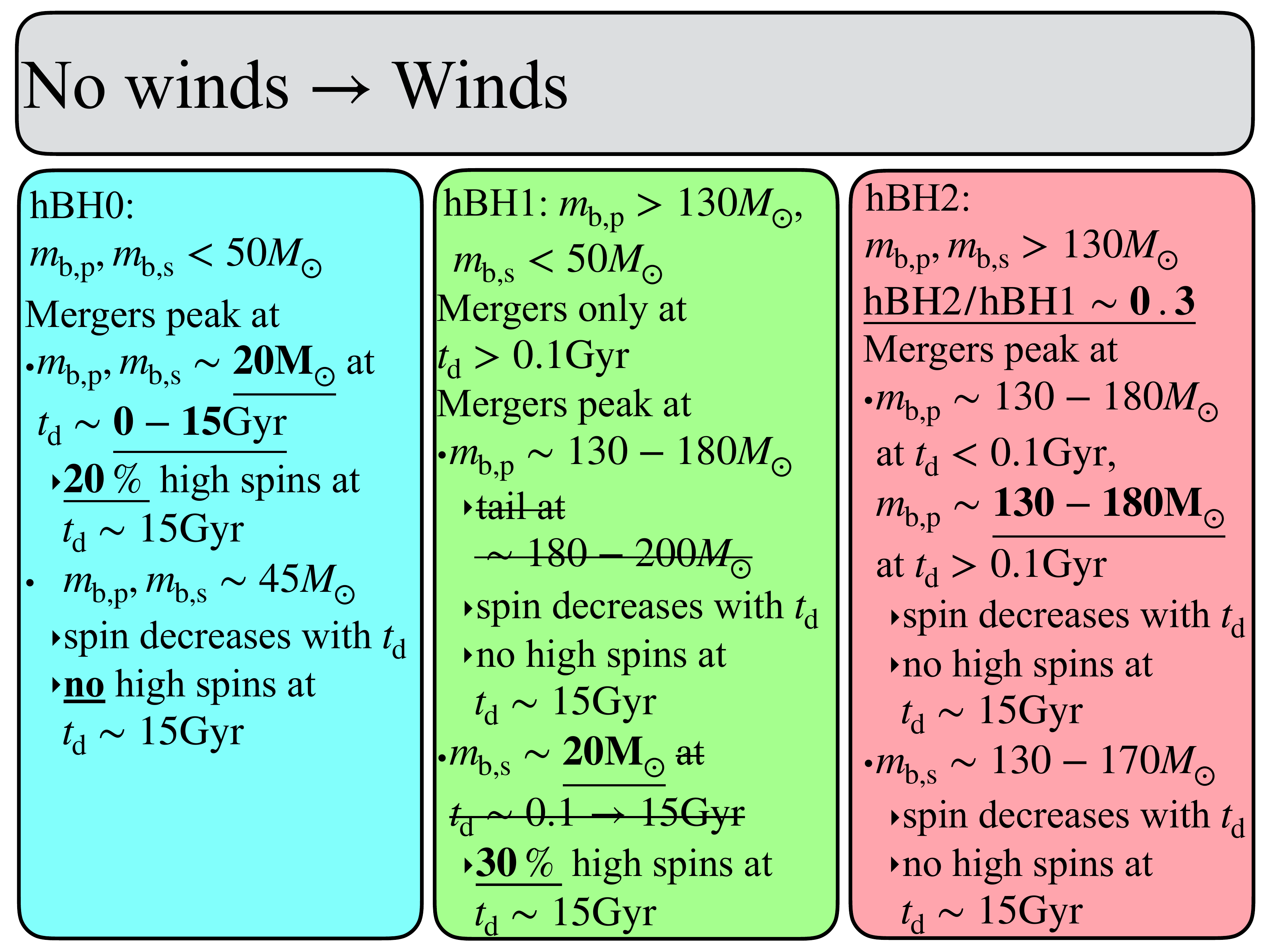}
  \caption{Features of hBH0s, hBH1s, and hBH2s for the optiq0.0a1e1
    model, and their variations with initial conditions and stellar
    evolution models. We regard spin magnitude of $\sim 0.3$ as the
    boundary between high and low spins. The ratio of hBH2s to hBH1s
    is indicated by ${\rm hBH2}/{\rm hBH1}$. We cross out disappearing
    features, and emphasize changed values with bold font and
    underline. For hBH2s, $\mbp \lesssim 200$ and $\lesssim 180 \msun$
    indicate the parallelogram and triangle shapes in 2D mass
    distribution, respectively. Note that these features consider only
    cases where only one of initial conditions and stellar models are
    changed.}
  \label{fig:feature}
\end{figure*}

\begin{figure*}[ht!]
  \begin{center}
  \includegraphics[width=0.65\textwidth]{\fdir/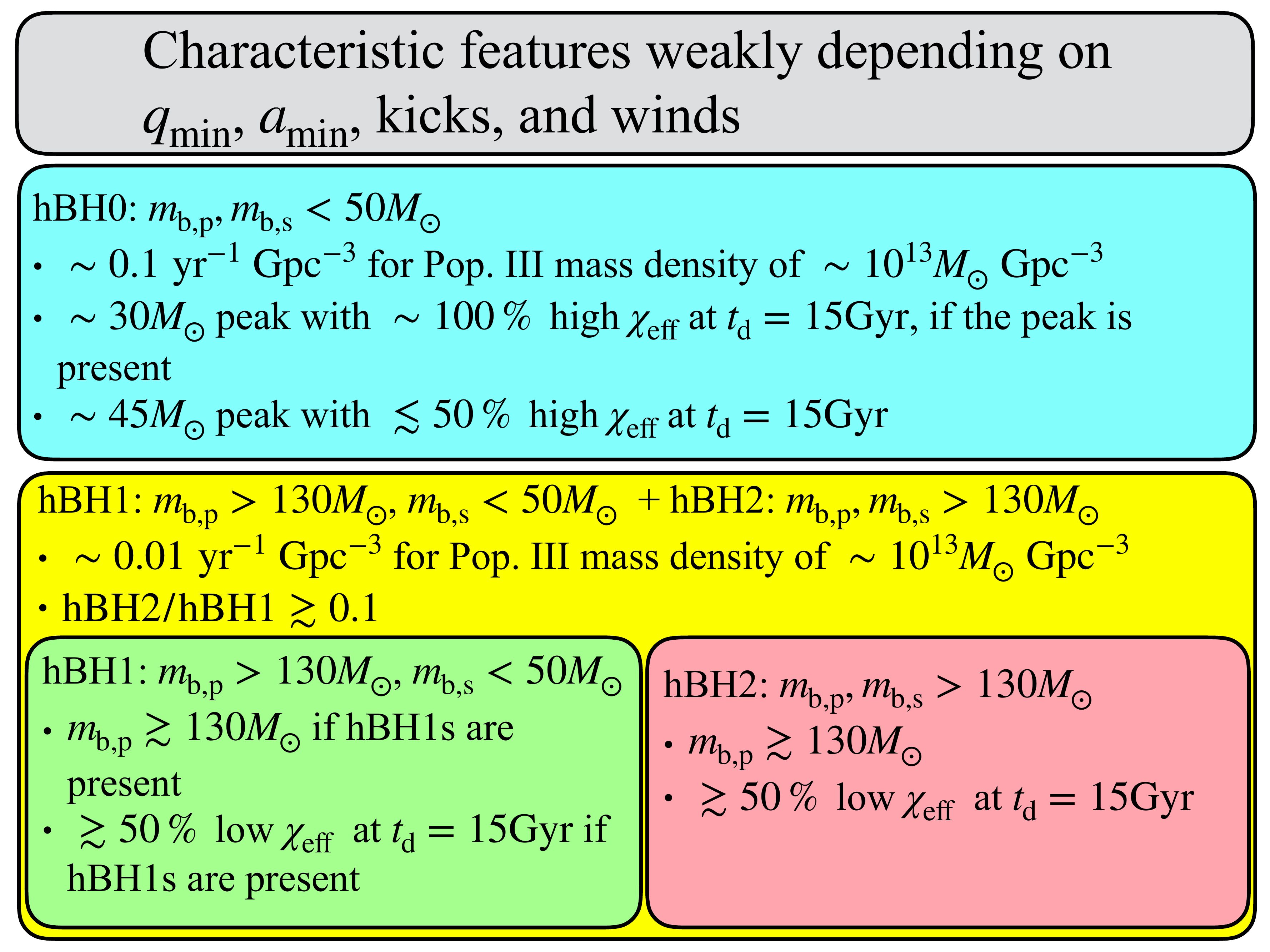}
  \end{center}
  \caption{Characteristic features weakly depending on initial
    conditions ($\qmin$ and $\amin$) and stellar evolution models
    (stellar winds and natal kicks).}
  \label{fig:robust}
\end{figure*}

In the following sections, we investigate in detail hBH0, hBH1, and
hBH2 in sections~~\ref{sec:hBH0}, \ref{sec:hBH1}, and \ref{sec:hBH2},
respectively. In advance, Figure~\ref{fig:feature} summarizes their
features, and Figure~\ref{fig:robust} shows their characteristic
features that have weak dependences on initial conditions ($\qmin$ and
$\amin$) and stellar evolution models (stellar winds and natal
kicks). They will help reading the following sections.

\subsubsection{Binaries with two low-mass BHs (hBH0s)}
\label{sec:hBH0}

We investigate first the optiq0.0a1e1 model, and next the other main
models. This is because hBH0s in the opti0.0a1e1 model have all the
features the other models have, and those in the other models have
a subset of them, as revealed later.

\begin{figure*}[ht!]
  \plottwo{\fdir/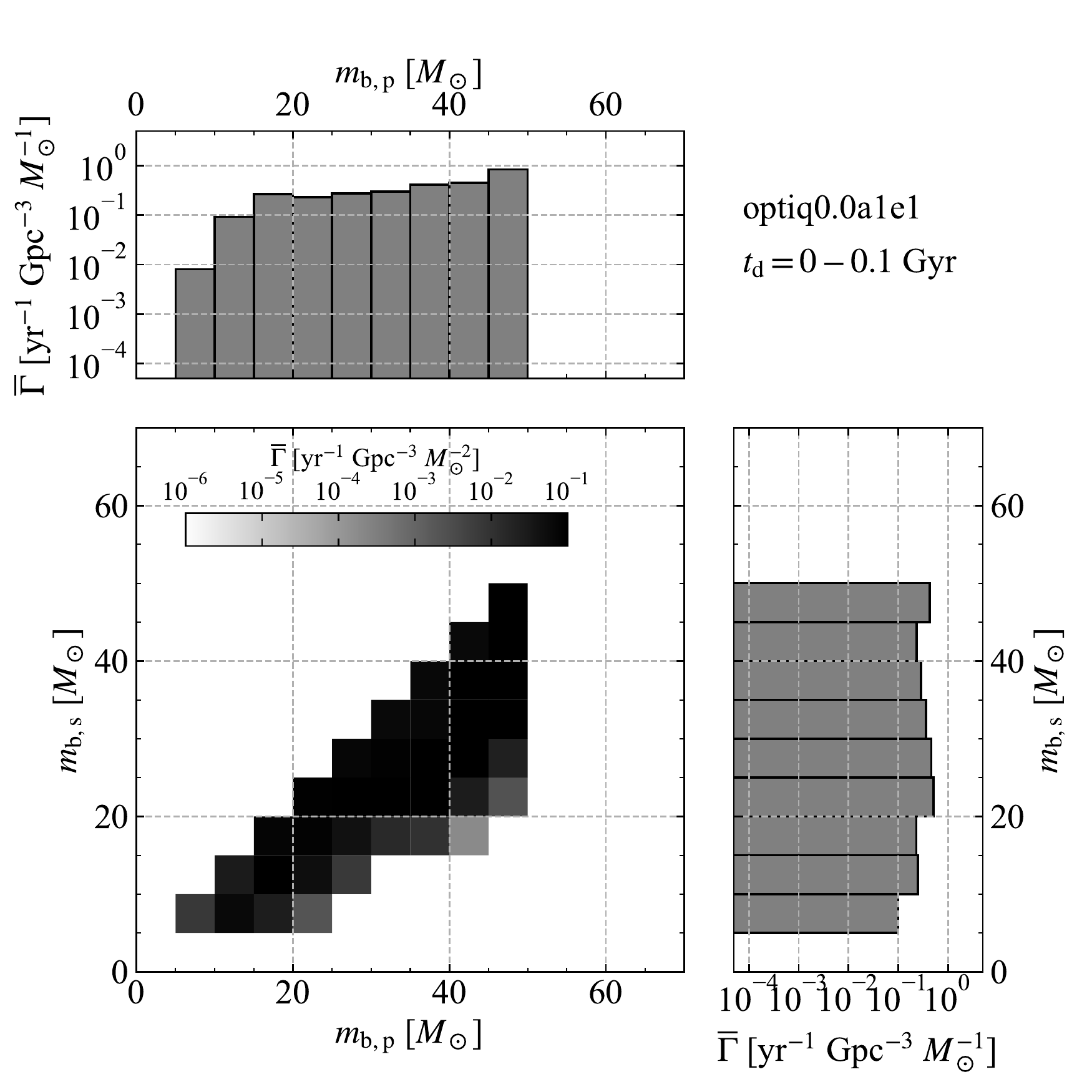}{\fdir/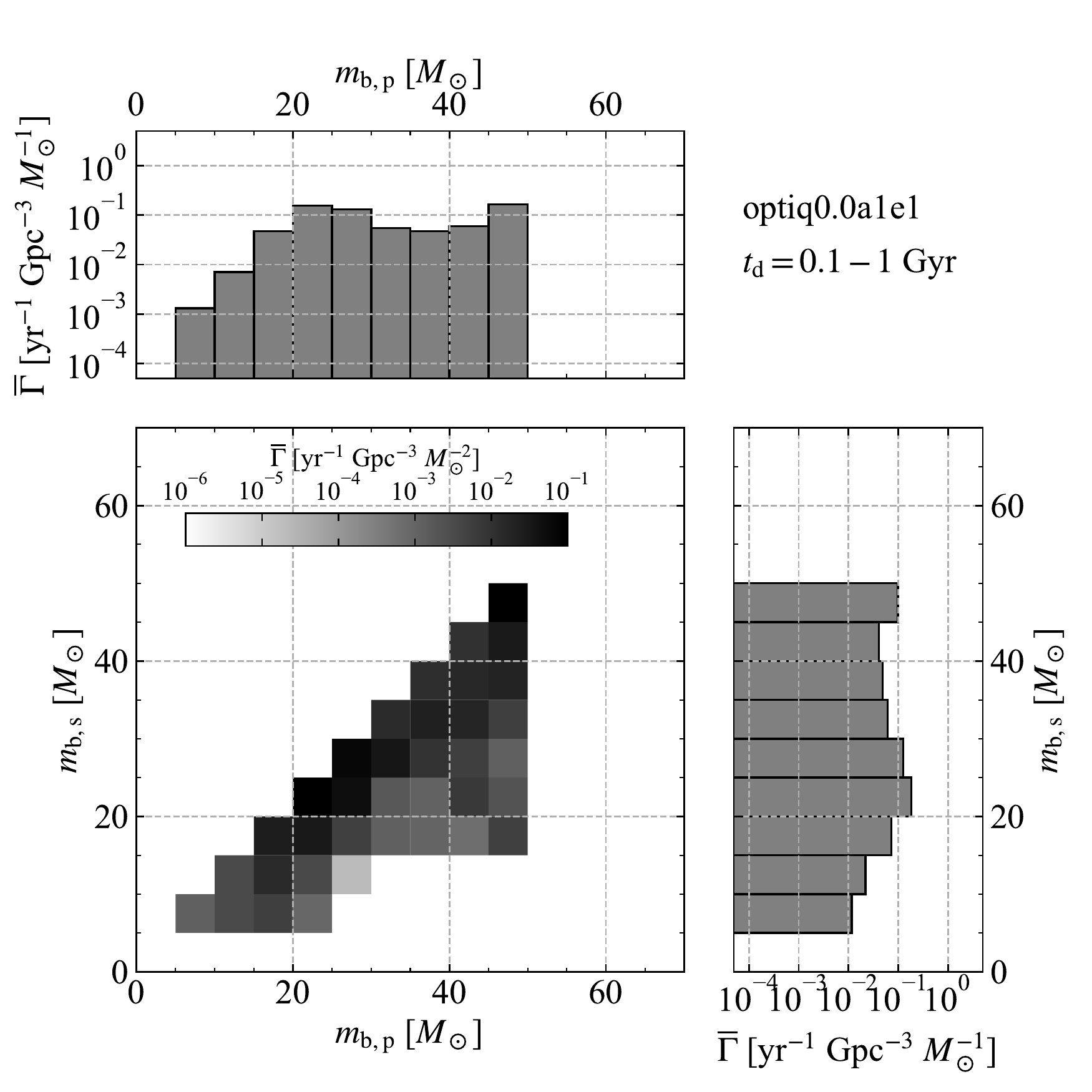}
  \plottwo{\fdir/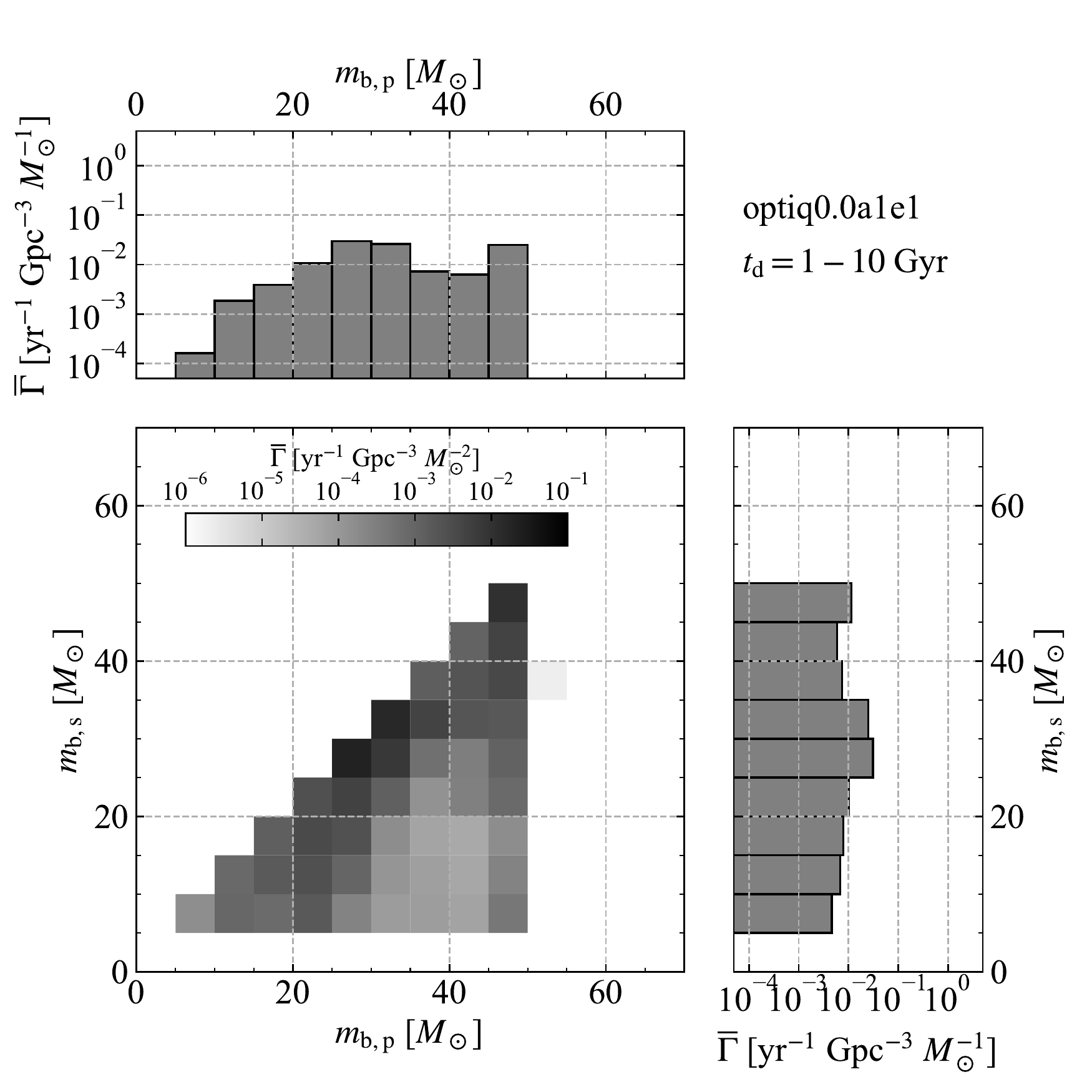}{\fdir/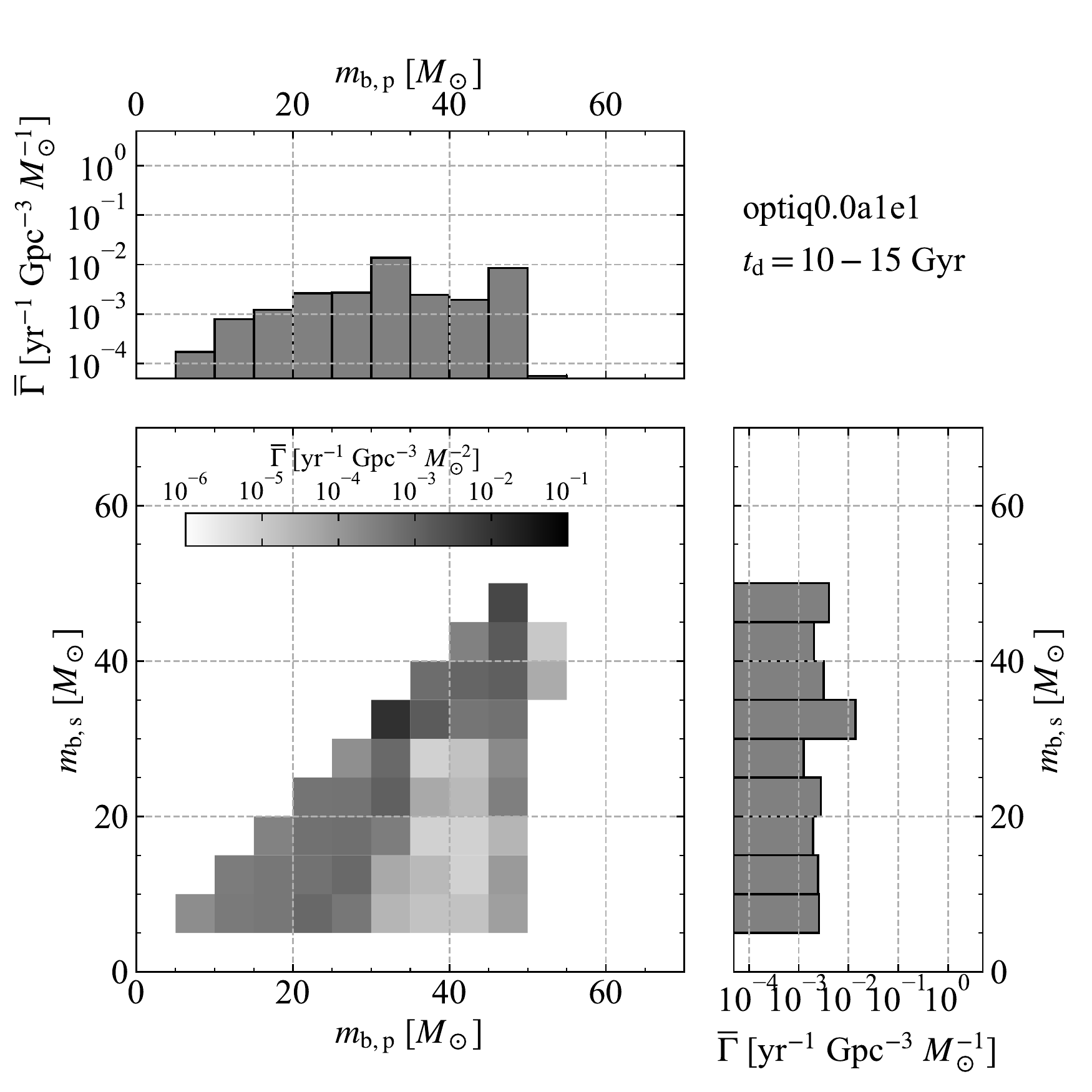}
  \caption{Merger rate densities of hBH0s for $\td = 0-0.1$ (top left),
    $0.1-1$ (top right), $1-10$ (bottom left), and $10-15$~Gyr (bottom
    right) in the optiq0.0a1e1 model.}
  \label{fig:hbh0MassDist_optiq0a1}
\end{figure*}

Figure~\ref{fig:hbh0MassDist_optiq0a1} shows the merger rate density
of hBH0s for each $\td$ in the optiq0.0a1e1 model. We can find two
peaks in the $\mbp$ and $\mbs$ distributions for $\td =
0.1-15$~Gyr. One peak is at $m_{\rm b,p} \sim m_{\rm b,s} \sim 45$ --
$50 \msun$. We call this peak ``the higher-mass peak''. The
higher-mass peak appears for all $\td$ (including $\td =
0-0.1$~Gyr). The other peak is at $m_{\rm b,p} \sim m_{\rm b,s} \sim
20 - 35 \msun$ and for $\td=0.1-15$~Gyr, named ``the lower-mass
peak''.  We do not regard that a population of $m_{\rm b,p}=15-20
\msun$ for $\td = 0-0.1$~Gyr is associated with the lower-mass peak,
since it is too weak.  The lower-mass peak shifts from $(m_{\rm
  b,p},m_{\rm b,s}) \sim (20 \msun, 20 \msun)$ to $(m_{\rm b,p},m_{\rm
  b,s}) \sim (35 \msun, 35 \msun)$ with $\td$. The merger rate density
is decreased sharply at both the ends of $\sim 5$ and $50
\msun$. There is no $3-5 \msun$ BHs, the so-called ``lower mass gap'',
since the rapid model never forms BHs in the lower mass gap. The
number of BHs in the PI mass gap is quite small due to PPI/PISN
effects. A few BHs in the PI mass gap can be formed from stars with
light He cores and heavy H envelopes (see
Figure~\ref{fig:remnantMass}), or through accretion from their
companion stars. These BHs have at most $55 \msun$, while BHs can have
$\sim 85 \msun$ if they evolve as single stars (see
Figure~\ref{fig:remnantMass}). This is because BH progenitors in
binaries are easy to lose their H envelopes through binary
interactions.

The higher-mass peak appears due to PPI effects. This peak is not
observed in the K14 model which does not account for PPI
effects. Stars with He core masses of $45 - 65 \msun$ experience PPI,
and are thus swept up together to BHs with $45 \msun$. Since PPI
effects are independent of binary star evolution, the higher-mass peak
always appears regardless of $\td$.

The lower-mass peak has two features; it shifts from $\sim 20 \msun$
to $\sim 35 \msun$, and the mass ratio favors nearly unity. The
reasons for the two features can be explained by their formation
channels. Here, we first explain the formation channel, and next the
reasons for the features. The lower-mass peak is formed through the
same mechanism as the peak of $\sim 30 \msun$ in the K14 model,
i.e. the CE0 channel. The detail processes are as follow. The
1st-evolving star enters into its CHeB phase, fills its Roche lobe,
and starts stable mass transfer to the 2nd-evolving star. The mass
transfer can keep stable, since a Pop.~III star with $\mzams = 10-50
\msun$ ends its life with a BSG star (see
Figure~\ref{fig:hrd}). Subsequently, the 1st-evolving star collapses
to a BH. After that, the 2nd-evolving star also fills its Roche lobe,
and begins stable mass transfer. Finally, the 2nd-evolving star
collapses to a BH.

The reason for the mass ratio of unity is as follows. The 1st mass
transfer makes the 2nd-evolving star heavier than the 1st BH. Since
mass transfer is more efficient when the donor star is heavier than
the accreting star, the 2nd mass transfer continues until the mass
ratio of the 2nd-evolving star to the 1st BH becomes close to
unity. After that, the 2nd-evolving star collapses to a BH with a
similar mass to the 1st BH. Thus, these BH-BHs tend to have the mass
ratio of unity.

The reason for the time shift of the lower-mass peak is as
follows. The merging timescale through GW radiation is expressed as
\begin{align}
  t_{\rm GW} &= \frac{5}{256} \frac{c^5}{G^3} \frac{a^4}{m_{\rm
      b,p}m_{\rm b,s}(m_{\rm b,p}+m_{\rm b,s})} g(e) \label{eq:tgw} \\
  g(e) &= \frac{(1-e^2)^{3.5}}{1+(73/24)e^2+(37/96)e^4}, \label{eq:ge}
\end{align}
where $a$ and $e$ are the semi-major axis and eccentricity of a BH-BH.
Since $m_{\rm b,p} \sim m_{\rm b,s}$ and $e \sim 0$ here, we reduce
these equations to $t_{\rm GW} \propto a^4m^{-3}$, where we use $m$
instead of $m_{\rm b,p}$ and $m_{\rm b,s}$. The semi-major axis is
proportional to the radius of the 2nd-evolving star, $a \propto R_2$,
when the 2nd mass transfer occurs. Since $R_2 \propto L_2^{1/2}$, and
nearly $L_2 \propto m^2$ for stars with $10-50 \msun$ in CHeB and
ShHeB phases (see Figure~\ref{fig:hrd}), $a \propto m$. Finally,
$t_{\rm GW} \propto m$. We now assume that the 2nd-evolving star with
different $m$ has the same effective temperature. If we relax this
assumption, the 2nd-evolving star with larger $m$ tends to have lower
effective temperature for $m \gtrsim 20 \msun$ (see
Figure~\ref{fig:hrd}). Then, $t_{\rm GW}$ can depend on $m$ more
strongly. Thus, the time shift of the lower-mass peak appears.

\begin{figure}[ht!]
  \plotone{\fdir/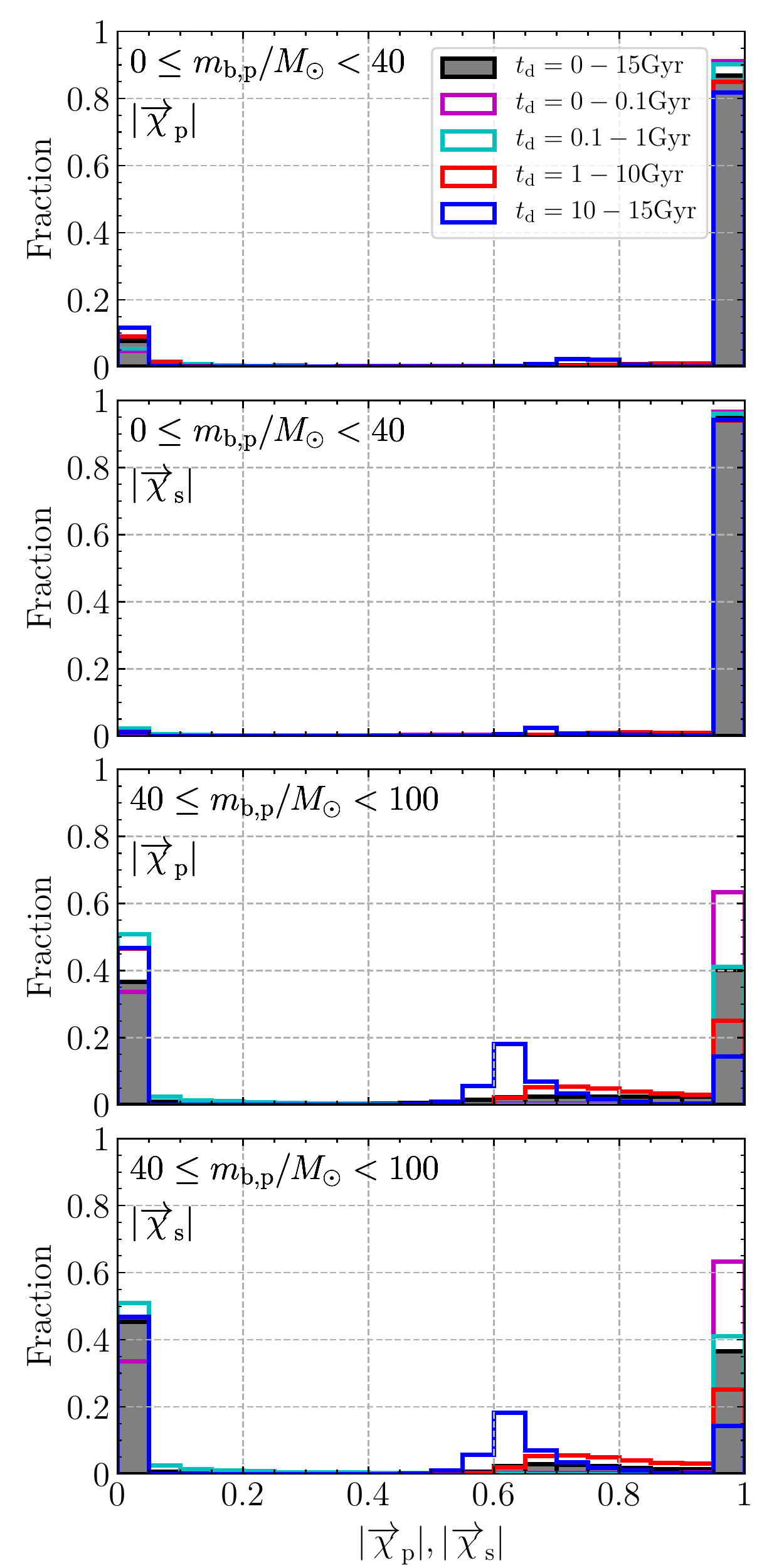}
  \caption{Spin distributions of hBH0s for each $\td$ in the
    optiq0.0a1e1 model. The panels indicate primary and secondary BH
    spins of hBH0 with $0 \le m_{\rm b,p}/\msun < 40$ (the top two
    panels) and $40 \le m_{\rm b,p}/\msun < 100$ (the bottom two
    panels).}
  \label{fig:hbh0SpinMagDistOpti}
\end{figure}

Figure~\ref{fig:hbh0SpinMagDistOpti} shows the spin distributions of
primary and secondary BHs of hBH0s in the optiq0.0a1e1 model. We can
regard that hBH0s with $0 \le m_{\rm b,p}/\msun < 40$ and $40 \le
m_{\rm b,p}/\msun < 100$ are associated with populations of the
lower-mass and higher-mass peaks, respectively. The top two panels
indicate that hBH0s in the lower-mass peak have two BHs with high
spins $\sim 1$ independently of $\td$. The reason is as follows. Their
progenitors can be highly spun up by tidal fields of their companion
stars, since they expand to their Roche-lobe radii. Moreover, they
keep their H envelopes more or less until they collapse to BHs.

On the other hand, half of hBH0s in the higher-mass peak have BHs with
nearly zero spins. This is due to PPI effects. We assume BH
progenitors undergoing PPI lose their spin angular momenta. The other
half have high spins with $\gtrsim 0.5$, and however have smaller
spins with $\td$ increasing. The reason for the time dependence is as
follows. BH progenitors gain and lose their spin angular momenta
through tidal interactions and mass transfer, respectively. As the
$\td$ increases, binary separation becomes larger, and tidal
interaction becomes weaker. Thus, a BH progenitor decreases its spin
angular momentum in the net, and its BH has a lower spin. The details
can be described in many studies
\citep{2016MNRAS.462..844K,2017ApJ...842..111H,2020ApJ...892...64P,2020ApJ...897L...7S}.
We call this relation the spin-$\td$ relation tentatively.

With respect to spins, the hBH0s have four subpopulations; both the
BHs have high spins ($\sim 1$), both the BHs have low spins ($\sim
0$), and the 1st (2nd) BH has high spin ($\sim 1$) but the other BH
has low spin ($\sim 0$). This is different from the results of
\cite{2016PTEP.2016c1E01K,2016PTEP.2016j3E01K}. In their model, the
hBH0s do not have a subpopulation with the 1st BHs with high spins and
the 2nd BHs with low spins. We obtain the subpopulation, since we
assume BHs after PPI have zero spins.

Hereafter, we assess the dependence of these features on initial
conditions of binaries ($\qmin$ and $\amin$), and single star models
(stellar winds and natal kicks). First, we focus on the dependence on
initial conditions.

\begin{figure*}[ht!]
  \plottwo{\fdir/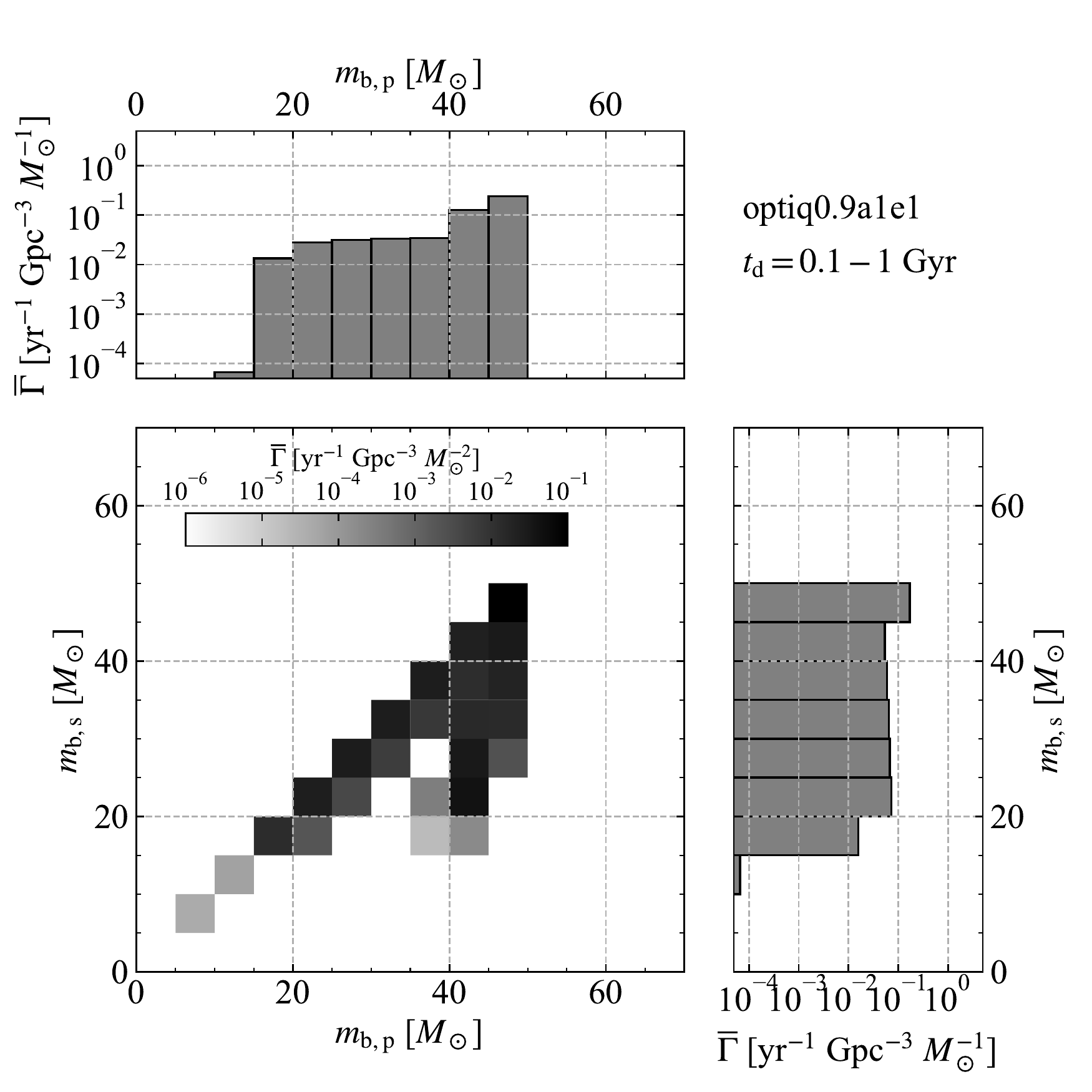}{\fdir/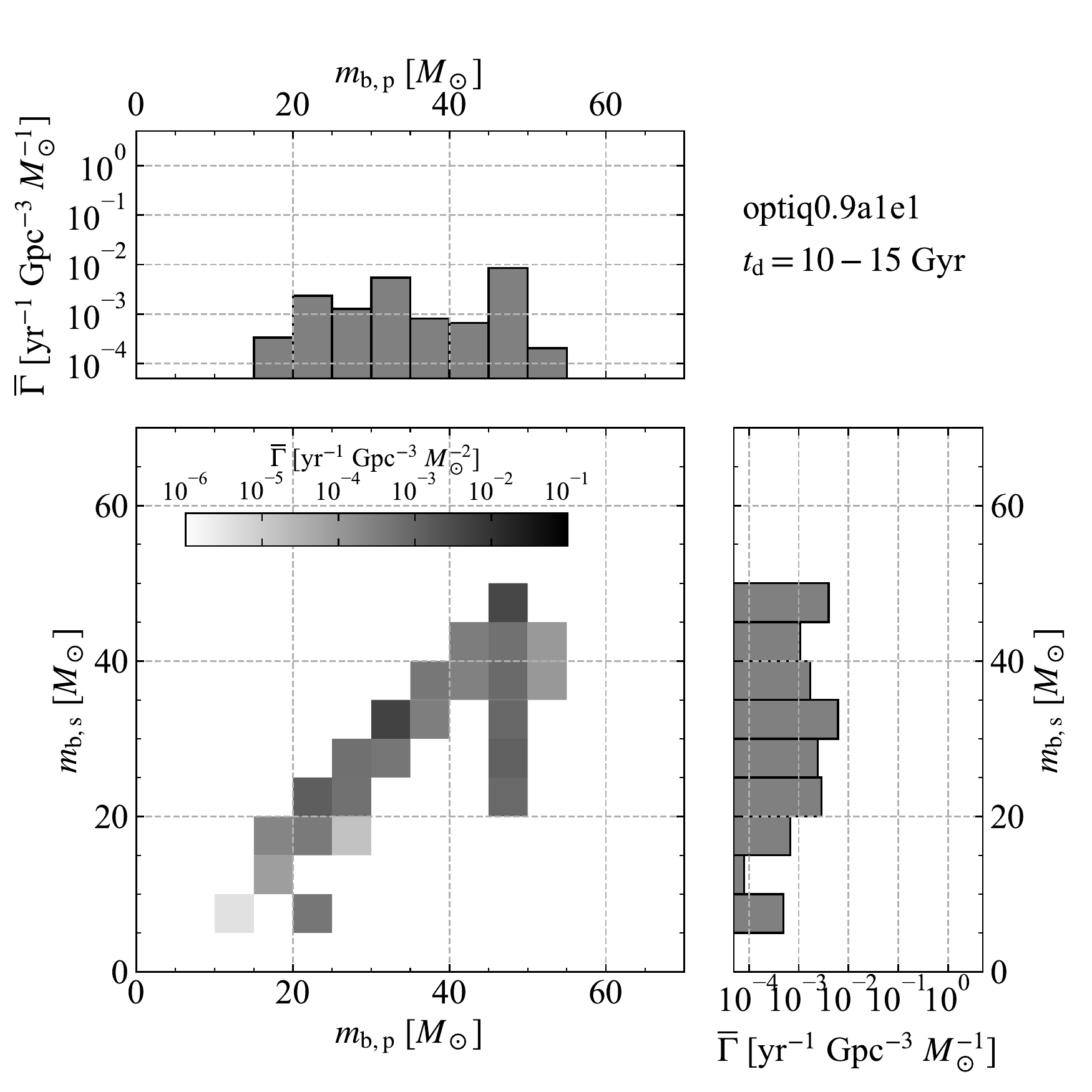}
  \plottwo{\fdir/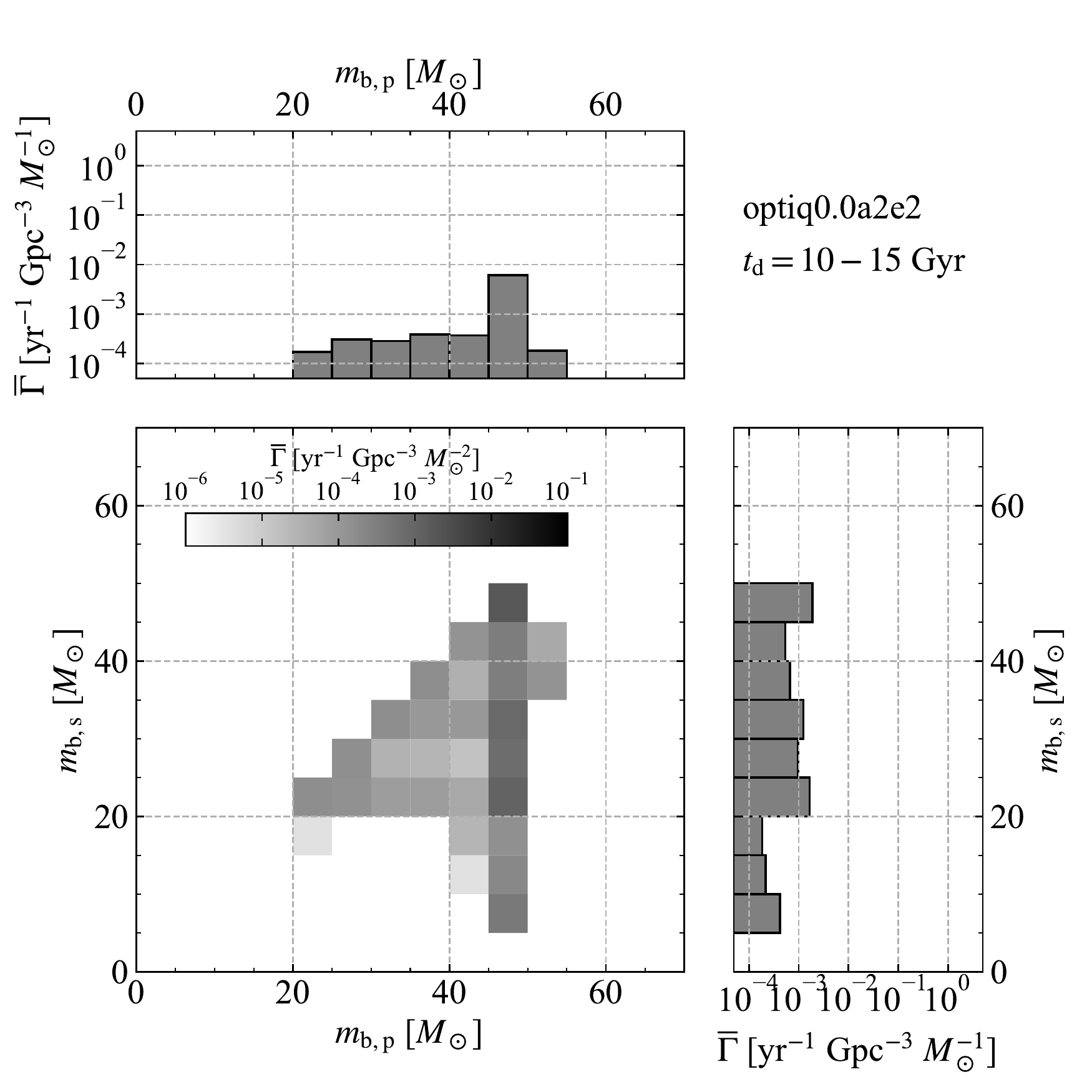}{\fdir/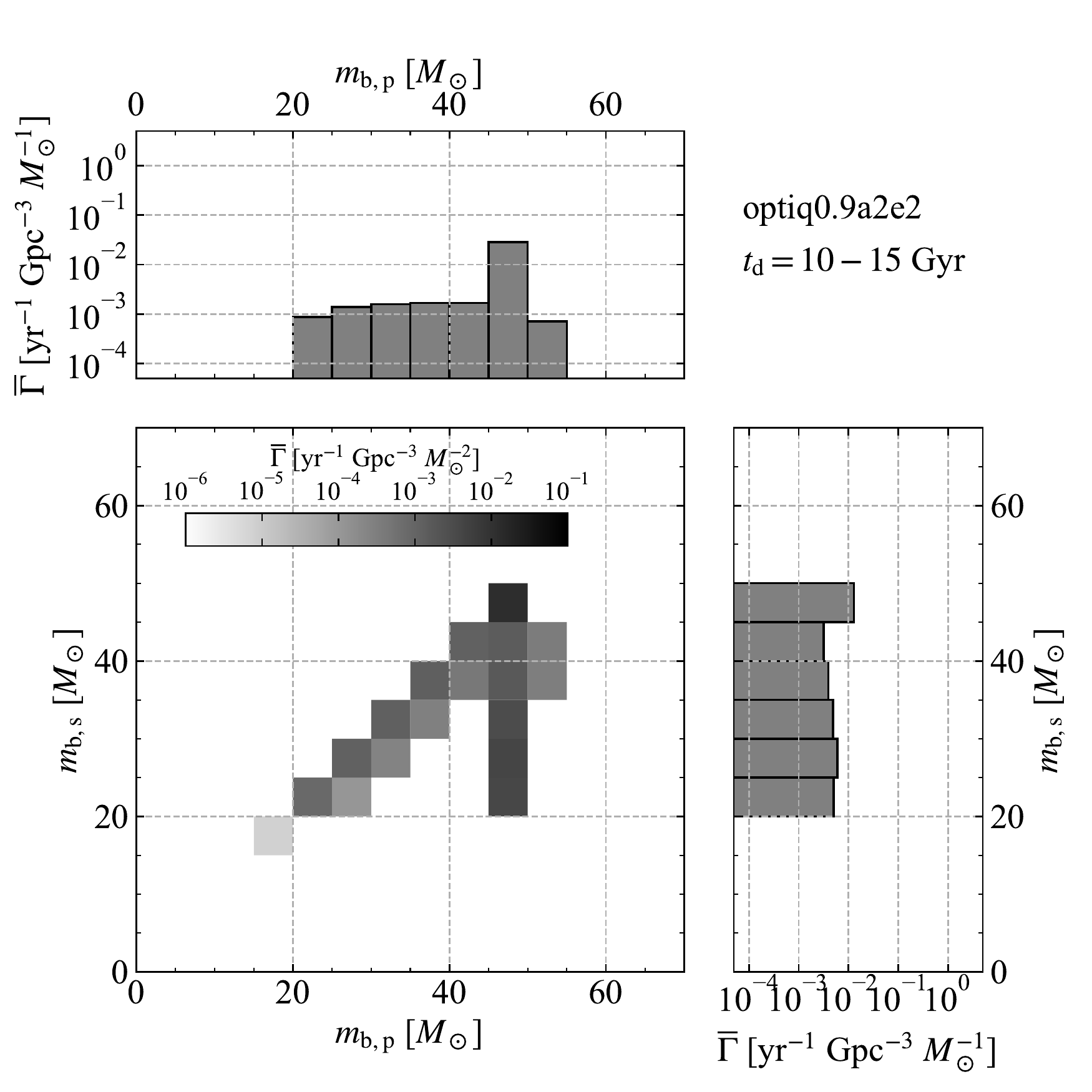}
  \caption{Merger rate densities of hBH0s. The top panels show the
    results of the optiq0.9a1e1 model for $\td = 0.1-1$ (left) and
    $10-15$~Gyr (right). The bottom panels show the results of the
    optiq0.0a2e2 (left) and optiq0.9a2e2 models (right) for $\td =
    10-15$~Gyr.}
  \label{fig:hbh0MassDist_icdependent}
\end{figure*}

Figure~\ref{fig:hbh0MassDist_icdependent} shows merger rate densities
of hBH0s in models with different initial conditions from the
optiq0.0a1e1 model. We can find peaks at $(m_{\rm b,p},m_{\rm b,s})
\sim (45 \msun, 45 \msun)$ in all the panels. These peaks correspond
to the higher-mass peak in the optiq0.0a1e1 model. Even in the
optiq0.0a2e2 and optiq0.9a2e2 models, there can be hBH0s with $\td \le
15$~Gyr. This is because stars with $\mzams \sim 100 \msun$ can expand
to $\gtrsim 10^3 \rsun$ (see Figure~\ref{fig:hrd}), interact with
their companion stars through common envelope evolutions, and leave
$45 \msun$ BHs (see Figure~\ref{fig:remnantMass}). Consequently, the
higher-mass peak in the optiq0.0a1e1 model is present regardless of
different initial conditions ($\qmin$ and $\amin$).

We cannot find the lower-mass peak in the optiq0.0a2e2 and
optiq0.9a2e2 model. In these models, hBH0s with $\td \le 15$~Gyr
cannot be formed through the CE0 channel. Since the CE0 channel does
not reduce binary separations, hBH0s formed through the CE0 channel
have binary separations $\gtrsim 100 \rsun$. Their merging timescales
through GW radiation are much larger than $15$~Gyr. We can find the
lower-mass peak in the optiq0.9a1e1 model for $\td = 10-15$~Gyr, which
is formed through the same mechanism in the optiq0.0a1e1 model. On the
other hand, we cannot find the lower-mass peak in this model for $\td
= 0.1-1$~Gyr. Since two stars with similar masses expand on similar
timescales, they fill their Roche lobes, and merge when their
separation is small. Eventually, the lower-mass peak in the
optiq0.0a1e1 model is sensitive to $\amin$, and its time shift is weak
against $\qmin$.

Although the presence of the higher-mass peak is insensitive to
initial conditions, spin distributions of hBH0s around the higher-mass
peak can be affected by initial conditions. Half of primary and
secondary BHs have nearly zero spins in the optiq0.0a1e1 model, while
more than half of them have nearly zero spins in the other models, and
moreover the fraction increases with time up to unity for $\td =
10-15$~Gyr. This is because the CE0 channel has little contribution to
the higher-mass peak in the other models. On the other hand, spin
distributions of hBH0s around the lower-mass peak are similar between
the optiq0.0a1e1 and optiq0.9a1e1 models if the lower-mass peak is
present in the optiq0.9a1e1 model.

\begin{figure*}[ht!]
  \plottwo{\fdir/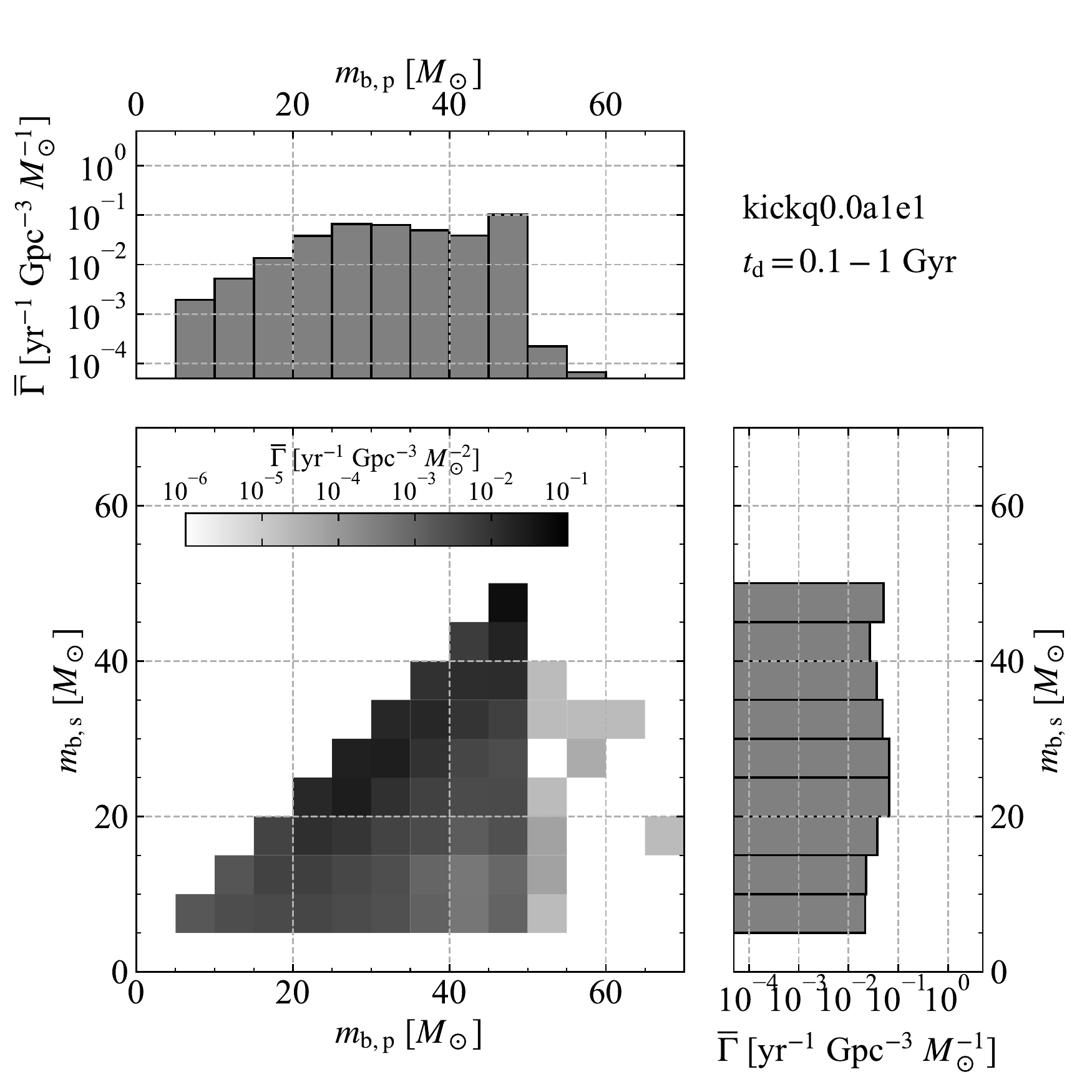}{\fdir/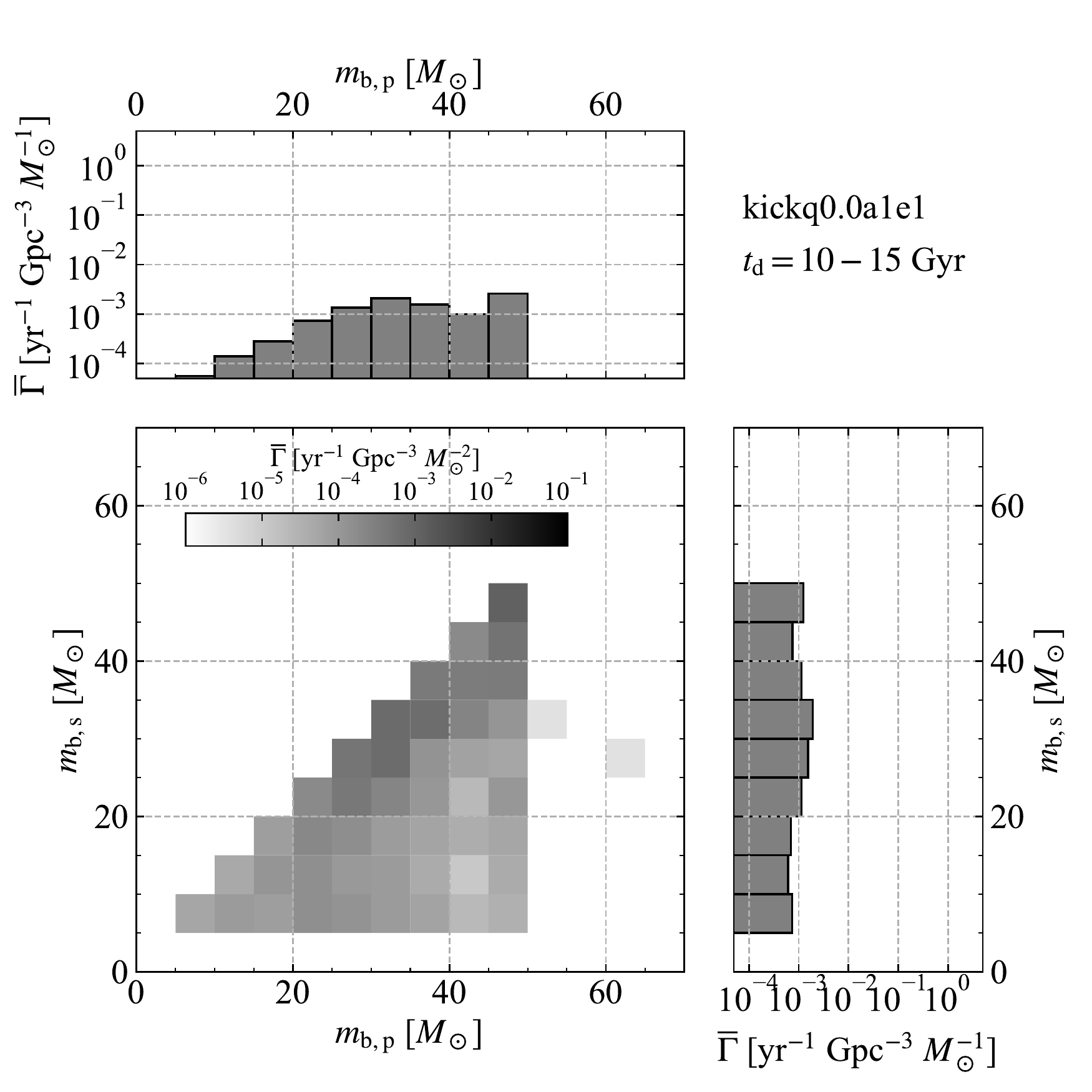}
  \plottwo{\fdir/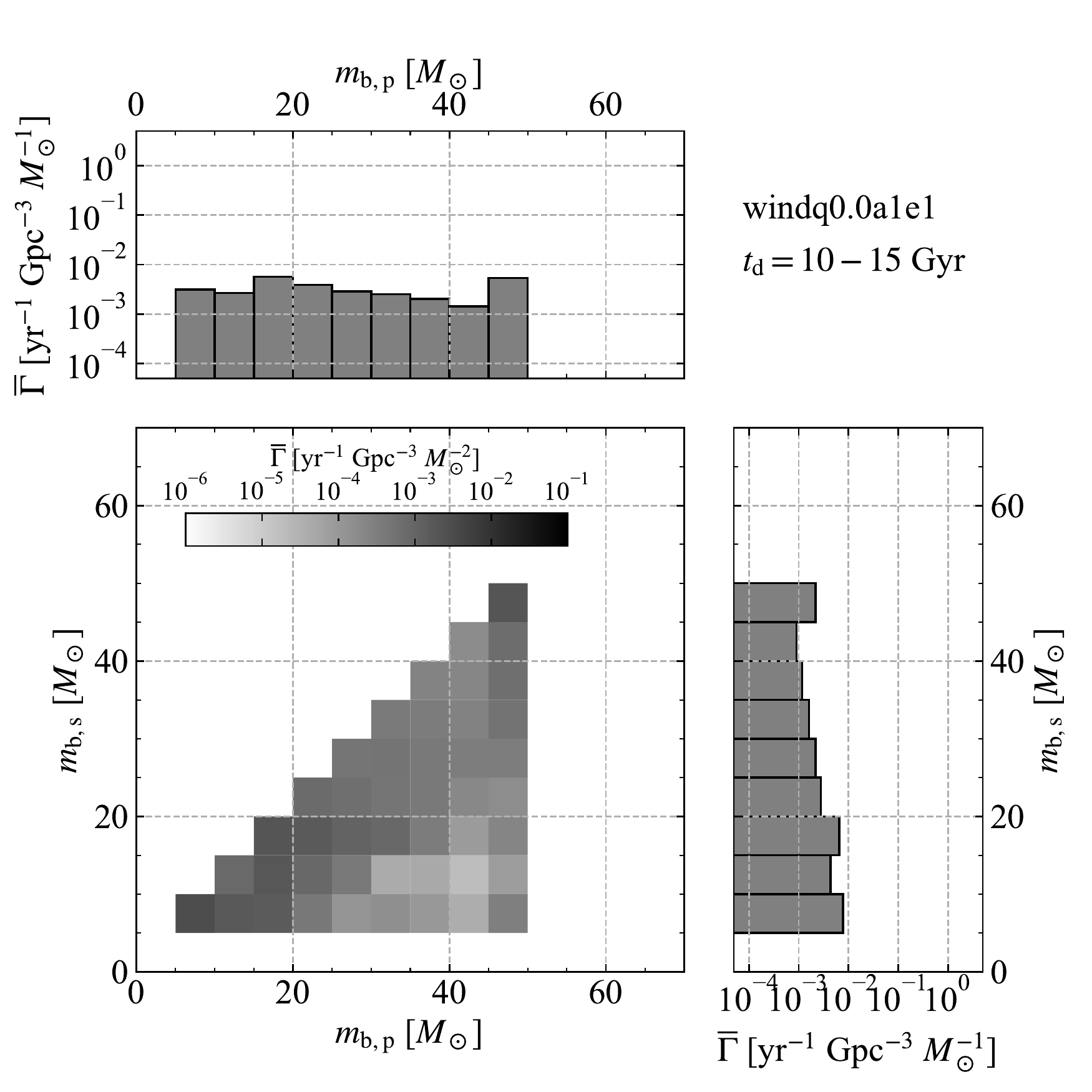}{\fdir/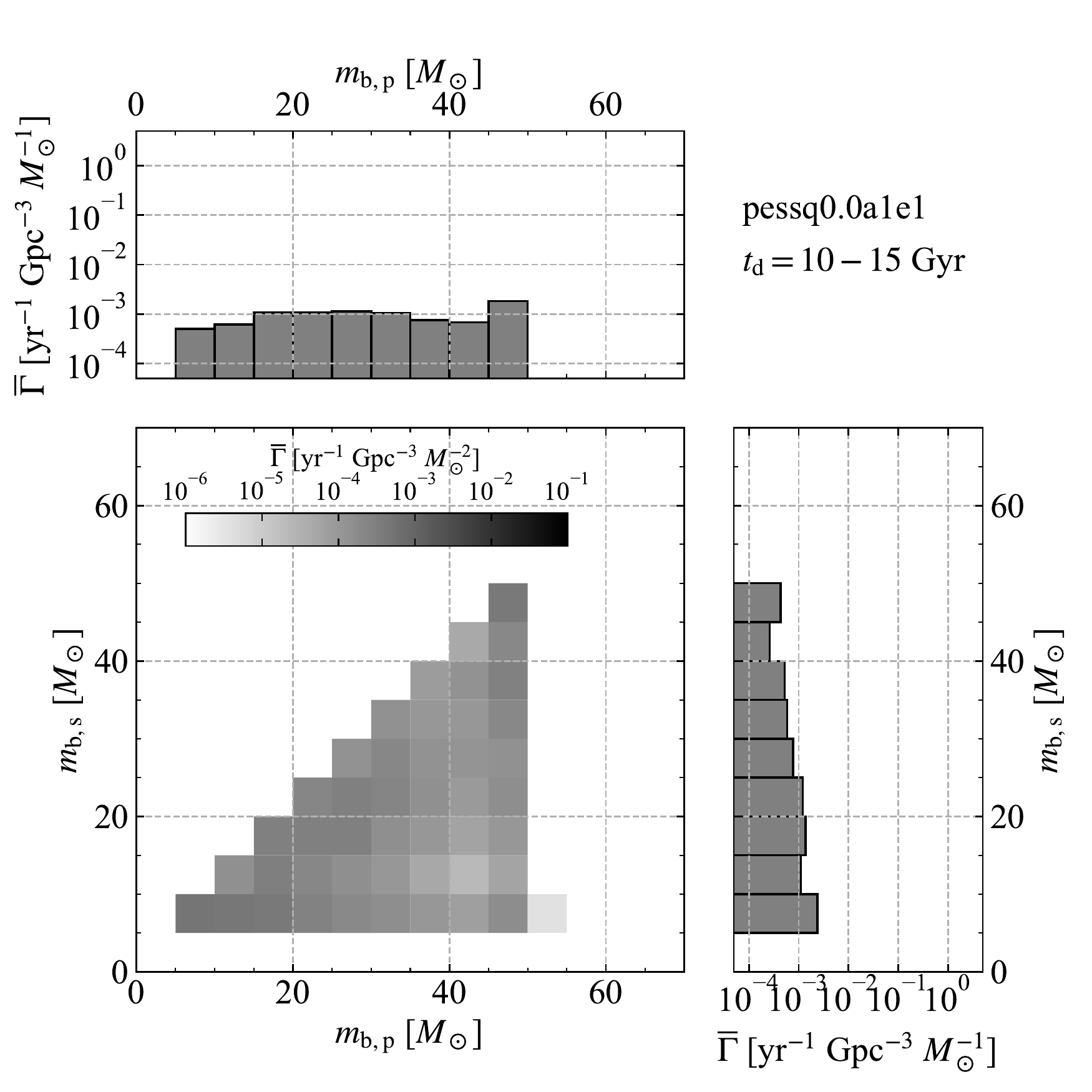}
  \caption{Merger rate densities of hBH0s. The top panels show the
    results of the kickq0.0a1e1 model for $\td = 0.1-1$ (left) and
    $10-15$~Gyr (right). The bottom panels show the results of the
    windq0.0a1e1 (left) and pessq0.0a1e1 models (right) for $\td =
    10-15$~Gyr.}
  \label{fig:hbh0MassDist_modeldependent}
\end{figure*}

Next, we focus on the dependence on single star
models. Figure~\ref{fig:hbh0MassDist_modeldependent} shows the merger
rate densities of hBH0s in models with different single star
models. We can find peaks at $(\mbp, \mbs) \sim (45 \msun, 45 \msun)$
in all the panels. These peaks correspond to the higher-mass peak in
the optiq0.0a1e1 model. Therefore, the higher-mass peak is present
with or without stellar winds and natal kicks.

The kickq0.0a1e1 model has the lower-mass peak of $(m_{\rm b,p},m_{\rm
  b,s}) \sim (30 \msun, 30 \msun)$ independently of $\td$ (see the top
panels). In other words, the lower-mass peak has no time shift, unlike
in the optiq0.0a1e1 model. This is because natal kicks shorten and
lengthen merging timescales of hBH0s, independently of their
masses. For example, if a BH progenitor gets a natal kick whose
direction is the same as and opposite to its traveling direction, the
resulting merging timescale becomes longer and shorter,
respectively. Whether the time shift can survive or not depends on the
magnitude of natal kicks. We can observe the time shift in the
half-kick model, although it is less distinct than in the optiq0.0a1e1
model. If we assume $\sigma_{\rm k}=265$~km~s$^{-1}$ for NSs and kick
velocities inversely proportional to remnant masses as in the commonly
used prescription, the time shift can be seen as prominent as in the
optiq0.0a1e1 model, since the kick velocities are $\sim
10$~km~s$^{-1}$ for BHs with several $10 \msun$.

Stellar winds significantly smooth the lower-mass peak, and shift it
to $\sim 20 \msun$ (see the bottom panels). The peak shift is simply
because BHs lose their masses through stellar winds. This is {\it not}
due to stellar winds of high metallicity, $10^{-3} \zsun$. In the
weak-wind model ($Z=10^{-5} \zsun$), hBH0s have merger rate densities
similar to those in the windq0.0a1e1 model. In fact, rotationally
enhanced winds are responsible. If we switch off the rotationally
enhanced winds in the no-rot-wind model, we find the merger rate
densities similar to those in the optiq0.0a1e1 model. It is natural
that the rotationally enhanced winds have strong effects in models
with $\amin=10 \rsun$, since most of stars are effectively spun up by
tidal fields of their companion stars. Eventually, the lower-mass peak
is fragile against stellar winds with rotational enhancement. Although
the lower-mass peak is present despite of the presence of natal kicks,
its shift with $\td$ is erased by the natal kicks.

\begin{figure}[ht!]
  \plotone{\fdir/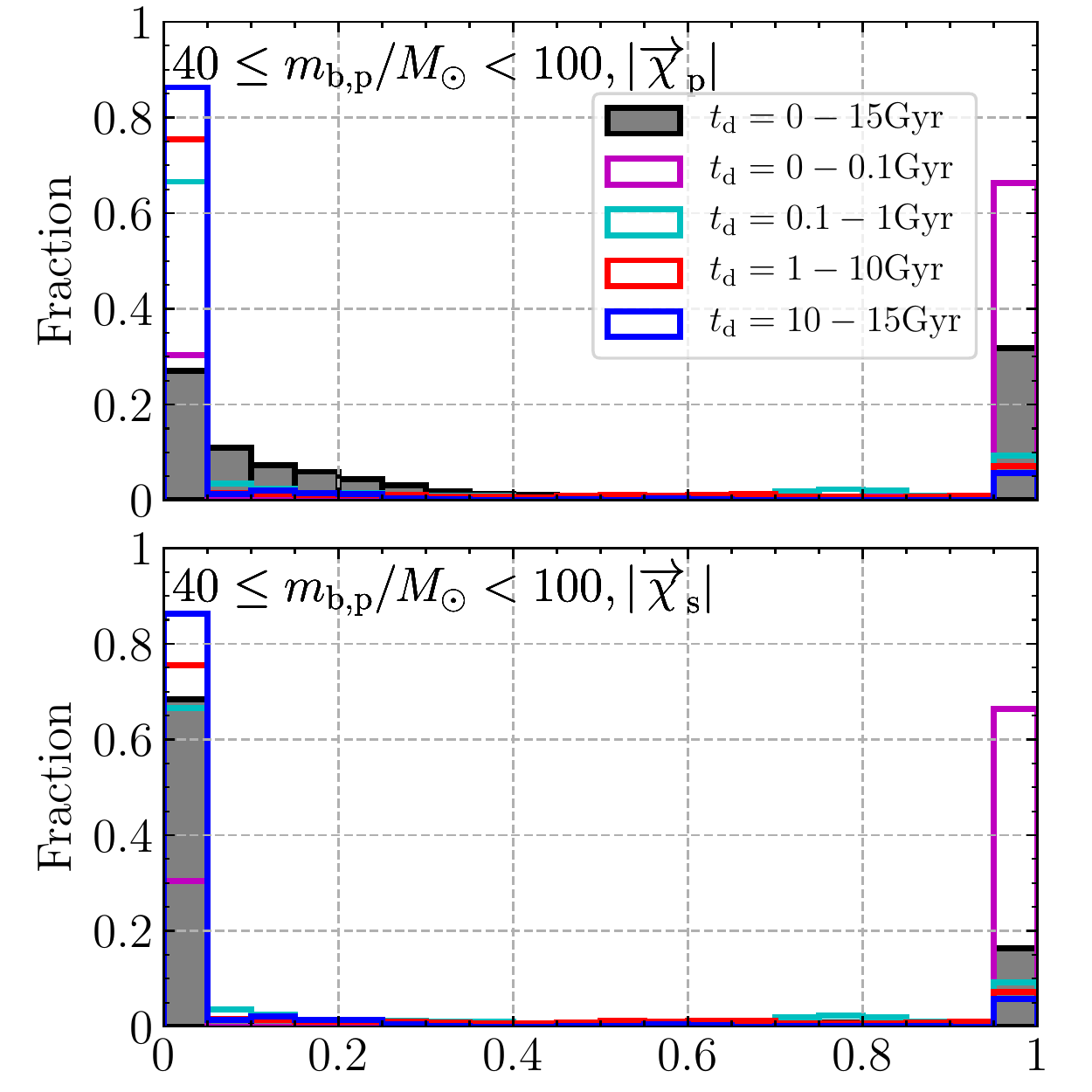}
  \caption{Spin distributions of hBH0s for each $\td$ in the
    windq0.0a1e1 model. The panels indicate primary and secondary BH
    spins of hBH0s with $40 \le \mbp/\msun < 100$.}
  \label{fig:hbh0SpinMagDistWind}
\end{figure}

The distribution of spin magnitudes in the kickq0.0a1e1 model is
similar to in the optiq0.0a1e1 model for both $0 \le \mbp/\msun < 40$
and $40 \le \mbs/\msun < 100$. On the other hand, the fraction of
hBH0s with nearly zero spins in the windq0.0a1e1 and pessq0.0a1e1
models is 2 times larger than in the optiq0.0a1e1 model for $40 \le
\mbs/\msun < 100$ (compare Figure~\ref{fig:hbh0SpinMagDistWind} with
Figure~\ref{fig:hbh0SpinMagDistOpti}). This is because stellar winds
carry away spin angular momenta. We do not show spin distributions of
hBH0s for $0 \le \mbp/\msun < 40$ in these models, since the
lower-mass peak is weak.

\begin{figure}[ht!]
  \plotone{\fdir/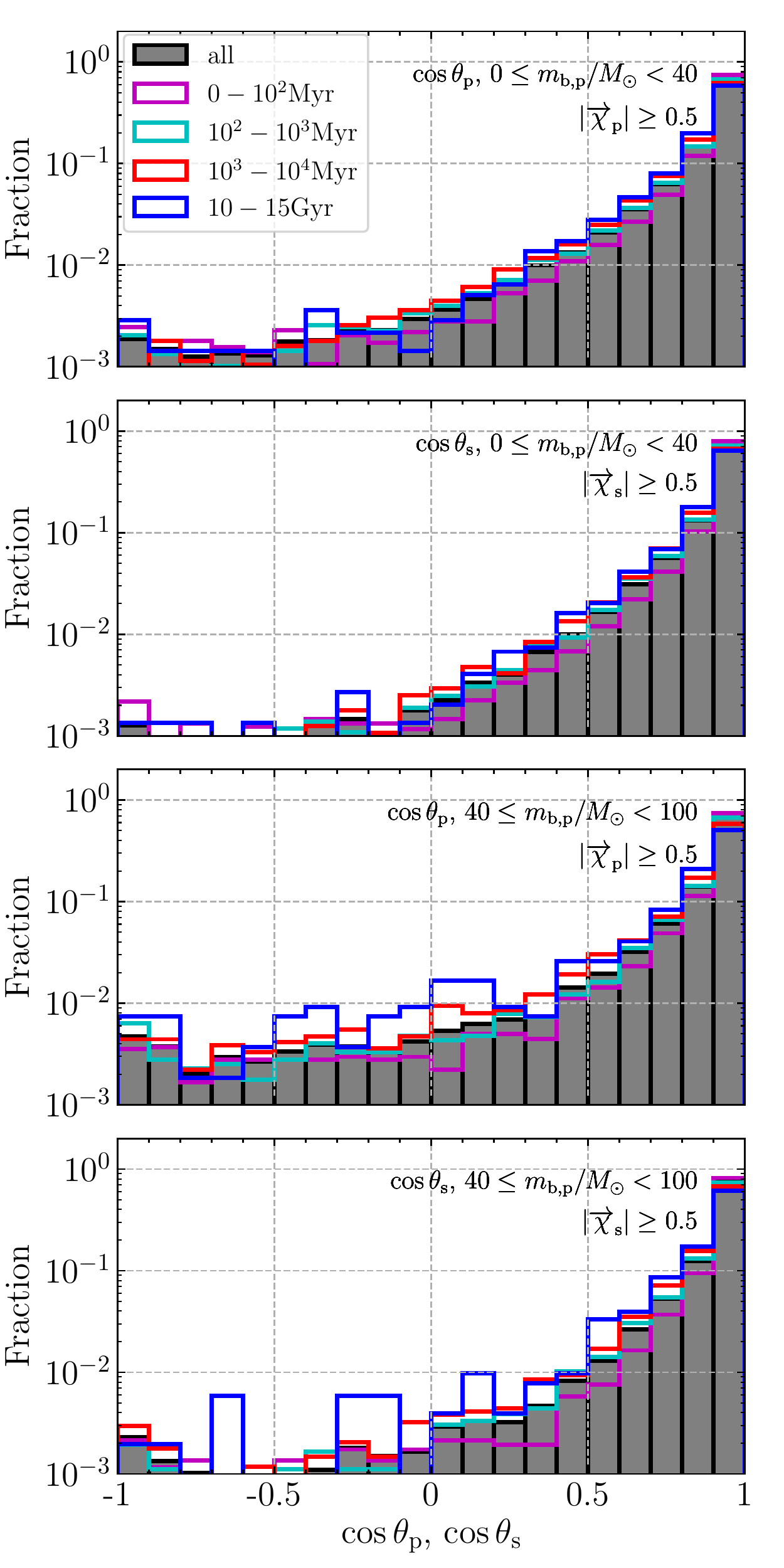}
  \caption{Distribution of angles between primary and secondary BH
    spin vectors and binary orbit vectors for $0 \le \mbp/\msun < 40$
    (the top two panels), and $40 \le \mbp/\msun < 100$ (the bottom
    two panels) in the kickq0.0a1e1 model. The BHs are limited to
    those with $|\xbp| \ge 0.5$ or $|\xbs| \ge 0.5$.}
  \label{fig:hbh0SpinTiltDist}
\end{figure}

Natal kicks can tilt BH spin vectors from binary orbit vectors of
hBH0s. Figure~\ref{fig:hbh0SpinTiltDist} shows the
tilt-angle distribution. Note that spin and binary orbit vectors are
parallel and anti-parallel for $\cos \theta = 1$ and $-1$,
respectively. The fractions of $\theta \gtrsim 45^\circ$ and $\gtrsim
90^\circ$ are $\sim 0.1$ and $\sim 0.01$, respectively, for both cases
of $m_{\rm b,p}$. Primary BHs have larger tilt angles than secondary
BHs. Since primary BHs tend to have 1st-evolving stars as their
progenitors, they have two chances to tilt their spin vectors from
binary orbit vectors.

We examine mass distribution of hBH0s in ``kick'', ``wind'', and
``pess'' models with $\qmin \neq 0.0$ or $\amin \neq 10 \rsun$. Then,
in all the models, we find peaks corresponding to the higher-mass peak
in the optiq0.0a1e1 model. On the other hand, there is no lower-mass
peak seen in the optiq0.0a1e1 model. Most of BHs have zero spins for
$\td = 10-15$~Gyr. This is because they have large $\td$, and because
most of them are not formed through the CE0 channel. If some of them
were formed through the CE0 channel, there would be the lower-mass
peak.

We summarize features of merging hBH0s with $\td \le 15$~Gyr. In the
optiq0.0a1e1 model, they have two peaks in their mass distribution:
$(m_{\rm b,p}, m_{\rm b,s}) \sim (45 \msun,45 \msun)$ (the higher-mass
peak) and $(m_{\rm b,p}, m_{\rm b,s}) \sim (30 \msun,30 \msun)$ (the
lower-mass peak). The lower-mass peak shifts from $20 \msun$ to $35
\msun$ from $\td=0.1-1$~Gyr to $\td=10-15$~Gyr. Half (most) of BHs
belonging to the higher-mass (lower-mass) peak has significantly high
spin. The higher-mass peak is present even in models with different
initial conditions ($\qmin$ and $\amin$), and different stellar
evolution models (stellar winds and natal kicks, while BHs around the
higher-mass peak tend to have zero spins in models other than the
optiq0.0a1e1 and kickq0.0a1e1 models due to small contribution of the
CE0 channel. The lower-mass peak and its $\td$ dependence are quite
fragile. It disappears when $\amin=200\rsun$ or is shifted to $\sim 20
\msun$ by stellar wind effects. Stellar winds have large effects on
the lower-mass peak due to the rotational enhancement, not due to the
high metallicity. The time-shift feature of the lower-mass peak is
erased by natal kicks.

\subsubsection{Binaries with high- and low-mass BHs (hBH1s)}
\label{sec:hBH1}

\begin{figure*}[ht!]
  \plottwo{\fdir/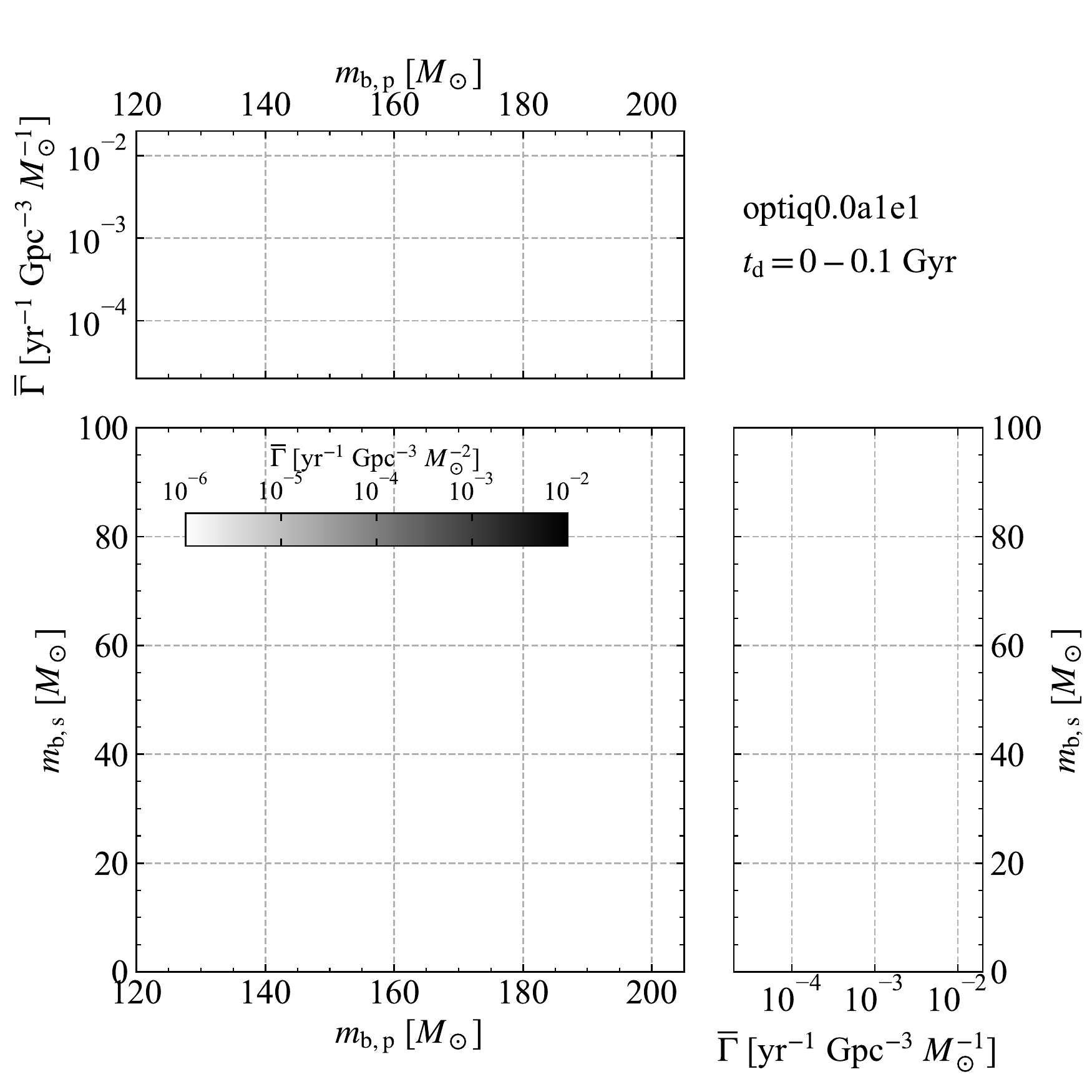}{\fdir/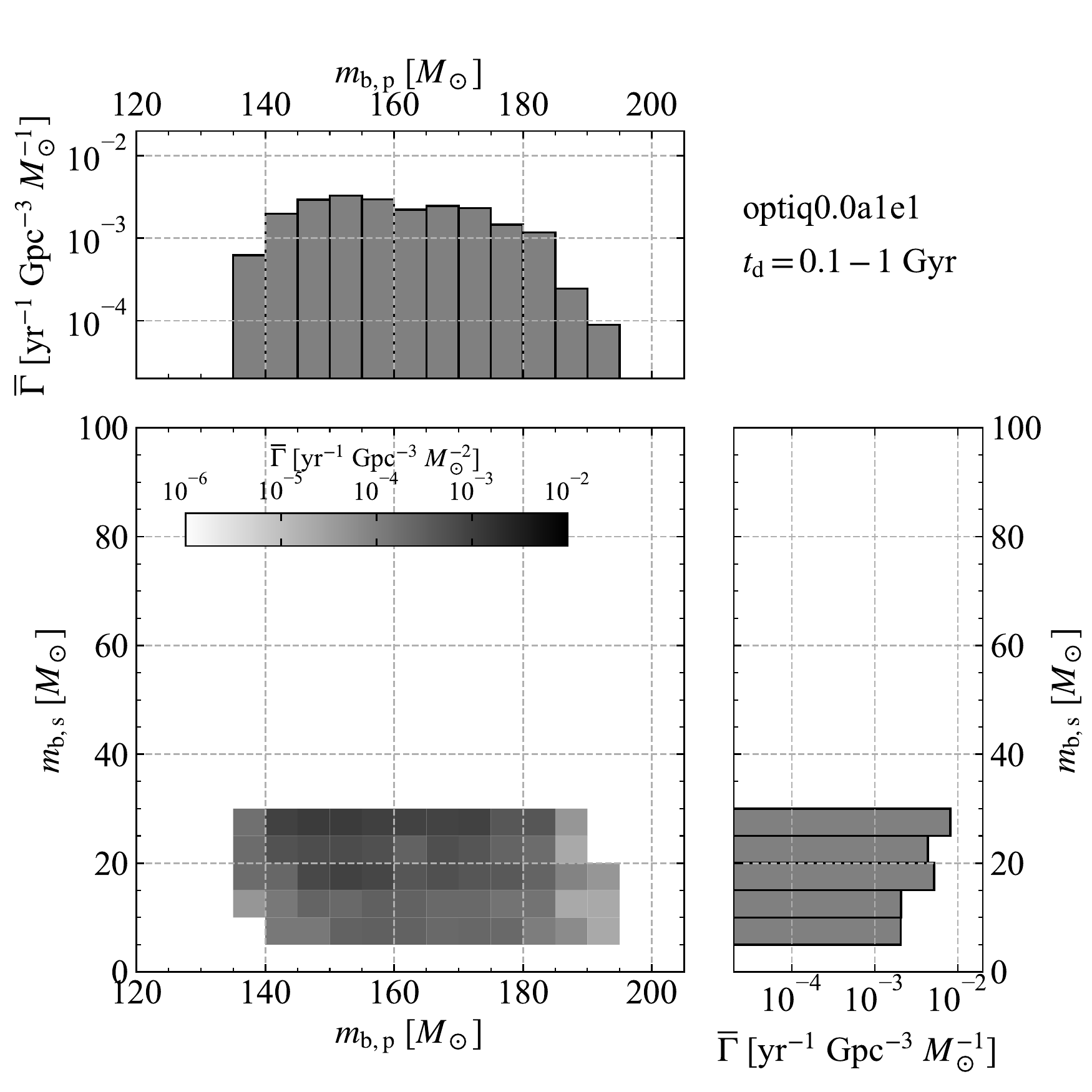}
  \plottwo{\fdir/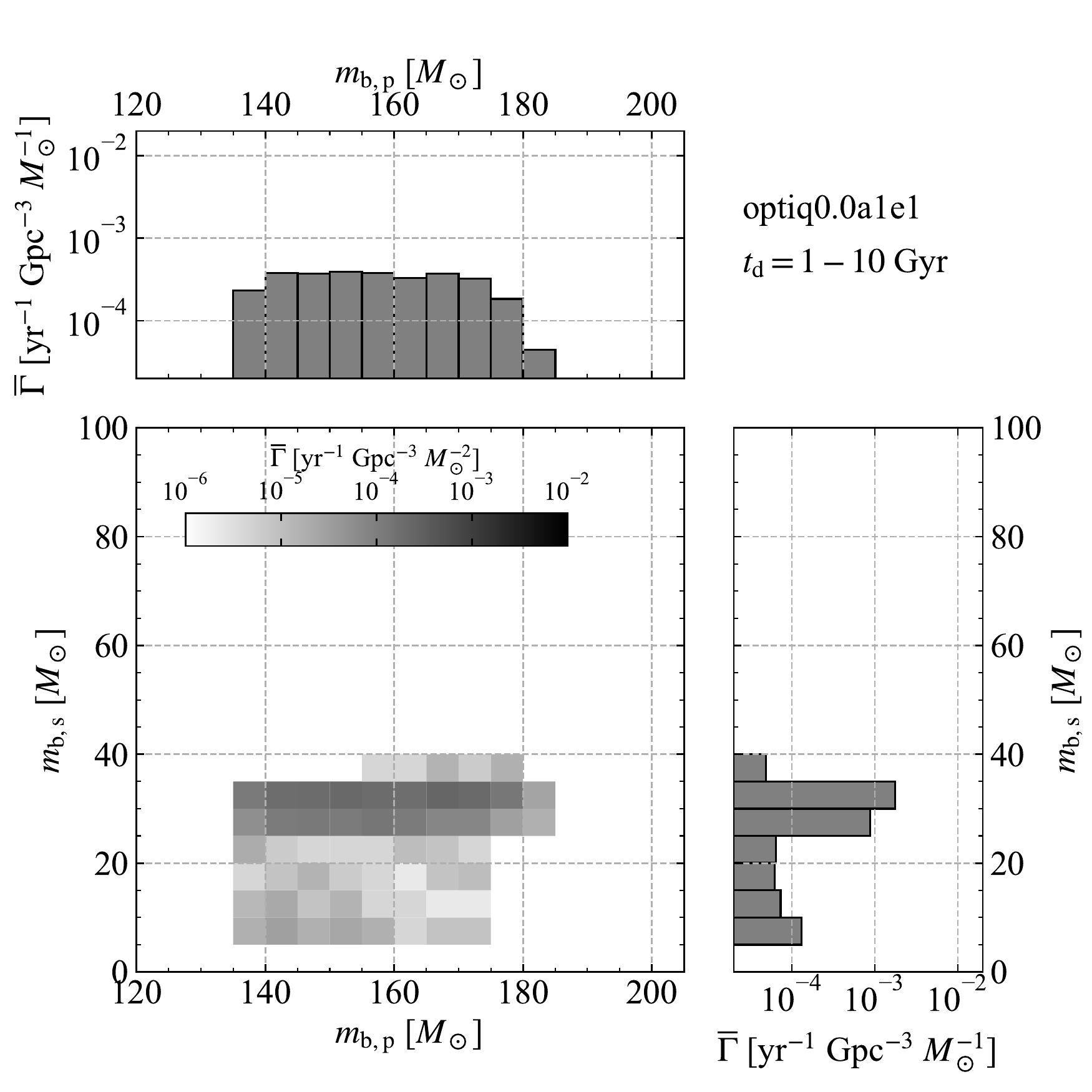}{\fdir/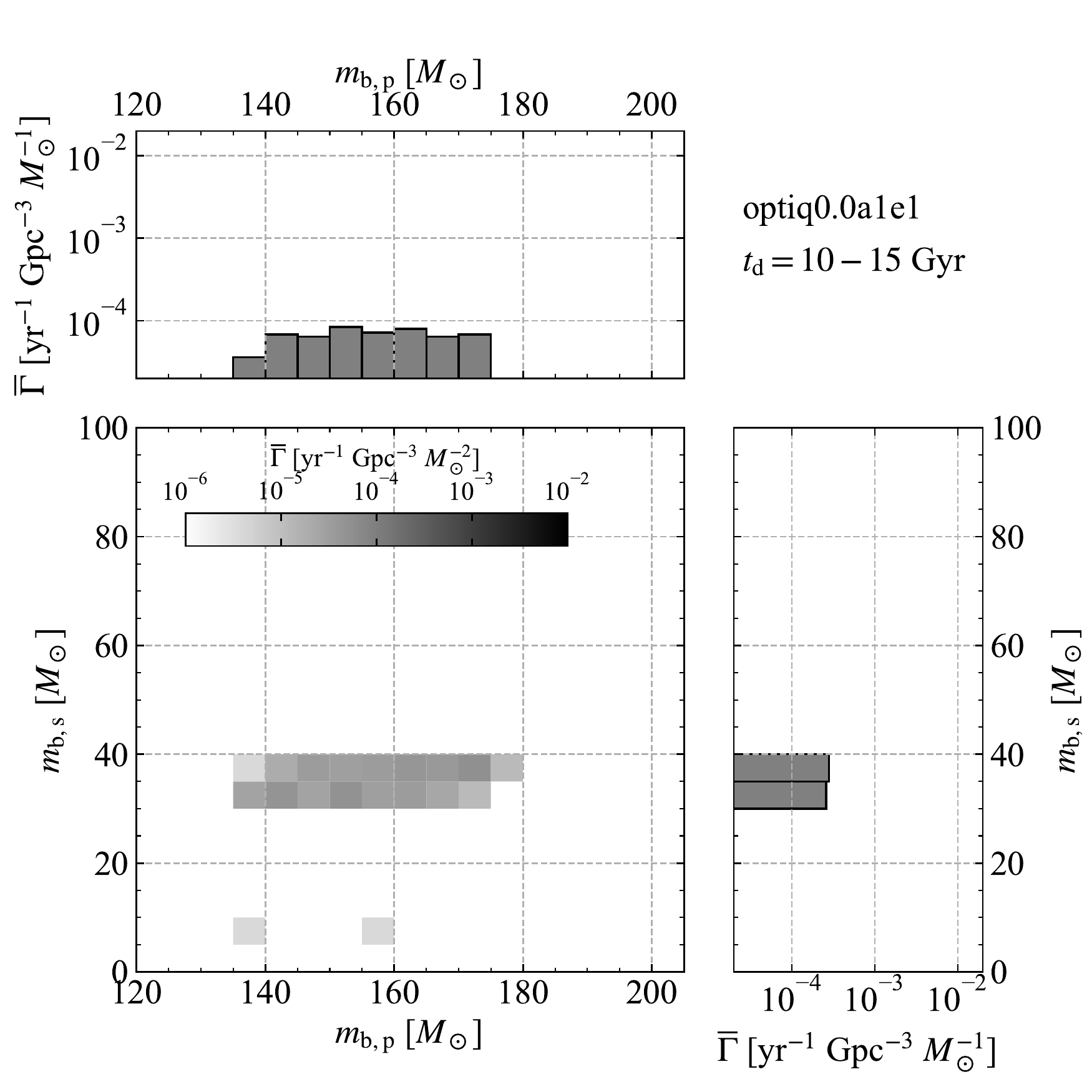}
  \caption{Merger rate densities of hBH1s in the optiq0.0a1e1 model.}
  \label{fig:hbh1MassDist}
\end{figure*}

First, we investigate the optiq0.0a1e1 model again for the same reason
in section~\ref{sec:hBH0}. Figure~\ref{fig:hbh1MassDist} shows the
merger rate densities of hBH1s for each $\td$. There is no hBH1 with
$\td \lesssim 0.1$~Gyr. Only binaries with $\rpi \gtrsim 10^2 \rsun$
can form hBH1s through the CE1 channel, and their minimum $\td$ is
$\sim 0.1$~Gyr. On the other hand, a binary with $\rpi \lesssim 10^2
\rsun$ experiences merger when the 1st-evolving star fills its Roche
lobe for the reason described in section~\ref{sec:MergingBH-BHs}.

For $\td = 0.1 -15$~Gyr, the $\mbp$ distribution has a plateau in
$\mbp \sim 130 - 180\msun$, and has a tail in $\mbp \sim 180 -
200\msun$. The plateau ranges from the minimum He core mass needed for
the direct collapse ($M_{\rm c,He,DC} = 135 \msun$) to the He core
mass of a star with $\mzams = 300\msun$, the maximum mass of $\mi$ in
our models. The merger rate density of $\sim 130\msun$ should be
larger than that of $\sim 180\msun$, since the $\mi$ distribution is
top-light ($\propto \mi^{-1}$). Nevertheless, the number of stars
forming $\mche \sim 180\msun$ ($\mi \sim 300 \msun$) is comparable to
that of stars forming $\mche \sim 135\msun$ ($\mi \sim 250
\msun$). This is because the number ratio of the former star to the
latter star is only $\sim 1.2$. The tail is formed through stable mass
transfer from the 2nd-evolving stars to the 1st BH (see
Figure~\ref{fig:hbh1channel}). The tail is more prominent in smaller
$\td$, since the 2nd-evolving stars are closer to the 1st BH, and give
larger mass to the 1st BH.

The $\mbs$ distribution has a peak at $25-40\msun$, and the peak
shifts from $25\msun$ to $40\msun$ with $\td$. This is the back
reaction of the formation of the $\mbp$ tail. As the 2nd-evolving
stars, progenitors of the 2nd BHs, give larger masses to the 1st BH,
they lose their masses.

We describe the three features of hBH1s: no merger for $\td \lesssim
0.1$~Gyr, the $\mbp$ distribution with the plateau and tail, and the
$\mbs$ distribution with the peak and its shift with $\td$. Hereafter,
we investigate the dependence of these features on initial conditions
($\qmin$ and $\amin$) and single star models (stellar winds and natal
kicks). Actually, these features are insensitive to $\amin$. As seen
in Figure~\ref{fig:icCeMergBh}, hBH1s can be formed through the CE1
channel even if binaries have $\ai \gtrsim 200\rsun$. On the other
hand, the formation of hBH1s is seriously affected by $\qmin$. For
$\qmin=0.9$, hBH1s are never formed (see
Figures~\ref{fig:delayTimeDist} and \ref{fig:icMtMergBh}). Since a
binary has stars with similar masses, it cannot form one lBH and one
hBH.

Natal kicks again erase $\td$ dependence. In the kickq0.0a1e1 model,
hBH1s with $\td \lesssim 0.1$~Gyr appear, and the time shift of the
peak in the $\mbs$ distribution disappears. The $\mbs$ peak is always
at $\sim 30\msun$. Stellar winds also affect the mass
distribution. The $\mbp$ distribution also has the plateau for the
following reason. The 1st-evolving stars always become nHe stars
regardless of the presence and absence of stellar winds due to the CE1
channel. The nHe stars lose little mass through stellar winds, since
they do not exceed the Eddington limit, and are not sufficiently spun
up. However, the tail disappears for the following reason. The
2nd-evolving stars are strongly spun up by the 1st-evolving star or
1st BH after the CE1 channel. Then, they lose a significant of their H
envelopes through stellar winds as seen in
Figure~\ref{fig:remnantMass}. Thus, they cannot transfer their masses
to the 1st BHs. The $\mbs$ distribution is sensitive to the stellar
winds, since the 2nd-evolving stars are spun up and lose their mass
rotationally enhanced stellar winds as described above. Then, the peak
of $\mbs$ moves down to $\sim 20\msun$. In the other models, we find
that mass distributions can be explained by simple combinations of
initial conditions, natal kicks and stellar winds.

\begin{figure}[ht!]
  \plotone{\fdir/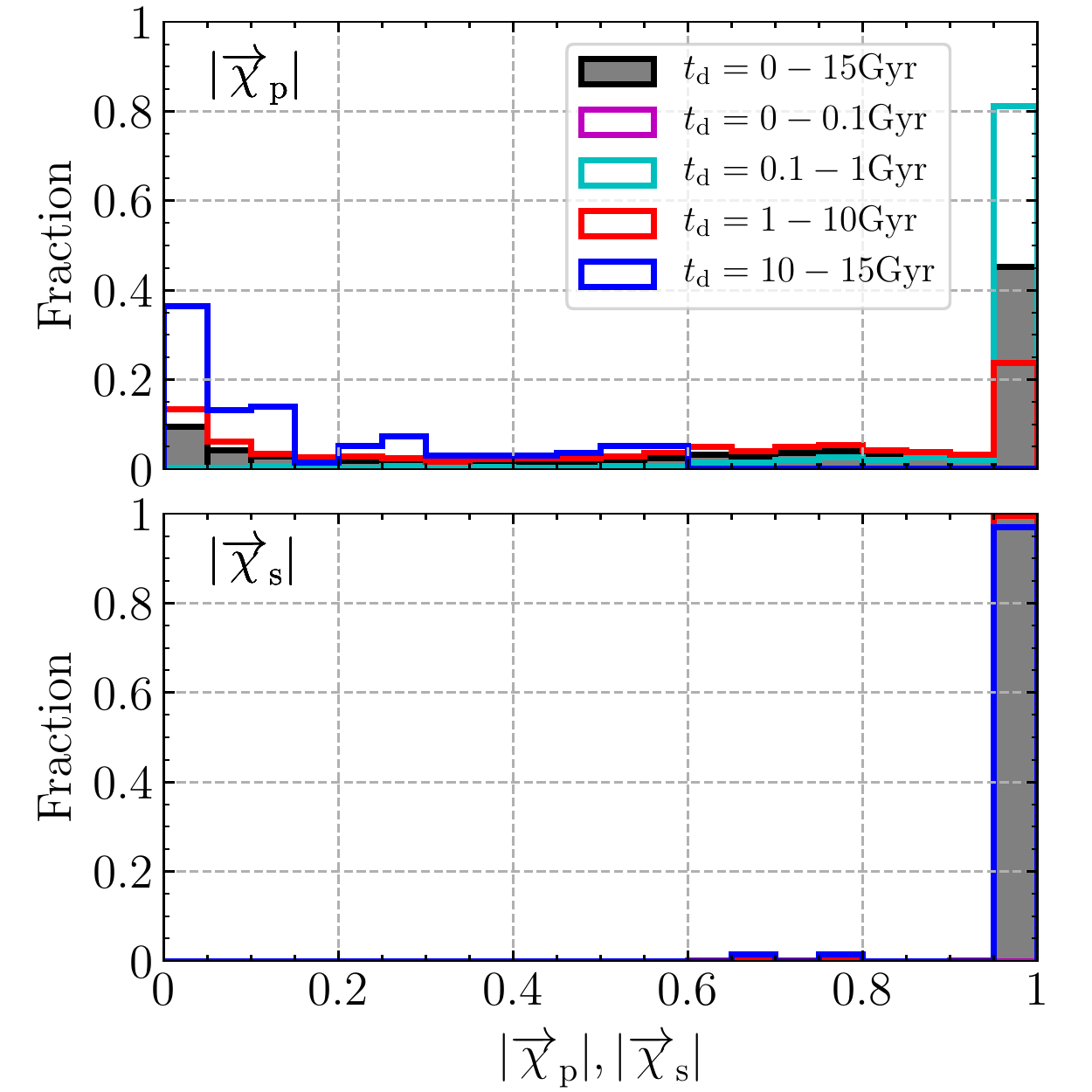}
  \caption{Spin distributions of primary and secondary BHs in hBH1s
    for each $\td$ in the optiq0.0a1e1 model.}
  \label{fig:hbh1SpinMagDistOpti}
\end{figure}

Figure~\ref{fig:hbh1SpinMagDistOpti} shows the spin distributions of
BHs in hBH1s. The fraction of primary BHs with nearly zero spins
increases with $\td$. The fraction of $|\xbp| \lesssim 0.2$ reaches to
$\sim 1$ for $\td = 10-15$~Gyr. This can be also explained by the
spin-$\td$ relation described in section~\ref{sec:hBH0}. On the other
hand, the secondary BHs have high spins $\sim 1$ for all $\td$. This
is because they are spun up by tidal fields of the 1st BHs, and keep
their H envelopes.

\begin{figure}[ht!]
  \plotone{\fdir/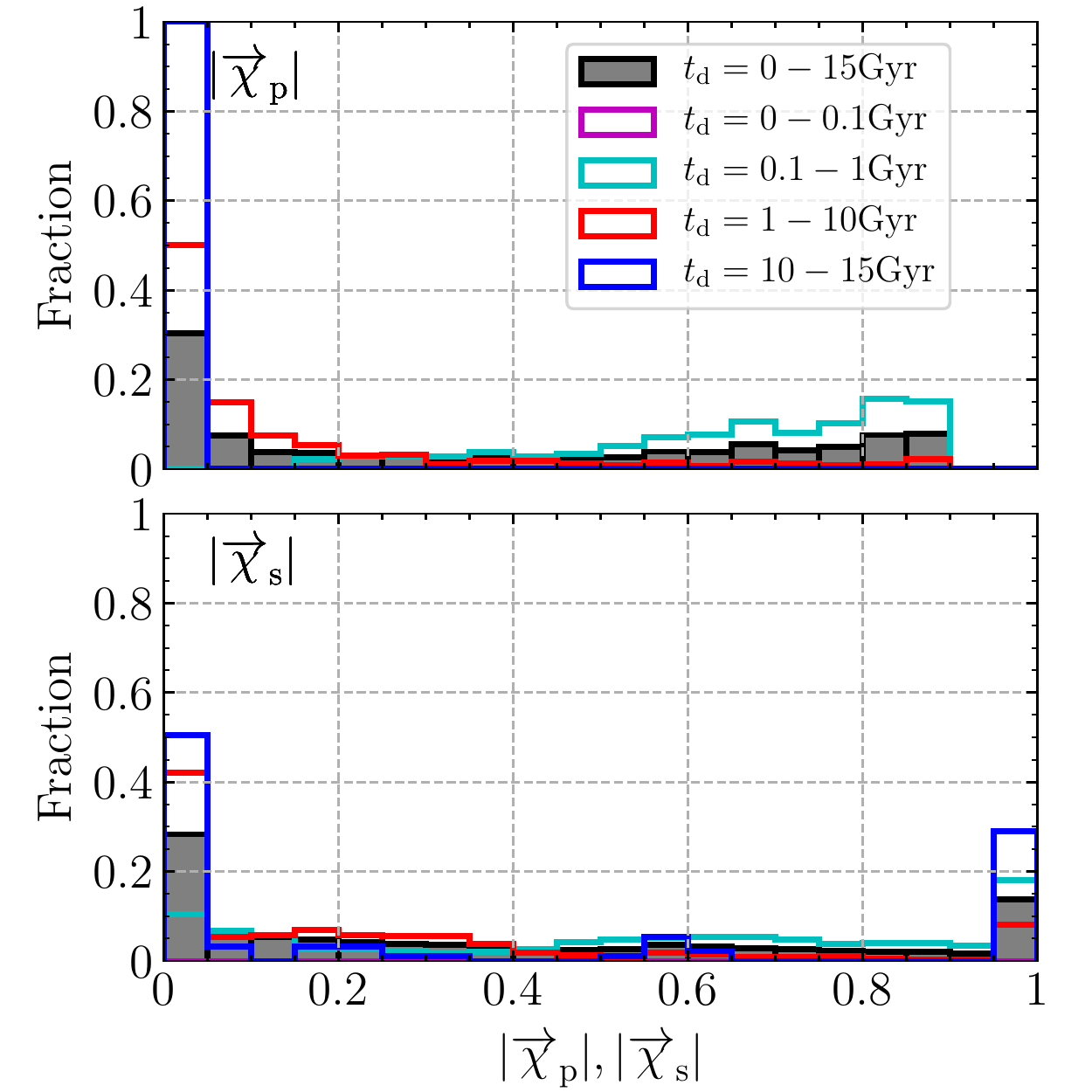}
  \caption{Spin distribution of primary and secondary BHs in hBH1s for
    each $\td$ in the windq0.0a1e1 model.}
  \label{fig:hbh1SpinMagDistWind}
\end{figure}

The spin distribution is insensitive to $\amin$ for the same reason as
the mass distribution. Natal kicks erase the $\td$ dependence, and
average the distribution to that for $\td = 0-15$~Gyr. This is because
natal kicks shorten and lengthen $\td$ at random as described in
section~\ref{sec:hBH0}. Figure~\ref{fig:hbh1SpinMagDistWind} shows the
spin distributions of BHs in hBH1s in the windq0.0a1e1 model. Stellar
winds largely reduce spins for both the primary and secondary
BHs. Although the fraction of non-spinning BHs for $\td=10-15$~Gyr in
the pessq0.0a1e1 model is smaller than in the windq0.0a1e1 due to
natal kicks, the fraction is still high ($\sim 0.6$) for both the
primary and secondary BHs.

\begin{figure}[ht!]
  \plotone{\fdir/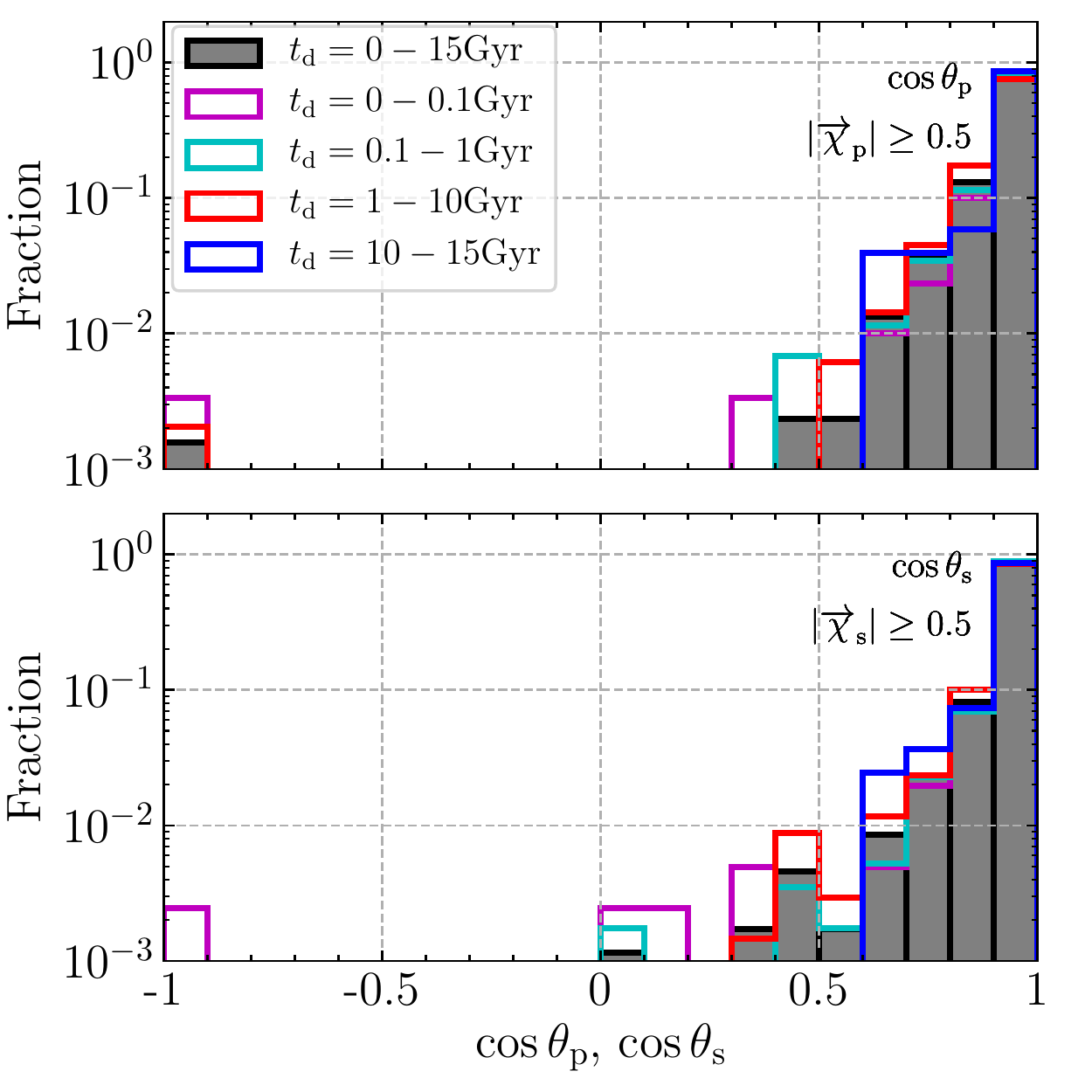}
  \caption{Distribution of angles between primary and secondary BH
    spin vectors and binary orbit vectors for hBH1s in the
    kickq0.0a1e1 model. The BHs are limited to those with $|\xbp| \ge
    0.5$ or $|\xbs| \ge 0.5$.}
  \label{fig:hbh1SpinTiltDist}
\end{figure}

Figure~\ref{fig:hbh1SpinTiltDist} shows the distribution of angles
between primary and secondary BH spin vectors and binary orbit vectors
for hBH1s in the kickq0.0a1e1 model. The angles are formed only due to
natal kicks. Comparing Figure~\ref{fig:hbh1SpinTiltDist} with
Figure~\ref{fig:hbh0SpinTiltDist}, we can find that spins of hBH1s are
more aligned than those of hBH0s. This is not due to the small number
of hBH1s, although the number of BHs in the top panel of
Figure~\ref{fig:hbh1SpinTiltDist} is $\sim 10^3$ of $10^6$ binaries,
and that in the top panel of Figure~\ref{fig:hbh0SpinTiltDist} is
$\sim 3 \times 10^4$ of $10^6$ binaries. For example, the fraction of
$\theta \gtrsim 45^\circ$ is a few $0.01$ for hBH1s, while it is $\sim
0.1$ for hBH0s. This reason can be interpreted as follows. We can
obtain the circular velocity of a BH-BH, rewriting Eq.~(\ref{eq:tgw})
as
\begin{align}
  v_{\rm circ} \propto t_{\rm GW}^{-1/8} \mbp^{-1/8} \mbs^{-1/8} (\mbp
  + \mbs)^{3/8}. \label{eq:vcirc}
\end{align}
As seen in Eq.~(\ref{eq:vcirc}), the internal velocity of a BH-BH
becomes larger with its total mass, if $t_{\rm GW}$ is fixed. Thus,
BH-BHs with larger total masses are less sensitive to natal
kicks. This is the reason why spins of hBH1s are more aligned than
those of hBH0s.

We summarize features of merging hBH1s. We first see the optiq0.0a1e1
model. They merge for $\td \gtrsim 0.1$~Gyr. The $\mbp$ distribution
has a plateau in $130-180 \msun$ and a tail in $180-200 \msun$. The
tail is more prominent in smaller $\td$. The $\mbs$ distribution has a
peak shifting from $20 \msun$ to $40 \msun$ with $\td$. The fraction
of hBH1s with $|\xbp| \lesssim 0.2$ increases with $\td$, and reaches
to unity for $\td = 10-15$~Gyr, while the fraction of hBH1s with
$|\xbs| \sim 1$ is unity for all $\td$.

The above features are changed by various effects. First, hBH1s
themselves vanish if $\qmin = 0.9$. The above features have weak
dependence on $\amin$. Natal kicks erase the $\td$ dependence, and
raise spin-orbit misalignments smaller than those of hBH0s. Stellar
winds erase the tail of the $\mbp$ distribution, and reduce the peak
location of the $\mbs$ distribution to $\sim 20 \msun$. Moreover,
stellar winds greatly increase the fraction of hBH1s with $|\xbp| \sim
0$ and $|\xbs| \sim 0$.

\subsubsection{Binaries with two high-mass BHs (hBH2s)}
\label{sec:hBH2}

\begin{figure*}[ht!]
  \plottwo{\fdir/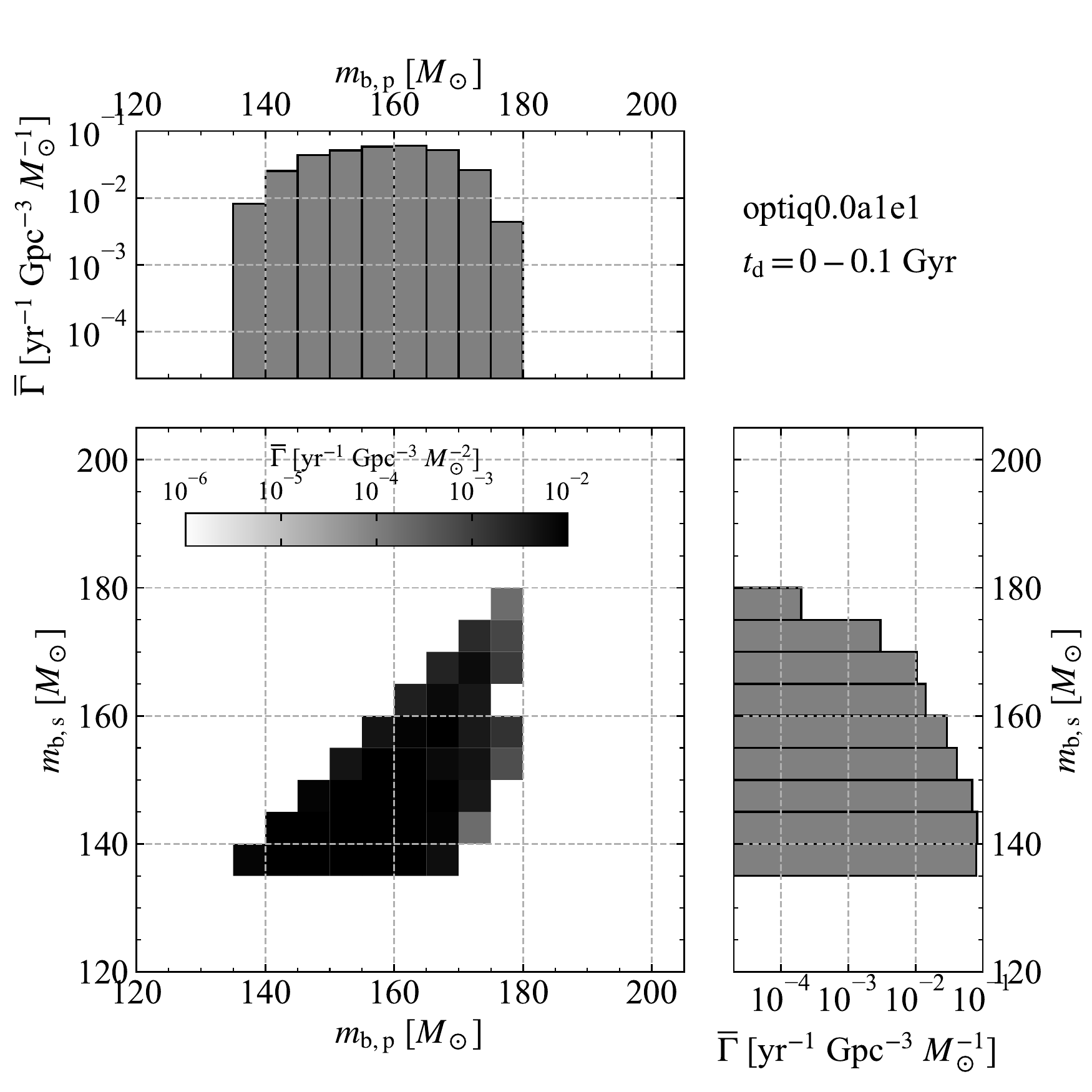}{\fdir/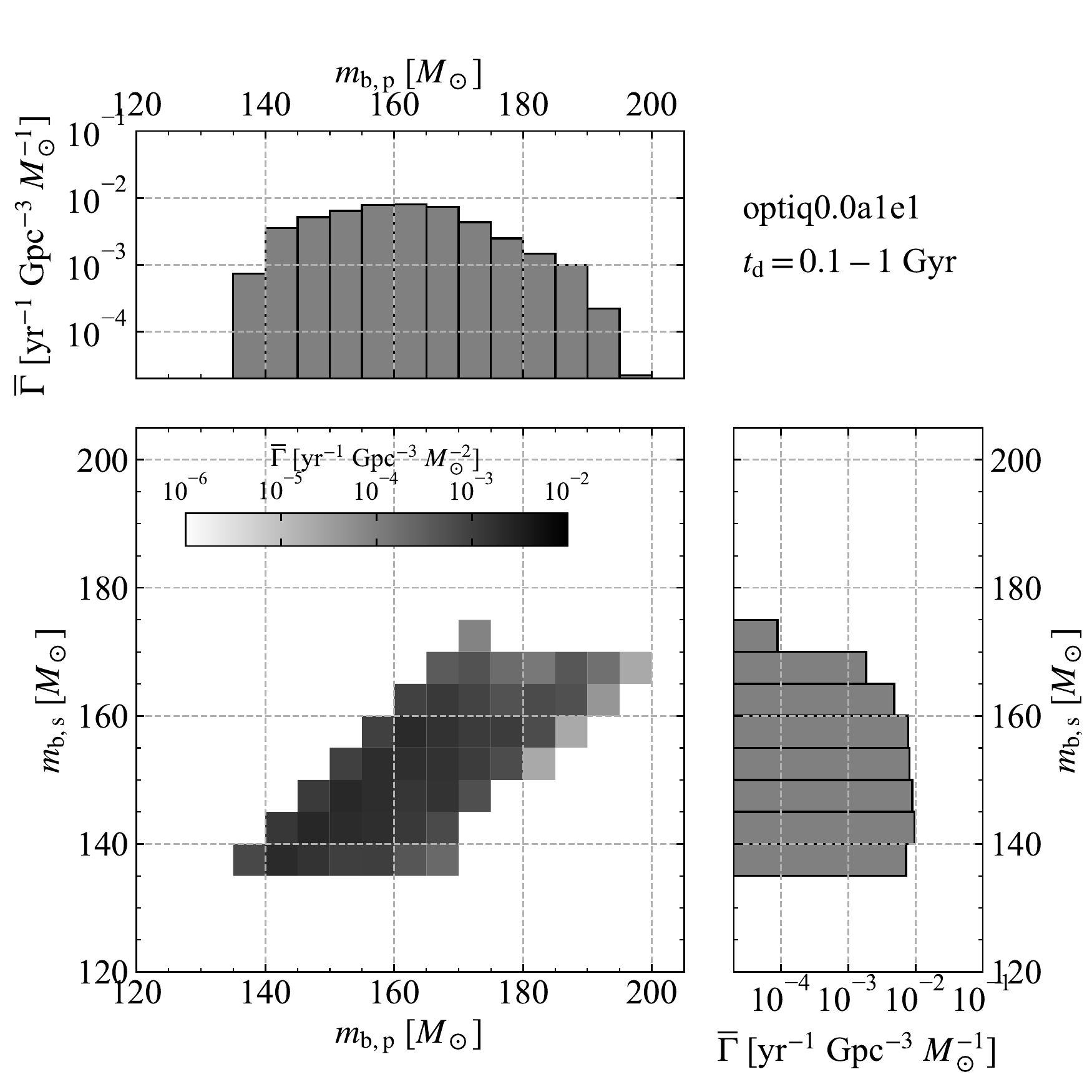}
  \plottwo{\fdir/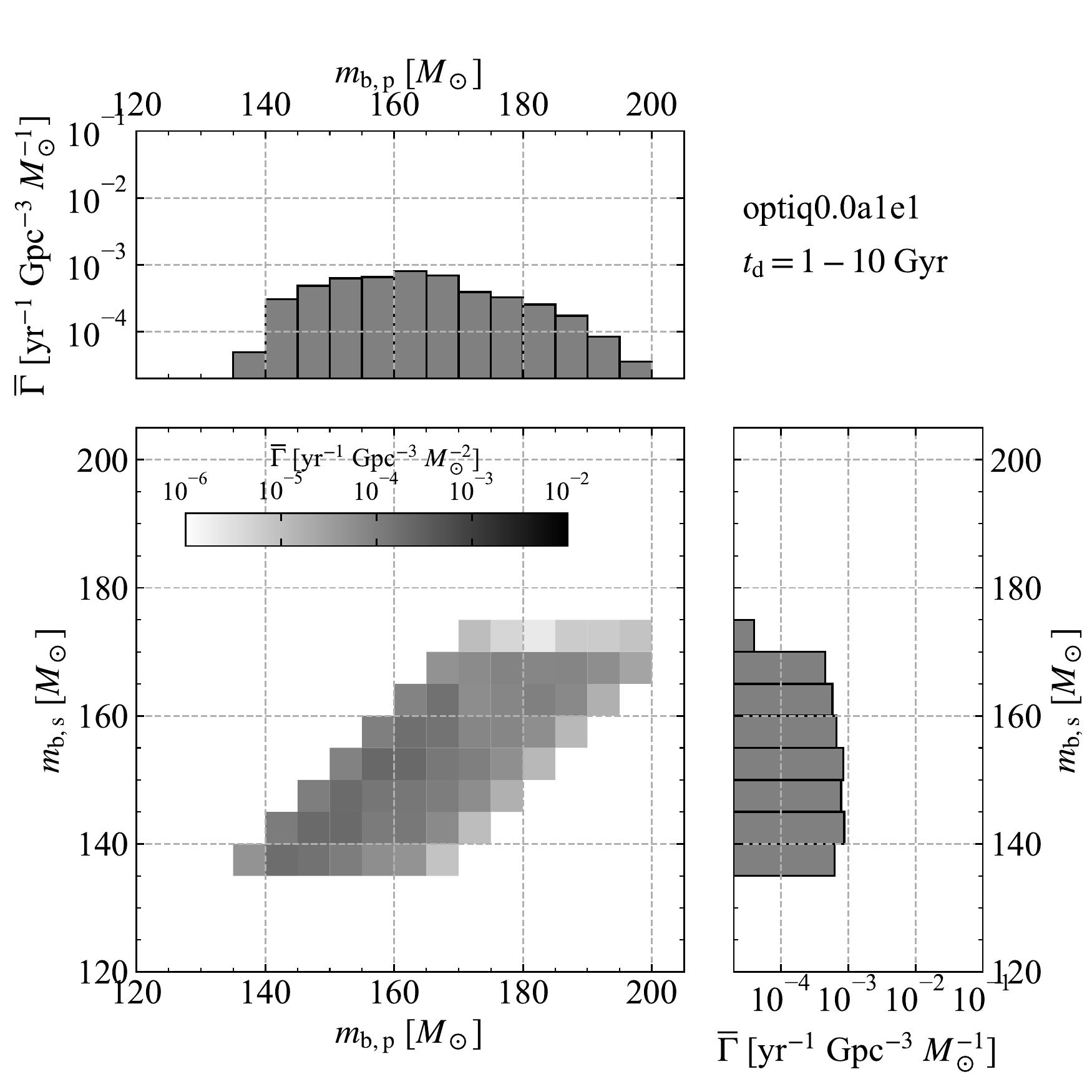}{\fdir/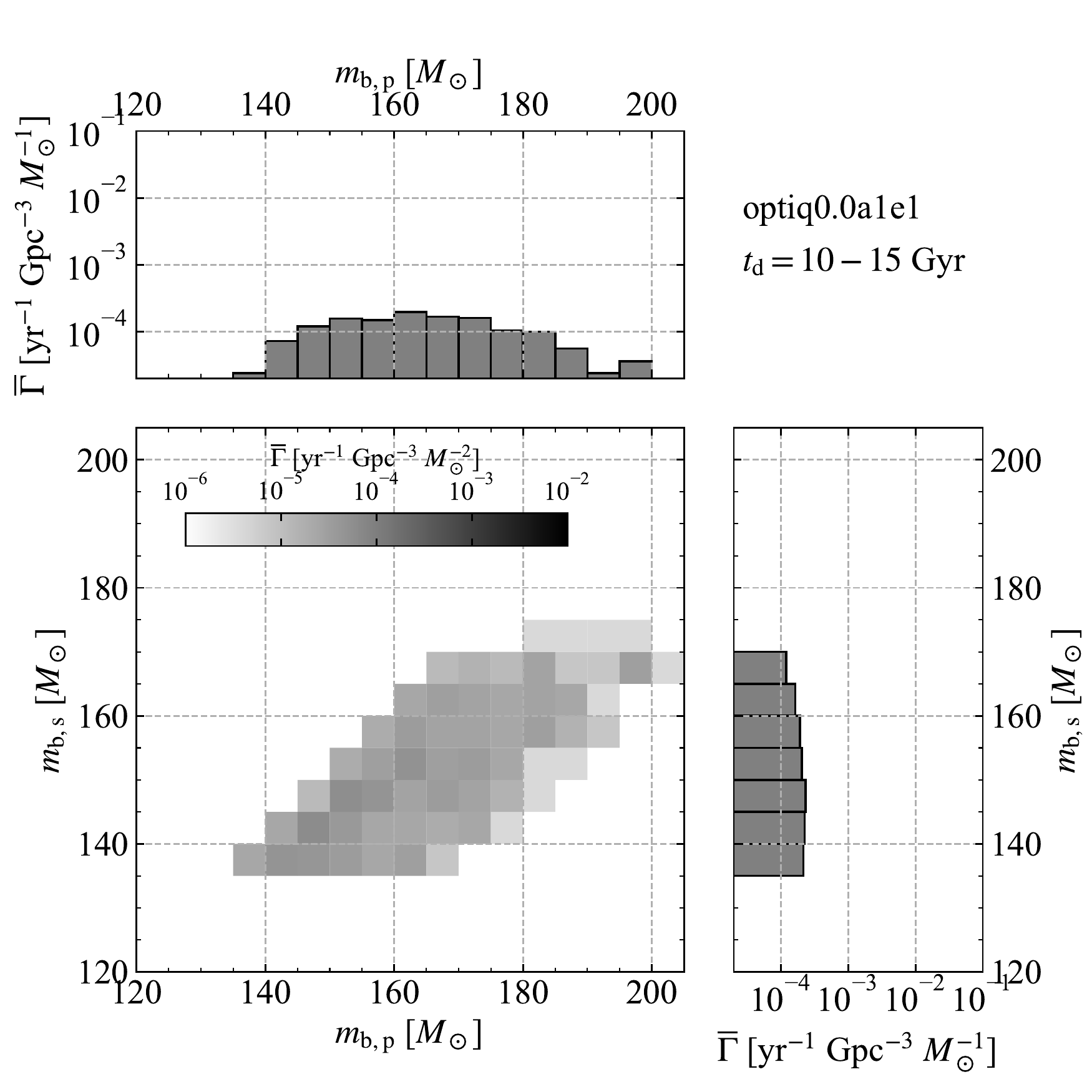}
  \caption{Merger rate densities of hBH2s in the optiq0.0a1e1 model.}
  \label{fig:hbh2MassDist}
\end{figure*}

We first investigate the optiq0.0a1e1 model, similarly to the previous
sections. Figure~\ref{fig:hbh2MassDist} shows the merger rate density
of hBH2s for each $\td$. The 2D distribution for $\td=0-0.1$~Gyr has a
triangle shape ($\mbp \lesssim 180 \msun$), while those for
$\td=0.1-15$~Gyr have a parallelogram shape ($\mbp \lesssim 200
\msun$). Their different shapes are due to the difference between
their formation channels. For $\td \lesssim 0.1$~Gyr, they are formed
through the CE4 channel (see also Figure~\ref{fig:icCeMergBh}). This
is the reason why the $\mbp$ distribution is flat in $130-180\msun$,
the same reason for the plateau of the primary BHs in hBH1s described
in section~\ref{sec:hBH1}. The $\mbs$ distribution monotonically
decreases with $\mbs$. In our initial conditions, as the 2nd-evolving
stars become more massive, they have less chances to have the
1st-evolving stars with $\mzams = 250-300\msun$ which overcome
PISN. For $\td \gtrsim 0.1$~Gyr, hBH2s are formed through the CE0
channel (see also Figure~\ref{fig:icCeMergBh}). This is the reason why
the primary BHs can have larger masses than He core masses formed from
stars with $\mzams = 300\msun$. The 2D distributions have shapes like
parallelograms. This means hBH2s have two BHs with similar
masses. This is because the 2nd mass transfer tends to stop when a
mass ratio of the binary becomes close to unity, similarly to hBH0s
around the lower-mass peak in the optiq0.0a1e1 model.

For $\qmin=0.9$ or $\amin=200\rsun$, the mass distributions in all $\td$
are similar to that in $\td \lesssim 0.1$~Gyr in the optiq0.0a1e1
model. In other words, all the hBH2s are formed only through the CE4
channel. This is because the CE0 channel does not work. As seen in
Figure~\ref{fig:icCeMergBh}, the CE0 channel works well only when $\qi
< 0.9$ (the top panel) and $\ai < 200\rsun$ (the left-hand side of the
vertical dashed line). For $\qi \ge 0.9$, two stars fill their Roche
lobes simultaneously. Then, they experience merger or common envelope
evolution. They merge when they are MS stars, and they experience
common envelope evolution when they are post-MS stars. Thus, they
cannot undergo the CE0 channel. For $\ai \ge 200 \rsun$, binaries can
undergo the CE0 channel. However, the resulting BH-BHs have too large
separations to merge for $\td \le 15$~Gyr, since the CE0 channel does
not shrink binary separations.

Natal kicks erase the $\td$ dependence. Thus, hBH2s formed through the
CE0 channel appear for $\td \lesssim 0.1$~Gyr, and the 2D
distributions have shapes like parallelograms for all $\td$. Stellar
winds erase hBH2s formed through the CE0 channel. The 1st-evolving
star fills its Roche lobe when it is an MS star. It is spun up, and
loses large mass due to rotationally enhanced stellar winds. Since it
makes a He core with smaller mass than $M_{\rm c,He,DC} (= 135 \msun)$
due to the stellar winds, it experiences PISN. Thus, the shapes of the
2D distributions for all $\td$ are triangle-like, similar to that for
$\td=0-0.1$~Gyr in the optiq0.0a1e1 model. This is the same if both of
natal kicks and stellar winds are taken into account (i.e. the pess
models). The mass distributions in models with different initial
conditions and single star models can be understood as simple
combinations of the effects described above.

\begin{figure}[ht!]
  \plotone{\fdir/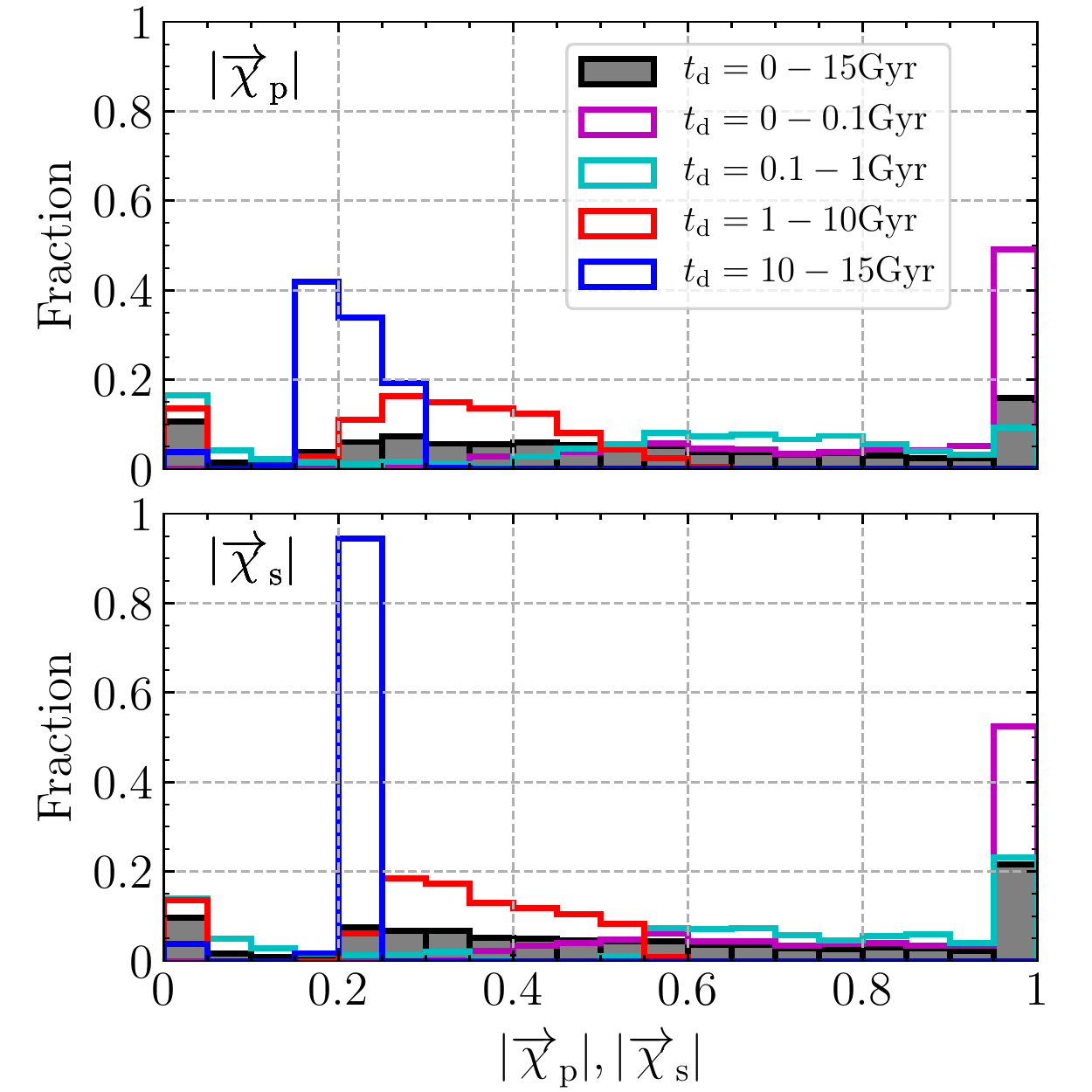}
  \caption{Spin distributions of hBH2s for each $\td$ in the
    optiq0.0a1e1 model.}
  \label{fig:hbh2SpinMagDistOpti}
\end{figure}

Figure~\ref{fig:hbh2SpinMagDistOpti} shows the spin distribution of
primary and secondary BHs of hBH2s in the optiq0.0a1e1 model. The
fraction of BHs with high spins decreases with $\td$ because of the
spin-$\td$ relation. A large number of BHs have low spins, $\sim 0.2$
for $\td=10-15$~Gyr. Note that their spins are non-zero. Since they
are formed through the CE0 channel, they can retain their H envelopes
and spin angular momenta until their collapses.

\begin{figure}[ht!]
  \plotone{\fdir/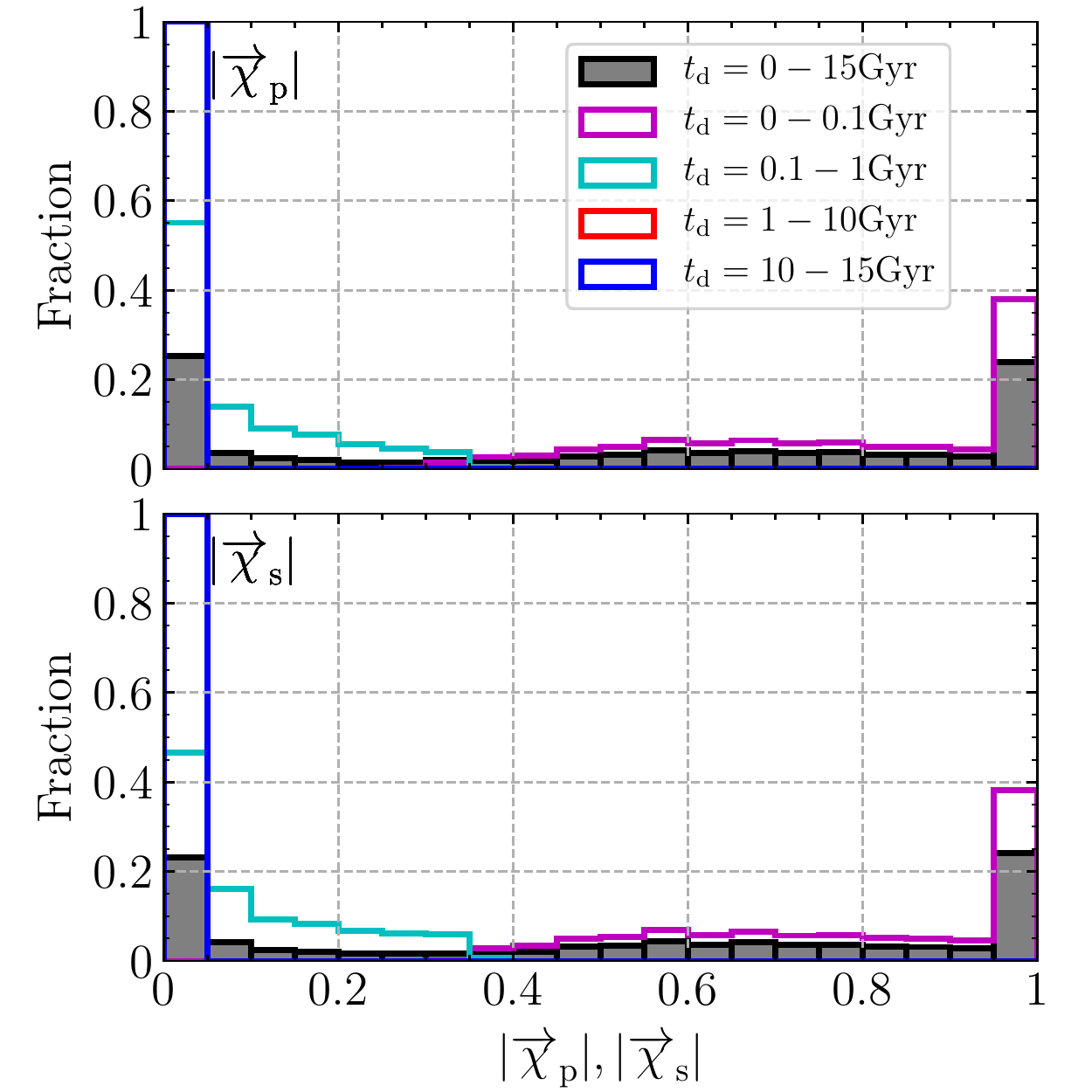}
  \caption{Spin distributions of hBH2s for each $\td$ in the
    optiq0.0a2e2 model.}
  \label{fig:hbh2SpinMagDistOpti_q0a2}
\end{figure}

Different initial conditions make the spin distribution smaller,
especially for
$\td=10-15$~Gyr. Figure~\ref{fig:hbh2SpinMagDistOpti_q0a2} shows the
spin distribution of hBH2s in optiq0.0a2e2 model. Most of primary and
secondary BH spins are zero for $\td=10-15$~Gyr in contrast to those
in the optiq0.0a1e1 model. This is because hBH2s for $\td=10-15$~Gyr
are formed only through the CE4 channel. Note that the corresponding
hBH2s in the optiq0.0a1e1 model are formed dominantly through the CE0
channel. Since the CE4 channel carries away H envelopes of progenitors
of both primary and secondary BHs, the resulting BHs have lower spins
than BHs formed through the CE0 channel.

Natal kicks erase the $\td$ dependence of the spin distribution, and
the spin distribution for each $\td$ is similar to that for
$\td=0-15$~Gyr. Stellar winds decrease spins of both the primary and
secondary spins. Most of BH spins are zero for $\td=10-15$~Gyr. When
both of initial conditions and single star models are different from
the optiq0.0a1e1 model, we can interpret the spin distributions as
simple combinations of the effects described above.

\begin{figure}[ht!]
  \plotone{\fdir/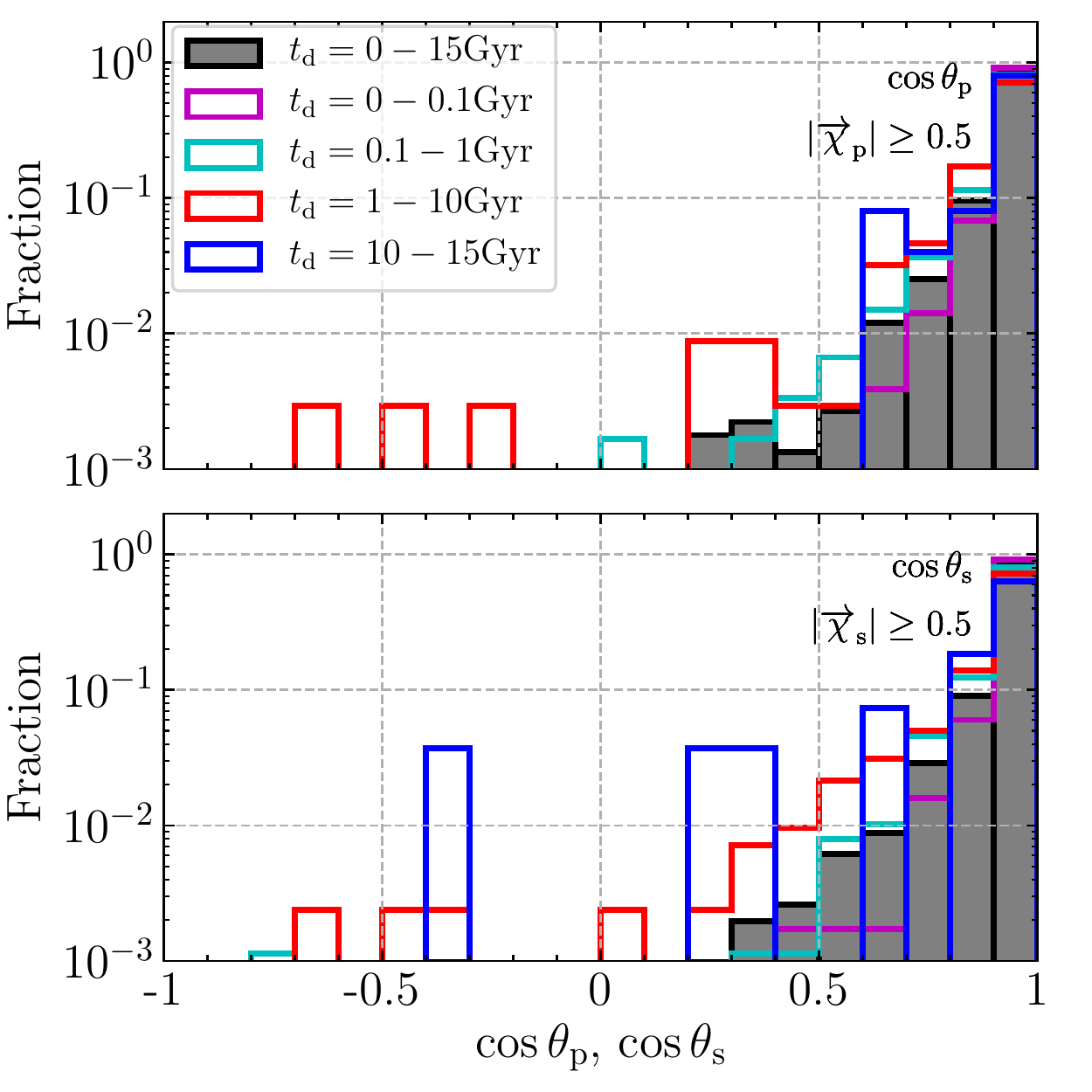}
  \caption{Distribution of angles between primary and secondary BH
    spin vectors and binary orbit vectors for hBH2s in the
    kickq0.0a1e1 model. The BHs are limited to those with
    $|\overrightarrow{\chi}_1| \ge 0.5$ or $|\overrightarrow{\chi}_2|
    \ge 0.5$.}
  \label{fig:hbh2SpinTiltDist}
\end{figure}

Natal kicks raise spin-orbit misalignments of hBH2s. We show in
Figure~\ref{fig:hbh2SpinTiltDist} the distribution of the
misalignments. The misalignments are clearly weaker than those of
hBH0s (see Figure~\ref{fig:hbh0SpinTiltDist}) because of their larger
internal velocities as discussed in section~\ref{sec:hBH1}. However,
the misalignments appear similar to those of hBH1s (see
Figure~\ref{fig:hbh1SpinTiltDist}). There can be two reasons. First,
the difference between internal velocities of hBH1s and hBH2s is
smaller than the difference between internal velocities of hBH0s and
hBH1s. Second, the numbers of hBH1s and hBH2s are small, $\sim 10^3$
of $10^6$ binaries.

We summarize the features of hBH2s. In the optiq0.0a1e1 model, they
are formed through the CE0 and CE4 channels. The CE4 and CE0 channels
are dominant for $\td = 0-0.1$ and $0.1-15$~Gyr, respectively. Thus,
the mass distribution is limited to He core masses in $130-180\msun$
for $\td=0-0.1$~Gyr, while it extends to $\sim 200\msun$, since BHs
keep their H envelopes. The 2D mass distributions have a triangle-like
shape for $\td = 0-0.1$~Gyr, and parallelogram-like shapes for $\td =
0.1-15$~Gyr. Primary and secondary BH spins decrease with $\td$ due to
the spin-$\td$ relation, and they have only $\sim 0.2$ for
$\td=10-15$~Gyr. If initial conditions are different or stellar winds
are switched on, hBH2s formed through the CE0 channel vanish. Their
mass distributions for all $\td$ have triangle-like shapes, similar to
those for $\td=0-0.1$~Gyr in the optiq0.0a1e1 model. BH spins decrease
with $\td$, and most of them are zero for $\td=10-15$~Gyr. Natal kicks
erase the $\td$ dependence, and raise spin-orbit misalignments smaller
than those of hBH0s, and similar to those of hBH1s. Mass and spin
distributions with different initial conditions and single star models
can be interpreted as simple combinations of their effects.

\section{Identification of Pop.~III BH-BHs}
\label{sec:Identification}

We discuss possibility to identify Pop.~III BH-BHs by the current and
future GW observatories. In section~\ref{sec:CurrentGwObservatories},
we compare our results with previous events of BH-BH mergers, and
present clues to identify Pop.~III BH-BHs if Pop.~III BH-BHs are
contained in previous and near-future events detected by the current
GW observatories. In section~\ref{sec:FutureGwObservatories}, we
examine how to find Pop.~III BH-BHs by future GW
observatories. Figures~\ref{fig:feature} and \ref{fig:robust} will be
helpful also in this section.

\subsection{Current GW observatories}
\label{sec:CurrentGwObservatories}

In sections~\ref{sec:IdentifyhBH0} and \ref{sec:IdentifyhBHs}, we
discuss about hBH0s and hBH1s/hBH2s, respectively.

\subsubsection{Binaries with two low-mass BHs (hBH0)}
\label{sec:IdentifyhBH0}

As described above, the merger rate of Pop.~III BH-BHs at the present
day (or $\td \sim 10$~Gyr) is $\sim 0.1$~yr$^{-1}$~Gpc$^{-3}$, much
smaller than inferred by LIGO/Virgo observations by two or three
orders of magnitude. However, the merger rate density of Pop.~III
BH-BHs should be largely influenced by the total mass density of
Pop.~III stars, which contains large uncertainties. Hereafter, we
compare Pop.~III hBH0's mass and spin distributions with observed
BH-BHs. These distributions should not be affected by the total mass
density of Pop.~III stars.

Here, we focus on the lower-mass ($\sim 30 \msun$) peak of Pop.~III
hBH0s. As seen in Figure~\ref{fig:robust}, a high fraction of hBH0s
have high $\xeff$ if the peak is present. Since this peak is unique to
Pop.~III hBH0s, it can be a clue to identify Pop.~III BH-BHs. We do
not mention the higher-mass ($\sim 45 \msun$) peak despite that it is
present regardless of different initial conditions ($\qmin$ and
$\amin$) and different stellar evolution models (stellar winds and
natal kicks). This is because it is also predicted for Pop.~I/II
hBH0s.

We compare the $\mbp$ distributions for $\td=10-15$~Gyr with observed
BH-BHs. From GW observations, \cite{2020arXiv201014527A} have inferred
that the merger rate density has a global maximum at
$7.8^{+2.2}_{-2.1}\msun$. No model in this paper indicates such a
global maximum as seen in Figures~\ref{fig:hbh0MassDist_optiq0a1},
\ref{fig:hbh0MassDist_icdependent}, and
\ref{fig:hbh0MassDist_modeldependent}. Pop.~III binaries may not cover
all BH-BHs observed by the current GW observatories.

We examine whether Pop.~III binaries can explain a possible
subcomponent of currently observed BH-BHs. Here, we focus on a peak at
$30-40\msun$ in the merger rate density shown by GW observations
\citep{2020arXiv201014527A}. If we increase the total mass density of
Pop.~III stars by $30-100$ times, Pop.~III merger rate density at the
lower-mass peak (see Figure~\ref{fig:hbh0MassDist_optiq0a1}) can be
matched with observed BH-BH merger rate density around at $\sim
30\msun$ ($\sim 0.3$~yr$^{-1}$~Gpc$^{-3}$~$\msun^{-1}$). However, most
of observed BH-BHs have $\xeff \lesssim 0.4$
\citep{2020arXiv201014527A}, while most of Pop.~III hBH0s around at
the lower-mass peak have $\xeff \sim 1$. It may be difficult to claim
that the observed $30-40\msun$ peak is formed from Pop.~III
hBH0s. Note that BH spins can depend on models of tidal interaction
and mass transfer \citep{2020MNRAS.498.3946K}.

Hereafter, we assess whether Pop.~III hBH0s can explain rare events
discovered in the first half of the third observing run by LIGO and
Virgo. The rare events are GW~190412, GW~190814, and GW~190521.

GW~190412 has $\mbp \sim 30 \msun$, $\mbs \sim 8 \msun$, $\xeff \sim
0.28$, and $|\xbp| \sim 0.31$ \citep{2020PhRvD.102d3015A}. Some
studies have pointed out that the secondary BH may have high spin, and
the primary BH may not have high spin
\citep[e.g.][]{2020ApJ...901L..39O,2020ApJ...897L...7S}. On the other
hand, reanalysis by \cite{2020ApJ...899L..17Z} have supported that the
primary BH is a spinning BH. In either case, Pop.~III hBH0s in our
results cannot explain this event, since Pop.~III hBH0s with $\mbp
\sim 30 \msun$ should have high spins \citep[but
  see][]{2020MNRAS.498.3946K}.

Our supernova model is the rapid model in which BHs in the lower-mass
mass gap are not formed. Thus, our models have no hBH0 like GW~190814
that contains a $2.6 \msun$ compact object
\citep{2020ApJ...896L..44A}. \cite{2020arXiv200713343K} have chosen a
supernova model with the lower mass gap, and have suggested that
GW~190814-like events can happen under Pop.~III environments.

GW~190521 has $\mbp \sim 85 \msun$ and $\mbs \sim 66 \msun$
\citep{2020PhRvL.125j1102A}. Unfortunately, our models cannot form
mass combinations like GW~190521 as seen in
Figure~\ref{fig:sampleMassDist} \citep[but
  see][]{2020MNRAS.tmpL.239F,2020MNRAS.tmpL.234K,2020ApJ...903L..21S,2020ApJ...903L..40L,2020arXiv201007616T}. However,
GW~190521 may be interpreted as a BH-BH with hBH and lBH
\citep{2020ApJ...904L..26F,2020arXiv201012558N}. We discuss about the
possibility that GW~190521 may be a Pop.~III hBH1 in
Section~\ref{sec:IdentifyhBHs}.

We summarize this section. We compare Pop.~III hBH0s with observed
BH-BHs. Even if we match the merger rate density of Pop.~III hBH0s
with that of observed BH-BHs by increasing the Pop.~III formation rate
by two or three orders of magnitude, Pop.~III hBH0s cannot explain a
global maximum at $\sim 8\msun$ in observed BH-BHs. The $\sim 30\msun$
peak of Pop.~III hBH0s may not be consistent with that of observed
BH-BHs, since Pop.~III hBH0s have much higher spins than observed
ones. GW~190412 and GW~190814 are unlikely to be Pop.~III hBH0s
according to our Pop.~III models.

\subsubsection{Binaries including hBHs (hBH1 and hBH2)}
\label{sec:IdentifyhBHs}

\cite{2020arXiv200602211M} have shown that the current GW
observatories are most sensitive to BH-BHs with the total masses of
$\sim 200 \msun$, which is similar to the total masses of hBH1s and
hBH2s. Hence, we discuss about properties of Pop.~III hBH1s and hBH2s,
and their difference from others. We focus on the characteristic
features of Pop.~III hBH1 and hBH2 weakly depending on initial
conditions ($\qmin$ and $\amin$) and stellar evolution models (stellar
winds and natal kicks), as described in Figure~\ref{fig:robust}.

First, we compare Pop.~III BH-BHs with those in globular clusters
(GCs) \citep{2019PhRvD.100d3027R,2020ApJ...900..177K}, nuclear star
clusters (NSCs) at galactic centers
\citep{2016ApJ...831..187A,2020A&A...640L..20B}, open clusters (OCs)
\citep{2020MNRAS.497.1043D,2019arXiv191206022B}, and triple and
quadruple systems (TS and QS, respectively)
\citep{2020ApJ...895L..15F}. In these scenarios, hBHs grow through
stellar or BH mergers. Since these mergers should not be frequent, the
ratio of hBHs to lBHs should not be high. Thus, the ratio of hBH2s to
hBH1s should be much smaller than the Pop.~III ratio. Many discoveries
of hBH1s and hBH2s can distinguish Pop.~III ones from the other
ones. Moreover, Since GC- and NSC-origin ones involve BH mergers, most
of them should have high $\xeff$. On the other hand, at least half of
Pop.~III ones have low $\xeff$.

Gas accretion can increase BH masses up to the mass range of hBHs
\citep[e.g.][]{2019A&A...632L...8R}. This should make hBHs highly
spinning. Thus, hBH1s and hBH2s in this scenario should have high
$\xeff$. On the other hand, at least half of Pop.~III hBH1s and hBH2s
have low $\xeff$. Thus, they can be identified from those formed
through gas accretion.

We compare our BH-BHs with Pop.~III BH-BHs formed dynamically
\citep{2020MNRAS.495.2475L}. Their BH-BHs have $\mbp \gtrsim 250
\msun$ in contrast to our BH-BHs with $\mbp \gtrsim 130 \msun$. This
is because each BH evolves as a single star, and do not lose its mass
through binary interactions, such as mass transfer, common envelope,
and rotationally enhanced stellar winds. Their BH-BHs should have high
$\xeff$ different from Pop.~III BH-BHs, since it gets extra mass
through gas accretion. Eventually, by hBH masses and spins, we can
distinguish whether Pop.~III BH-BHs are formed through binary
evolution or dynamical interaction.

\cite{2019ApJ...883L..27M} have examined merger rates of Pop.~I/II
hBH1s and hBH2s. Their merger rates are $\sim
10^{-2}$~yr$^{-1}$~Gpc$^{-3}$, comparable with merger rates of
Pop.~III hBH1s and hBH2s we estimate. It may be difficult to
distinguish Pop.~III hBH1s and hBH2s from Pop.~I/II ones if we confine
ourselves to discussing about merger events in the current
universe. Although \cite{2019ApJ...883L..27M} have not presented
detail properties, such as $\mbp$, $\mbs$, $\xbp$, and $\xbs$
distribution, these properties may be similar among Pop.~I/II/III
hBH1s and hBH2s, since all of them are formed through binary
evolution.

\cite{2019Natur.575..618L} \citep[see also][]{2020ApJ...900...42L}
have discovered a $70 \msun$ BH in LB-1, and have raised the
possibility that massive BHs can be formed more easily than previously
thought \citep{2020ApJ...900...98G,2020ApJ...890..113B}. Then,
Pop.~I/II hBH1s and hBH2s may be also formed more easily. However,
there are many opposite opinions against the presence of a $70 \msun$
BH in LB-1
\citep{2020A&A...633L...5I,2020A&A...634L...7S,2020MNRAS.493L..22E,2020arXiv200611974E,2020MNRAS.496L...6Y,2020Natur.580E..11A,2020MNRAS.495.2786E,2020PASJ...72...39T,2020A&A...641A..43B}. Thus,
we do not take into account the possibility here.

Finally, we discuss about GW~190521. \cite{2020ApJ...904L..26F} and
\cite{2020arXiv201012558N} have suggested that GW~190521 may be a
hBH1. \cite{2020arXiv201012558N} have also indicated that the heavier
BH is most likely to have high spin magnitude, and large spin-orbit
misalignment ($\gtrsim 90^\circ$). Pop.~III hBH1s are difficult to
explain the spin features of GW~190521. If we include BH natal kicks,
hBHs in hBH1s can have high spins even in the current universe, since
BH natal kicks relaxes the spin-$\td$ relation. However, their
spin-orbit misalignments are $\lesssim 60^\circ$ (see
Figure~\ref{fig:hbh1SpinTiltDist}). We have to adopt $\sigma_{\rm k}
\gg 265$~km~s$^{-1}$ for BH natal kicks in order to form Pop.~III
hBH1s with spins as large as GW~190521.

We summarize this section. Pop.~III BH-BHs with hBHs (hBH1s and hBH2s)
can be identified from those formed through stellar or BH mergers in
GCs, NSCs, OCs, and TS/QSs, or through gas accretion by the ratio of
hBH1s to hBH2s and hBH spins. On the other hand, it is difficult to
distinguish between Pop.~III and Pop.~I/II BH-BHs. Although GW~190521
may consist of hBH and lBH similar to Pop.~III hBH1s, the spin-orbit
misalignment of GW~190521 cannot be explained by Pop.~III hBH1s,
unless we adopt BH natal kicks with $\sigma_{\rm k} \gg
265$~km~s$^{-1}$.

\subsection{Future GW observatories}
\label{sec:FutureGwObservatories}

Future ground-based GW observatories, such as Einstein telescope
\citep{2010CQGra..27s4002P,2020JCAP...03..050M} and Cosmic explorer
\citep{2019BAAS...51g..35R}, can detect hBH0s at the high-redshift
universe. Moreover, future space-borne GW observatories, such as
DECIGO \citep{2006CQGra..23S.125K}, may observe hBH1s and hBH2s at the
high-redshift universe. In this section, we discuss about the $\td$
dependence of Pop.~III hBH0s, hBH1s, and hBH2s, and suggest clues to
identify Pop.~III BH-BHs from other BH-BHs.

First, we state a unique feature to Pop.~III BH-BHs, which are common
with Pop.~III hBH0s, hBH1s, and hBH2s. Since Pop.~III stars are formed
much earlier than Pop.~I/II stars, Pop.~III BH-BHs can be observed at
the higher-redshift universe (say $z \sim 10$) than Pop.~I/II BH-BHs
($z \sim 2$, when stars are most actively formed
\citep{2014ARA&A..52..415M}). If Pop.~III BH-BHs dominate observed
BH-BHs contrary to our results, the BH-BH merger rate density should
have $\Gamma \propto \td^{-1}$ beyond $z \sim 2$ regardless of initial
conditions and stellar models (see Figure~\ref{fig:delayTimeDist}).

Pop.~III hBH0s have two time-evolving features in the optiq0.0a1e1
model (see also Figure~\ref{fig:feature}):
\begin{itemize}
\item Time evolution of the lower-mass peak in BH mass distribution
  from $20\msun$ to $35\msun$.
\item High fraction of high effective spins ($\sim 1$) around at the
  lower-mass peak.
\end{itemize}

These features can be strong evidence for the presence of Pop.~III
BH-BHs, even if Pop.~III BH-BHs are not dominant GW sources. This
$\td$ dependence is quite different from BH-BHs formed in other
places. \cite{2020ApJ...898..152S} do not find the cosmic evolution of
mass distributions of BH-BHs formed through binary evolution and open
clusters. Although BH-BH masses can evolve if GC-origin BH-BHs are
dominant, their masses should decrease with time, since more massive
BHs join in BH-BHs and merge at an earlier time
\citep[e.g.][]{2013MNRAS.435.1358T}.

Pop.~III hBH1s and hBH2s have two time-evolving features in the
optiq0.0a1e1 model (see also Figure~\ref{fig:feature}):
\begin{itemize}
\item Decreasing of the maximum $\mbp$ with $\td$ for hBH1s.
\item Larger maximum $\mbp$ mass for $\td \gtrsim 0.1$~Gyr than for
  $\td \lesssim 0.1$~Gyr for hBH2s.
\end{itemize}

These features can be clues to identify Pop.~III hBH1s and
hBH2s. Since the Pop.~III formation epoch is much earlier than those
of NSCs, GCs, OCs, TC/QCs, and Pop.~I/II binaries, the Pop.~III BH
mass evolves earlier.

We summarize this section. The time evolution of the merger rate
densities of hBH0s, hBH1s, and hBH2s can be a clue to identify
Pop.~III BH-BHs. If Pop.~III BH-BHs are dominant, the BH-BH merger
rate density keeps $\Gamma \propto \td^{-1}$ beyond $z \sim 2$
regardless of initial conditions and stellar models. For an ideal
condition (i.e. the optiq0.0a1e1 model), the time evolution of BH mass
distributions can be clues to identify Pop.~III BH-BHs. Pop.~III hBH0s
have lBHs with larger mass for larger $\td$, and with high spin $\sim
1$. The maximum $\mbp$ mass of Pop.~III hBH1s and hBH2s evolves
earlier than those of other scenarios. No other scenario has suggested
such time evolution.

\section{Discussion}
\label{sec:Discussion}

\subsection{Chemically homogeneous Pop.~III stars}

So far, we assume that Pop.~III stars have small stellar rotations at
their ZAMS times. However, they can have large stellar rotations.
In fact, poloidal magnetic field around Pop.~III stars is
  weaker than the turbulent field and toroidal field
  \citep{2020MNRAS.497..336S}. Thus the magnetic breaking effect to
  spin down the protostar could be inefficient. Furthermore, the
  abundant Nitrogen in EMP stars suggests that the Pop~III stars are
  fast rotators \citep[e.g.][]{2019A&A...632A..62C}.  If they have
large stellar rotations at their ZAMS times, they can switch on CHEs
\citep[e.g.][for Pop.~III stars]{2012A&A...542A.113Y}. Here, we make
two simple models for Pop.~III binary evolutions under CHEs, and
estimate BH-BH merger rates. In the first model, we assume that a
whole star evolves to a nHe star without stellar wind mass loss, and
that a binary does not change its orbit without any interactions, such
as tidal interaction, mass transfer and common envelope. The
assumption of no binary interaction can be justified by the fact that
the stars enter into CHEs due to their initial rotations (not due to
tidal spinup), and do not expand so much. Such a resulting nHe star
experiences supernovae and PPI/PISN described in
section~\ref{sec:SingleStarModel}, and leaves NS, BH, or no
remnant. Since a nHe star with $\lesssim 10\msun$ and $45-65 \msun$
loses a part of its mass at supernova, the binary orbit should be
changed. However, we ignore this effect for simplicity. In the second
model, we assume that a star evolves to a nHe star, losing $30$~\% of
its ZAMS mass due to stellar winds, which is roughly consistent with
results of \cite{2012A&A...542A.113Y}. We adiabatically change binary
orbital parameters through the stellar winds so that the binary period
is inversely proportional to square of the total mass, and the
eccentricity is constant. The nHe star experiences supernovae in the
same way as the first model, and the supernovae do not change the
binary orbital parameters.

\begin{figure}[ht!]
  \plotone{\fdir/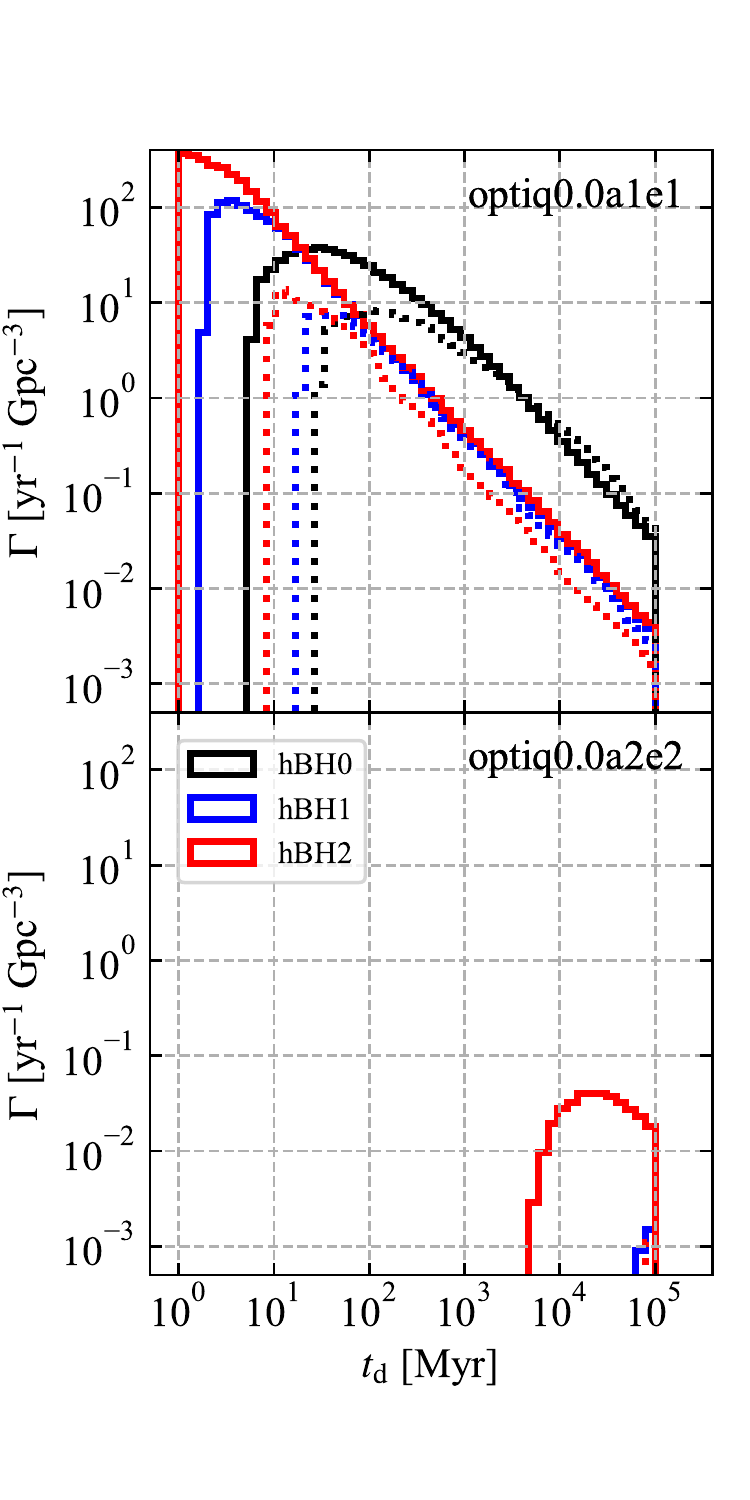}
  \caption{Merger rate densities of Pop.~III hBH0s, hBH1s, and hBH2s
    as a function of $\td$ when all the Pop.~III stars experience CHEs
    with initial conditions of the optiq0.0a1e1 and optiq0.0a2e2
    models (the top and bottom panels, respectively). Solid and dashed
    curves indicate the first and second models, respectively.}
  \label{fig:cheDelayTime}
\end{figure}

Figure~\ref{fig:cheDelayTime} shows merger rate densities of Pop.~III
hBH0s, hBH1s, and hBH2s as a function of $\td$ when all the Pop.~III
stars are CHEs. If $\amin=10\rsun$, the hBH0, hBH1, and hBH2 merger
rates for the CHE case are comparable to, or more than those for the
non-CHE cases in the present-day universe ($\td = 10-15$~Gyr). Since
we ignore stellar mergers before stars collapse to BHs, we
overestimate the merger rates for $\td \lesssim 0.1$~Gyr. However,
binaries forming BH-BHs with $\td \sim 10$~Gyr have much larger
semi-major axes than nHe star radii, and thus they should not
experience stellar mergers before their members collapse to BHs. Thus,
the merger rates for $\td \sim 10$~Gyr should be correct. If
$\amin=200\rsun$, hBH0s and hBH1s never merge at any time. If we take
into account stellar wind mass loss, even hBH2s have no chance to
merge for $\td \le 15$~Gyr. This shows that $\amin$ plays a crucial
role if Pop.~III stars have large stellar rotations, and enter into
CHEs. Here, we stop discussing about CHEs. It is beyond of the scope
of this paper to determine $\amin$ and whether Pop.~III stars enter
into CHEs or not.

\begin{figure*}[ht!]
  \plotone{\fdir/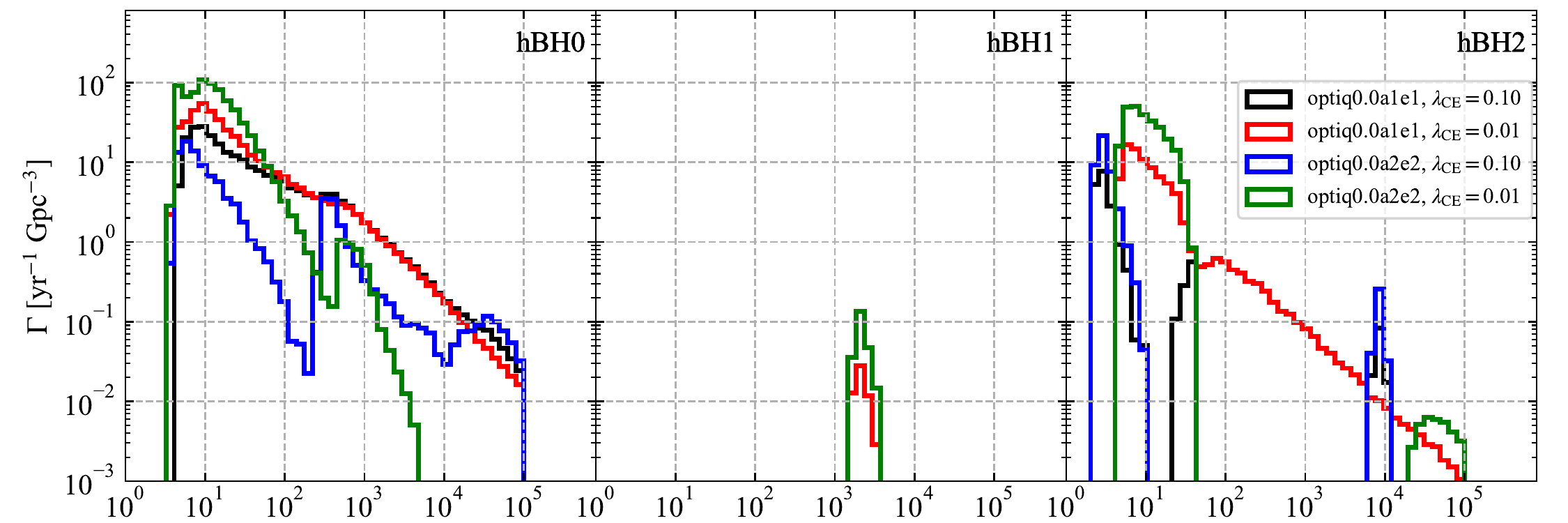}
  \caption{Merger rate densities of hBH0, hBH1, and hBH2 as a function
    of the delay time in the optiq0.0a1e1 and optiq0.0a2e2 models,
    where we set $\lambda_{\rm CE}=0.1$ and $0.01$. The left, middle,
    and right panels indicate the merger rate densities of hBH0s,
    hBH1s, and hBH2s, respectively.}
  \label{fig:delayTimeLamb}
\end{figure*}

\subsection{Sensibility to $\lambda_{\rm CE}$}

We set $\lambda_{\rm CE}=1$ in section~\ref{sec:Results}. However,
$\lambda_{\rm CE}$ is quite uncertain, and \cite{2020arXiv200611286K}
have shown $\lambda_{\rm CE}$ can be $0.01$. For the optiq0.0a1e1 and
optiq0.0a2e2 models, we change $\lambda_{\rm CE}$ to $0.1$ and $0.01$,
and perform binary population synthesis
calculations. Figure~\ref{fig:delayTimeLamb} shows their merger rate
densities. We can find that $\lambda_{\rm CE}$ has small effects on
the merger rates of hBH0s and hBH2s in the optiq0.0a1e1 models (see
also Figure~\ref{fig:delayTimeDist}). This is because a significant
part of hBH0s and hBH2s in these models are formed through the CE0
channel not involving common envelope evolution. On the other hand,
hBH1s in all the models, and hBH0s and hBH2s in the optiq0.0a2e2
models are drastically decreased with $\lambda_{\rm CE}$
decreasing. This is because they are formed through channels involving
common envelope evolution. \cite{2020arXiv200611286K} have also
reported that common envelope evolution is harder to take place than
previously thought, when a donor star is massive ($\ge 40 \msun$), and
a BSG star. If so, the dependence on $\lambda_{\rm CE}$ would become
more strong.

\subsection{BHs in the PI mass gap}

In section~\ref{sec:Results}, we obtain BHs in the PI mass gap,
however their masses are at most $\sim 55 \msun$. We examine initial
conditions and single star models in which BHs with $\sim 80 \msun$
emerge. Note that the possible maximum mass of BHs in the PI mass gap
is $\sim 85 \msun$ in our single star model (see
Figure~\ref{fig:remnantMass}). We modify the optiq0.0a1e1 model, and
prepare the following initial conditions. We extend the maximum
semi-major axis from $2000 \rsun$ to $10^5 \rsun$, and account for
natal kicks with $\sigma_{\rm k}=50$~km~s$^{-1}$. We call this
parameter set ``PI-mass-gap'' model.

\begin{figure}[ht!]
  \plotone{\fdir/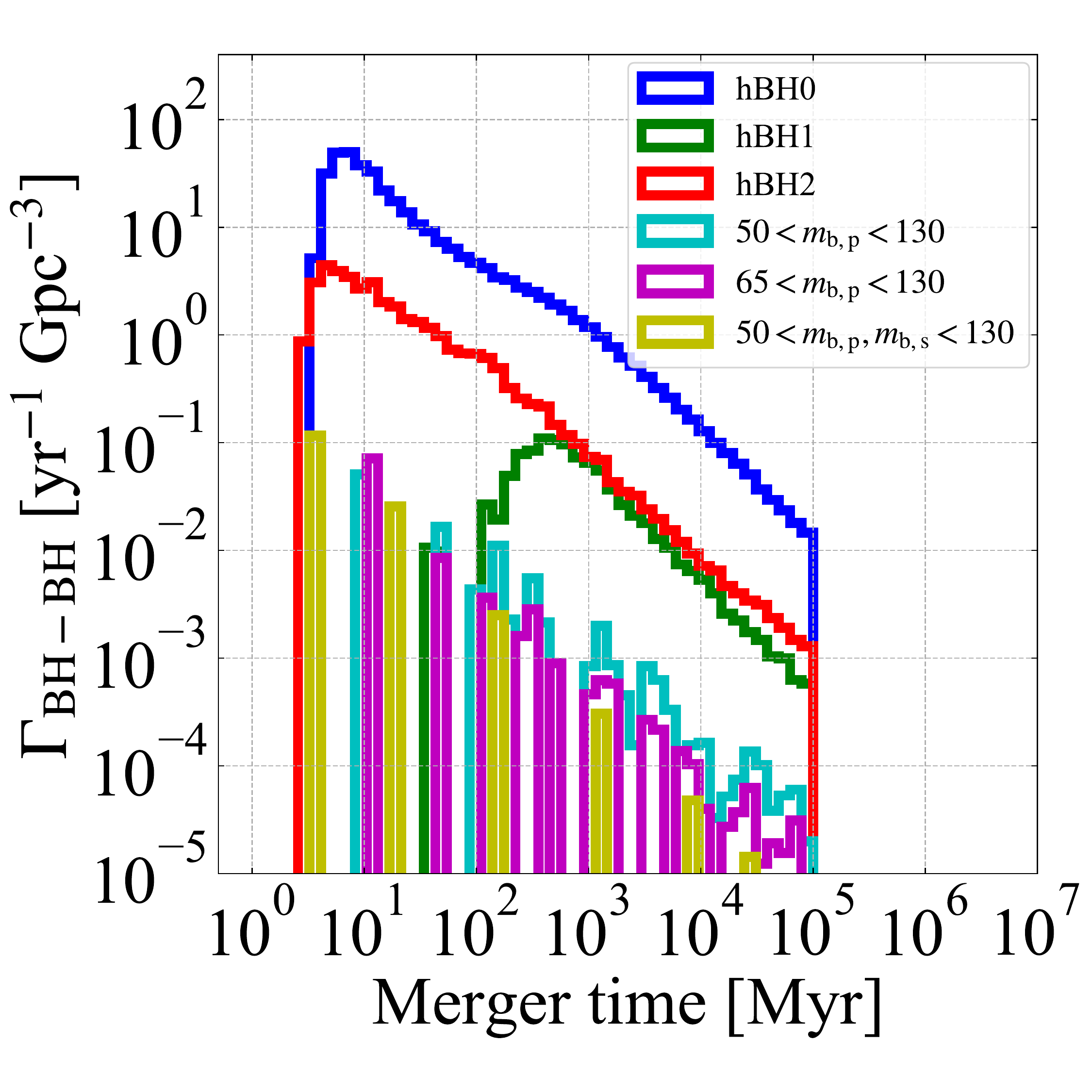}
  \caption{Merger rate densities of Pop.~III BH-BHs as a function of
    $\td$ in the PI-mass-gap model. Here, the maximum mass of lBHs is
    defined as $50 \msun$, and the minimum mass of hBHs is defined as
    $130 \msun$.}
  \label{fig:hmgDelayTime}
\end{figure}

Figure~\ref{fig:hmgDelayTime} shows merger rate densities of Pop.~III
BH-BHs as a function of $\td$ in the PI-mass-gap model. The merger
rate densities of hBH0s, hBH1s, and hBH2s are similar to those in the
optiq0.0a1e1 model (see black curves in the upper panels of
Figure~\ref{fig:delayTimeDist}). We can find BH-BHs with BHs in the PI
mass gap. Their merger rate densities of $\mbp>50 \msun$ and $\mbp>65
\msun$ are $\sim 10^{-4}$~yr$^{-1}$~Gpc$^{-3}$ for $\td =
10$~Gyr. Even Pop.~III BH-BHs with two BHs in the PI mass gap are
present, although their merger rate is quite small. Among the main
$16$ model, the kickq0.9a2e2 model has the most BH-BHs with BHs in the
PI mass gap. The merger rate density of $\mbp>50 \msun$ is comparable
to that of the PI-mass-gap model. On the other hand, the merger rate
density of $\mbp>65 \msun$ is much smaller than that of the
PI-mass-gap model. Actually, the number of the BH-BHs is too small to
calculate the merger rate density. In contrast to the main $16$
models, the PI-mass-gap model can have plenty of BH-BHs with BHs in
the PI mass gap ranging from $50 \msun$ to $85 \msun$.

\begin{figure}[ht!]
  \plotone{\fdir/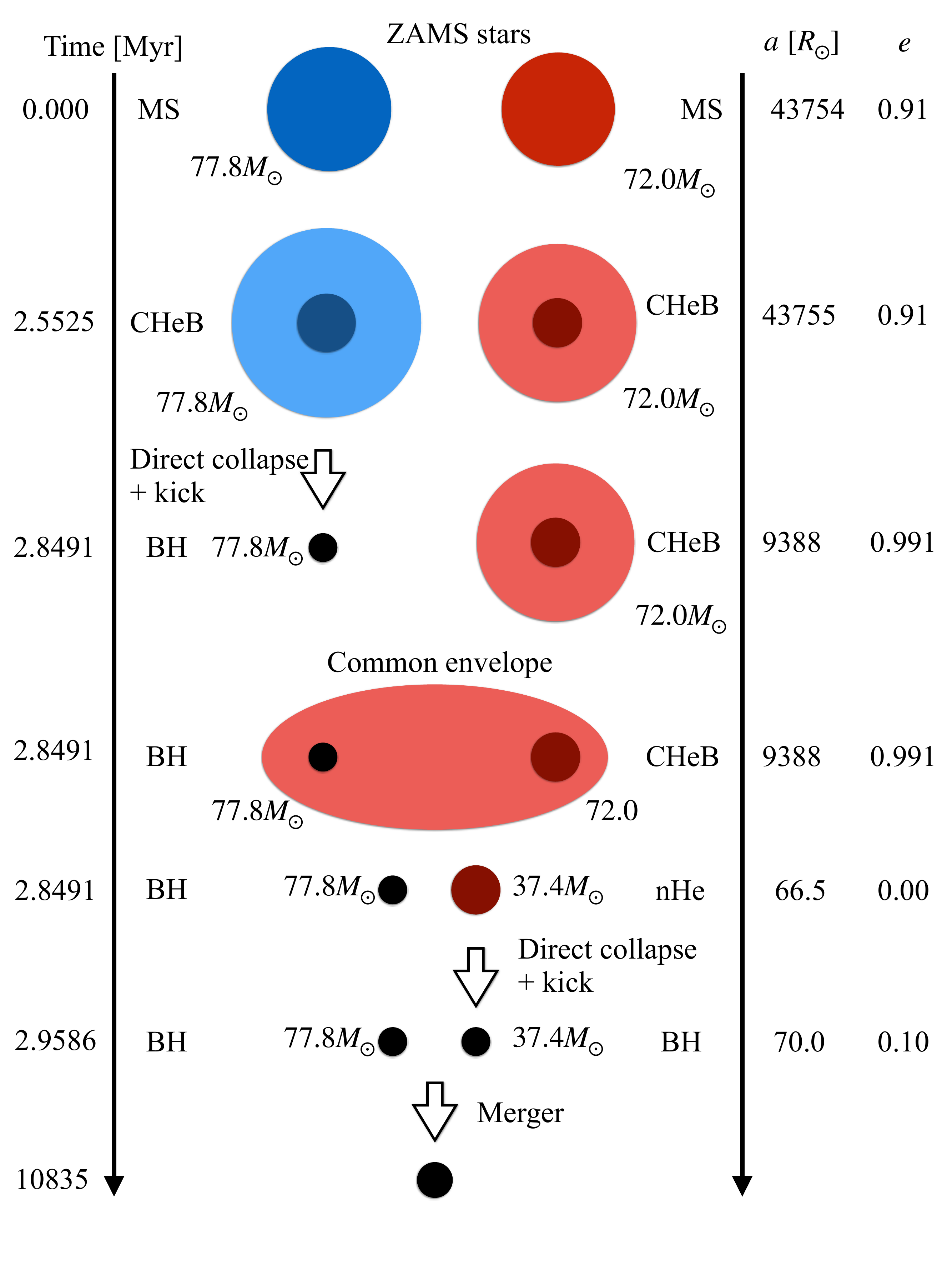}
  \caption{Example binary evolution leading to a BH-BH containing a BH
    in the PI mass gap.}
  \label{fig:hmgchannel}
\end{figure}

We draw a schematic picture of a formation channel for BH-BHs with BHs
in the PI mass gap in Figure~\ref{fig:hmgchannel}. There are two key
points. First, the two stars do not fill their Roche lobes until the
1st-evolving star collapses to the 1st BH. Thus, the 1st-evolving star
keeps its H envelope, avoids PPI/PISN effects owing to its small He
core with $< 45 \msun$, and leaves a BH in the PI mass gap (see also
Figure~\ref{fig:remnantMass}). Note that the 1st-evolving star is
mildly spun up by the tidal field of the 2nd-evolving star, however it
loses little mass even if we consider rotationally enhanced stellar
winds. Second, a natal kick exerted on the 1st BH makes the binary
eccentricity extremely high ($\sim 0.99$), since the natal kick
velocity is comparable to the orbital velocity of the binary. Then,
the 2nd-evolving star expands, touches the 1st BH, and drives common
envelope evolution. These two points finally make compact BH-BHs.

The PI-mass-gap model can form BH-BHs whose masses are similar to
GW~190521's masses. However, their merger rate density of $\sim
10^{-4}$~yr$^{-1}$~Gpc$^{-3}$ is much smaller than GW~190521-like one
$\sim 10^{-1}$~yr$^{-1}$~Gpc$^{-3}$ \citep{2020PhRvL.125j1102A}. If we
increase the total Pop.~III mass density in the local universe by 3
orders of magnitude, we can match their merger rate density with the
observed one. However, as a side effect, the merger rate densities of
hBH0s, hBH1s, and hBH2s are also increased to $\sim 10^2$, $\sim 10$,
and $\sim 10$~yr$^{-1}$~Gpc$^{-3}$. These are not consistent with the
observed merger rate density of hBH0s, and the upper limit of the
merger rate densities of hBH1s and hBH2s derived by
\cite{2020arXiv200602211M}.  We conclude that the PI-mass-gap model
cannot explain GW~190521-like events.

\subsection{Pop.~III hierarchical triple and quadruple systems}

\cite{2020ApJ...892L..14S} have shown possibility that Pop.~III stars
can be usually born as stable multiple systems. There may be many
interesting respects in stable multiple systems: the formation of BHs
in the PI mass gap \citep{2020ApJ...895L..15F,2020PASJ...72...39T},
excitation of binary eccentricity through Kozai-Lidov mechanism
\citep{1962AJ.....67..591K,1962P&SS....9..719L}, and so on. However,
initial conditions of Pop.~III multiple systems are much less clear
than those of binary stars. We will study Pop.~III multiple systems as
GW sources in future.

\section{Summary}
\label{sec:Summary}

We perform binary population synthesis calculations to obtain the
merger rate density of Pop.~III BH-BHs by means of the {\tt BSE} code
with extensions to very massive Pop.~III stars. Our {\tt BSE} code has
several novel features. The extensions enable us to follow Pop.~III
stars up to $300 \msun$, which include stars overcoming PISN and
leaving BHs above the PI mass gap. We take into account PPI/PISN
effects for binary population synthesis of Pop.~III stars. We
implement rotationally enhanced stellar wind into our {\tt BSE}
code. In our {\tt BSE} code, natal kicks tilt BH spin vectors from
binary orbit vectors, and the angles can be recorded. We use our {\tt
  BSE} code, and investigate properties of Pop.~III BH-BHs, and their
dependence on single star models (natal kicks and stellar winds), and
initial conditions (the minimum mass ratio $\qmin$ and the minimum
binary separation $\amin$, see Section~\ref{sec:InitialConditions}).

We can categorize Pop.~III BH-BHs into three subpopulations: hBH0s,
hBH1s, and hBH2s (see Figure~\ref{fig:sampleMassDist}). This is
because the PPI/PISN effects make the PI mass gap between $\sim 50 -
130 \msun$. Provided that the mass density of Pop.~III stars is $\sim
10^{13} \msun$~Gpc$^{-3}$, the merger rates of hBH0s and the sum of
hBH1s and hBH2s are $\sim 0.1$~yr$^{-1}$~Gpc$^{-3}$ and $\sim
0.01$~yr$^{-1}$~Gpc$^{-3}$, respectively (see
Figure~\ref{fig:delayTimeDist}). These merger rates are independent of
single star models and initial conditions. Since Pop.~III BH-BHs have
several formation channels, initial conditions have small effects on
their merger rates (see Figure~\ref{fig:icCeMergBh}).  The natal kicks
we adopt have velocities comparable to or smaller than internal
velocities of Pop.~III BH-BHs, and thus they do not affect much the
merger rates. Pop.~III stars lose their H envelopes through common
envelope evolution as well as stellar winds. The presence of stellar
winds has minor effects on Pop.~III BH masses.

We summarize features of Pop.~III hBH0s (see also
Figures~\ref{fig:feature} and \ref{fig:robust}). The merger rate
density of Pop.~III hBH0s has the higher-mass ($\sim 45 \msun$) and
lower-mass ($\sim 20 - 35 \msun$) peaks in the optiq0.0a1e1 model. The
lower-mass peak shifts from ${\sim} 20 \msun$ to ${\sim} 35 \msun$
with $\td$. Most of BHs in the lower-mass peak have high
spins. Although the lower-mass peak is unique to Pop.~III hBH0s, it is
fragile against initial conditions and stellar evolution models. The
characteristic feature of the lower-mass peak is that BHs around the
peak have high spins if the peak is present.

Although their merger rate may be much smaller than inferred from the
LIGO/Virgo observations, we could identify Pop.~III hBH0s from
observed BH-BHs, in an ideal case where the initial conditions are
similar to the optiq0.0a1e1 model. The clue is the lower-mass peak,
and highly-spinning BHs around at the lower-mass peak.  Future GW
observatories may detect conclusive Pop.~III BH-BHs, a group of BH-BHs
with high spins, and with peaks which shift from light to heavy with
$\td$. The above discussions can be also seen in
Section~\ref{sec:IdentifyhBH0}.

We describe characteristic features of Pop.~III hBH1s and hBH2s weakly
depending on initial conditions ($\qmin$ and $\amin$) and stellar
evolution models (stellar winds and natal kicks) as seen in
Figures~\ref{fig:feature} and \ref{fig:robust}. The rate ratio of
Pop.~III hBH2s to Pop.~III hBH1s is $\gtrsim 0.1$. Both of Pop.~III
hBH1s and hBH2s have two features: (i) $\mbp \gtrsim 130 \msun$, and
(ii) a high fraction of low effective spins ($\lesssim 0.2$).

Pop.~III hBH1s and hBH2s can be discovered soon by the current GW
observatories. The above features can be clues to identify Pop.~III
hBH1s and hBH2s from others. Future GW observatories should find their
merger rates are clearly $\propto \td^{-1}$ beyond redshift of $\sim
2$ if Pop.~III hBH1s and hBH2s dominate BH-BHs in this mass range. We
make detail discussions in Sections~\ref{sec:IdentifyhBHs} and
\ref{sec:FutureGwObservatories}.

Pop.~III BHs in the PI mass gap can merge if Pop.~III binaries have
initial semi-major axes of $\sim 10^4 \rsun$, and BHs have natal kicks
of $\sim 50$~km~s$^{-1}$ (see Figure~\ref{fig:hmgchannel}). However,
their merger rate is smaller than that of GW~190521-like BH-BHs
inferred by GW observations by $3$ orders of magnitude. If we match
their merger rate with the observed merger rate of GW~190521-like
BH-BHs, the total merger rate of Pop.~III BH-BHs exceeds the total
merger rate of observed BH-BHs. Thus, this formation process is not
appropriate for the formation of GW~190521-like BH-BHs.

In the worst scenario, the merger rates of Pop.~III hBH0s, hBH1s, and
hBH2s can be nearly zero (see Figure~\ref{fig:cheDelayTime}). The
worst scenario is realized if the two following conditions are
satisfied at the same time. First, Pop.~III stars form only wide
binaries with separations of $\gtrsim 100 \rsun$, which is plausible,
since Pop.~III stars expand to $\sim 100 \rsun$ in their protostellar
phases.  Second, Pop.~III stars have enough fast rotations to excite
CHEs from the beginning time, which is supported by the theoretical
study of Pop~III star formation as well as the observation of EMP
stars. In order to assess whether these conditions are satisfied, or
not, we need extremely high-resolution and long-term hydrodynamical
simulation of Pop.~III star formation. Such simulations should
consider chemical evolution, radiative transfer, magnetic fields, and
so on, and would be computationally prohibitive at the current state.
We expect that computational technology rapidly evolves to achieve the
above simulations, or that conclusive detections of Pop.~III BH-BHs
conversely rule out the worst scenario.

Aside from the worst scenario, we can expect that Pop.~III hBH1s and
hBH2s can be detected within a few years. Conversely, the
non-detection would mean that the number of Pop.~III stars overcoming
PISNe is quite small. Thus, we may constrain the Pop.~III IMF in the
mass range $\gtrsim 100 \msun$ in the near future.

\acknowledgments

We thanks the anonymous referee for helpful comments. AT thanks
Tilman Hartwig and Christopher Berry for fruitful discussions.  This
research has been supported by Grants-in-Aid for Scientific Research
(17H01101, 17H01130, 17H02869, 17H06360, 17K05380, 19K03907) from the
Japan Society for the Promotion of Science.


\end{document}